\newcommand*{\ATLASLATEXPATH}{latex/}
\pdfoutput=1
\documentclass[cernpreprint,UKenglish,texlive=2011,txfonts]{\ATLASLATEXPATH atlasdoc}
\usepackage{\ATLASLATEXPATH atlaspackage}
\usepackage{\ATLASLATEXPATH atlasbiblatex}

\usepackage[utf8]{inputenc} %% ERAM added this due to bibtex entry for Atlas Z' paper Aad:2014cka

\usepackage{rotating} % allow sideways table

\usepackage{\ATLASLATEXPATH atlascontribute}

\usepackage{\ATLASLATEXPATH atlasphysics}
\usepackage{lldefs}
\usepackage{verbatim}

\graphicspath{{logos/}{figures/}}

\AtlasTitle{
Measurement of the
double-differential high-mass Drell--Yan cross section
in $pp$ collisions at $\sqrt{s}= 8$~TeV with the ATLAS detector}

\author{The ATLAS Collaboration}

\AtlasContributor{Markus Zinser}{Electron channel analysis,electron / muon combination}
\AtlasContributor{Rob Hickling}{Muon channel analysis}
\AtlasContributor{Lewis Armitage}{Muon cross checks, ntuple production}
\AtlasContributor{Frank Ellinghaus}{Electron channel lead, editor}
\AtlasContributor{Eram Rizvi}{Muon channel lead, editor}
\AtlasContributor{Misha Lisovyi}{Theory predictions}
\AtlasContributor{Uta Klein}{NNLO $K$ factors}
\AtlasContributor{Kristin Lohwasser}{PDF sensitivity study}
\AtlasRefCode{STDM-2014-06}

\AtlasNote{ATL-COM-PHYS-2014-376}

\PreprintIdNumber{CERN-PH-2016-079}

\AtlasJournal{JHEP}
\AtlasAbstract{
This paper presents a measurement of the double-differential cross
section for the Drell--Yan $Z/\gamma^*\rightarrow \ell^+\ell^-$ and
photon-induced $\gamma \gamma \rightarrow \ell^+\ell^-$ processes
where $\ell$ is an electron or muon.  The measurement is performed for
invariant masses of the lepton pairs, $m_{\ell\ell}$, between
$116$~GeV and $1500$~GeV using a sample of $20.3$~fb$^{-1}$ of $pp$
collisions data at centre-of-mass energy of $\sqrt{s}=8$~TeV collected
by the ATLAS detector at the LHC in 2012.  The data are presented
double differentially in invariant mass and absolute dilepton rapidity
as well as in invariant mass and absolute pseudorapidity separation of
the lepton pair. The single-differential cross section as a function
of $m_{\ell\ell}$ is also reported. The electron and muon channel
measurements are combined and a total experimental precision of better
than $1\%$ is achieved at low $m_{\ell\ell}$.  A comparison to
next-to-next-to-leading order
perturbative QCD predictions using several recent parton distribution
functions and including next-to-leading order electroweak effects indicates the
potential of the data to constrain parton distribution functions.  In
particular, a large impact of the data on the photon PDF is
demonstrated.  }

\AtlasCoverSupportingNote{Measurement of the high-mass Drell--Yan double-differential cross section in $pp$ collisions at $\sqrt{s}$ = 8 TeV}{http://cds.cern.ch/record/1697711}
\AtlasCoverCommentsDeadline{24 March 2016}

\AtlasCoverAnalysisTeam{Markus Zinser , Rob Hickling , Lewis Armitage
, Frank Ellinghaus , Eram Rizvi , Misha Lysovyi , Uta Klein, Kristin Lohwasser, Juan Rojo}

\AtlasCoverEdBoardMember{Samira~Hassani~(chair), Kunihiro Nagano ,
Noam Hod , Benjamin Trocme}

\AtlasCoverEgroupEditors{atlas-stdm-2014-06-editors@cern.ch}

\AtlasCoverEgroupEdBoard{atlas-stdm-2014-06-editorial-board@cern.ch}

\hypersetup{pdftitle={ATLAS draft},pdfauthor={The ATLAS Collaboration}}

\begin{document}

\maketitle

\tableofcontents

% List of contributors - print here or after the Bibliography.
%\PrintAtlasContribute{0.30}
%\clearpage

%-------------------------------------------------------------------------------
\section{Introduction}
\label{sec:intro}
%-------------------------------------------------------------------------------

The Drell--Yan (DY) process~\cite{DrellYan} of lepton pair production
in hadronic interactions, $pp \rightarrow Z/\gamma^* + X$ with
$Z/\gamma^*\rightarrow \ell^+\ell^-$, is a powerful tool in
understanding the nature of partonic interactions and of hadronic
structure in detail. The study of this process has been fundamental in
developing theoretical perturbative calculations of
quantum chromodynamics (QCD) which are now performed at
next-to-next-to-leading-order (NNLO)
accuracy~\cite{Hamberg:1990np,Catani:2009sm,Catani:2007vq,FEWZ4}.
Measurements from the Large Hadron Collider (LHC) of neutral- and
charged-current Drell--Yan processes mediated by $Z/\gamma^*$ and $W$
exchange respectively at centre-of-mass energies of
$\sqrt{s}=7$~\TeV\ and $8$~\TeV\ have been recently published by the
ATLAS~\cite{STDM-2011-06,STDM-2012-10,STDM-2011-41},
CMS~\cite{CMS-EWK-10-005,CMS-EWK-10-007,CMS-SMP-13-003,CMS-SMP-14-003}
and
LHCb~\cite{LHCb-wz,Aaij:2012mda,Aaij:2015vua,Aaij:2015gna,Aaij:2015zlq}
collaborations. These data provide new constraints on the parton
distribution functions (PDFs) of the proton, some of which have been
used in recent global PDF
fits~\cite{Dulat:2015mca,Ball:2014uwa,Harland-Lang:2014zoa}.

Although on-shell $Z$ and $W$ boson measurements provide the greatest
experimental precision, they are restricted in the kinematic range of
partonic momentum fraction $x$, and four-momentum transfer
$Q=m_{\ell\ell}$, the invariant mass of the dilepton pair. Off-shell
measurements provide complementary constraints in a wider range of $x$
and $Q$. In the neutral-current case, the off-shell measurements are
dominated by the electromagnetic quark couplings to the virtual photon
$\gamma^*$, whereas the on-shell measurements are dominated by the
weak axial and vector couplings of the quarks to the $Z$
boson. Therefore, the measurements have different sensitivity to the
up-type and down-type quarks. At large $m_{\ell\ell}$ the measurements
offer constraints on the large-$x$ antiquark PDFs which are poorly
known. In addition, off-shell measurements may also be sensitive to
the largely unconstrained photon
PDF~\cite{MRST2004QED,STDM-2011-41,Ball:2013hta,STDM-2012-10} through
the photon-induced (PI) process $\gamma \gamma \rightarrow
\ell^+\ell^-$.

Neutral-current DY data at higher masses can also be used to determine
the running of the electroweak (EW) gauge couplings above the weak
scale, and to set model-independent limits on new states with
electroweak quantum numbers~\cite{Alves:2014cda}.  In particular, at
the highest invariant masses accessible at the LHC, the observed
dilepton spectrum may be sensitive to new physics, which could
manifest itself as a resonance or
a broad modification to the continuum spectrum.
Such searches performed by the ATLAS and CMS
experiments~\cite{CMS-EXO-12-061,EXOT-2012-23,EXOT-2013-19} have so
far not found any significant deviations from the Standard Model, and
the largest systematic uncertainty on the derived exclusion limits
arises from the lack of knowledge of the PDFs at high $x$. Since at
leading order the parton momentum fractions from the two protons (1 or
2) are given by $x_{1,2}=\left(m_{\ell\ell}/\sqrt{s}\right){\rm
  e}^{\pm y_{\ell\ell}}$, where $y_{\ell\ell}$ is the dilepton
rapidity, it can be seen that the large $x$ region is accessible at
large $m_{\ell\ell}$ in the case of central production
$(y_{\ell\ell}=0)$, as well as at lower $m_{\ell\ell}$ and large
$y_{\ell\ell}$. Therefore, a double-differential measurement of the
Drell--Yan cross section in $m_{\ell\ell}$ and $y_{\ell\ell}$ provides
PDF constraints in a new kinematic region which is expected to be
unaffected by the manifestation of potential new physics at the
highest invariant mass.

This article reports two inclusive double-differential cross-section
measurements for the process $pp\rightarrow\ell^+\ell^- + X$. The
first measurement is reported as a function of $m_{\ell\ell}$ and
absolute dilepton rapidity $|y_{\ell\ell}|$, and the second as a
function of $m_{\ell\ell}$ and absolute dilepton pseudorapidity
separation $|\Delta\eta_{\ell\ell}|$. These measurements are
sensitive to the proton PDFs, the PI process, and higher-order
electroweak corrections, which have different kinematic
dependencies. In particular, the $t$-channel PI process is expected to
contribute at large $|\Delta\eta_{\ell\ell}|$, small $|y_{\ell\ell}|$
and large $m_{\ell\ell}$. Therefore, measurements as a function of
various kinematic distributions are needed to disentangle the
different contributions~\cite{Boughezal:2013cwa}.  For completeness
the inclusive single-differential measurement ${\rm d}\sigma/{\rm
  d}m_{\ell\ell}$ is also provided. The measurements are performed
using $pp$ collision data collected at $\sqrt{s}=8$~\TeV\ in both
electron and muon channels. The data cover the kinematic region of
$116 \leq m_{\ell\ell} \leq 1500$~\GeV\, and access partonic momentum
fractions from $10^{-3}$ up to $x\sim 1$. The integrated luminosity of
the data sample is $20.3$~fb$^{-1}$, a factor five larger than used in
the previous ATLAS measurement~\cite{STDM-2012-10} at
$\sqrt{s}=7$~\TeV\ performed in the electron channel only. Therefore,
the results reported here have a substantially better precision than
earlier results.

%-------------------------------------------------------------------------------
\section{ATLAS detector}
\label{sec:detector}
%-------------------------------------------------------------------------------

The \mbox{ATLAS} detector~\cite{PERF-2007-01} consists of an inner
tracking detector (ID) surrounded by a thin superconducting
solenoid, electromagnetic and hadronic calorimeters, and a muon
spectrometer (MS).  Charged particles in the
pseudorapidity\footnote{ATLAS uses a right-handed coordinate system
  with its origin at the nominal interaction point in the centre of
  the detector and the $z$-axis along the beam pipe. The $x$-axis
  points from the interaction point to the centre of the LHC ring, and
  the $y$-axis points upward. Cylindrical coordinates $(r,\phi)$ are
  used in the transverse plane, $\phi$ being the azimuthal angle
  around the beam pipe. The pseudorapidity is defined in terms of the
  polar angle $\theta$ as $\eta=-\ln\tan(\theta/2)$.}  range $|\eta| <
2.5$ are reconstructed with the ID, which consists of layers of
silicon pixel and microstrip detectors and a straw-tube
transition-radiation tracker having coverage within $|\eta| < 2.0$.
The ID is immersed in a 2~T magnetic field provided by the solenoid.
The latter is surrounded by a hermetic calorimeter that covers $|\eta|
< 4.9$ and provides three-dimensional reconstruction of particle showers.
The electromagnetic calorimeter is a liquid-argon sampling
calorimeter, which uses lead absorbers for $|\eta| < 3.2$ and copper
absorbers in the very forward region.  The hadronic sampling
calorimeter uses plastic scintillator tiles as the active material and
steel absorbers in the region $|\eta| < 1.7$.  In the region $1.5 <
|\eta| < 4.9$, liquid argon is used as active material, with copper
or/and tungsten absorbers.  Outside the calorimeters, air-core toroids
supply the magnetic field for the MS.  There, three stations of
precision chambers allow the accurate measurement of muon track
curvature in the region $|\eta| < 2.7$.  The majority of these
precision chambers are composed of drift tubes, while cathode-strip
chambers provide coverage in the inner stations of the forward region
for $2.0 < |\eta| < 2.7$.  Additional muon chambers installed between
the inner and middle stations of the forward region and commissioned
prior to the 2012 run improve measurements in the transition region of
$1.05<|\eta|<1.35$ where the outer stations have no coverage.  Muon
triggering is possible in the range $|\eta| < 2.4$, using
resistive-plate chambers in the central region 
that also provide a measurement of the coordinate out of
 the bending plane,
and thin-gap chambers
in the forward region.  A three-level trigger
system~\cite{PERF-2011-02} selects events to be recorded for offline
analysis.

%-------------------------------------------------------------------------------
\section{Simulated event samples}
\label{sec:MC}
%-------------------------------------------------------------------------------

Monte Carlo (MC) simulation samples are used to model the expected
signal and background yields, with the exception of certain
data-driven background estimates.  The MC samples are normalised using
the highest-order cross-section predictions available in perturbation
theory.

The DY process is generated at next-to-leading order (NLO) using
\powheg~\cite{Nason:2004rx,Frixione:2007vw,Alioli:2008gx,Alioli:2010xd}
and the CT10 PDF~\cite{CT10}, with \pythia~8~\cite{pythia8} to model
parton showering and hadronisation.  To estimate systematic
uncertainties in the event modelling an
alternative sample is simulated using the same PDF but the
\mcatnlo~\cite{mcatnlo,Frixione:2003ei,Frixione:2008yi} generator with
\herwig++~\cite{Bahr:2008pv}. The $Z/\gamma^*$ differential cross
section as a function of mass has been calculated at
next-to-next-to-leading order (NNLO) in perturbative QCD (pQCD) using
FEWZ~3.1~\cite{fewznnlo,fewz2,FEWZ4} with the MSTW2008NNLO
PDF~\cite{mstw}. The calculation includes NLO electroweak (EW)
corrections beyond final-state photon radiation (FSR).
A mass-dependent $K$-factor used to scale the $Z/\gamma^{\ast}$ MC
sample is obtained from the ratio of the calculated NNLO pQCD cross
section with the additional EW corrections, to the cross section from
the \powheg\ sample. It is found to deviate from unity by $3.5$--$2.0\%$
across the measured range in $m_{\ell\ell}$.

The photon-induced (PI) process, $\gamma \gamma \to \ll$, is simulated
at leading-order using \pythia~8 and the MRST2004qed
PDF~\cite{MRST2004QED}.  The MC yield is scaled by a factor of 0.7 in
order to match the NLO calculations of
SANC~\cite{Bardin:2012jk,Bondarenko:2013nu}.

The background from \ttbar\ production is the dominant background with
isolated prompt leptons from electroweak boson decays.  It is
estimated at NLO using \powheg\ and the CT10 PDF, with
\pythia~6~\cite{Sjostrand:2006za} for parton showering and
hadronisation. 
Two further MC samples for \ttbar\ and single top ($Wt$) production in
association with a \w\ boson are modelled by \mcatnlo~and the CT10
PDF, with \herwig~\cite{herwig,Corcella:2002jc} for parton showering
and hadronisation. The \mcatnlo~\ttbar\ sample is used for estimating
systematic uncertainties only.
The $t\bar{t}$ MC samples are normalised to a cross section of
$\sigma_{t\bar{t}}= 253^{+13}_{-15}$~pb for a top-quark mass of
$172.5$~\GeV.  This is calculated at NNLO in QCD including resummation
of next-to-next-to-leading logarithmic soft-gluon terms with {\sc
  Top}++2.0
\cite{Cacciari:2011,Baernreuther:2012ws,Czakon:2012zr,Czakon:2012pz,Czakon:2013goa,TopPP:2011}.
The PDF and $\alpha_{\rm S}$ uncertainties on the $t\bar{t}$ cross section
are calculated using the PDF4LHC prescription~\cite{Botje:2011sn} with
the MSTW2008 68\% CL NNLO~\cite{mstw,mstw_alphas}, CT10
NNLO~\cite{CT10,Gao:2013xoa} and NNPDF2.3~\cite{nnpdf23} PDF error
sets added in quadrature to the scale uncertainty. Varying the
top-quark mass by $\pm$1~\GeV\ leads to an additional systematic
uncertainty of +8~pb and --7~pb, which is also added in quadrature.
The single-top background in association with a $W$ boson has a cross
section of $\sigma_{Wt}= 22.4 \pm 1.5$~pb \cite{Kidonakis:2010ux}.
Given that the $Wt$ contribution is small compared to the $t\bar{t}$
cross section, an overall uncertainty of 6\% is estimated on the
top-quark background.

Further important background contributions are due to diboson ($WW$,
$WZ$ and $ZZ$) production with decays to final states with at least
two leptons.  The diboson processes are generated at leading order
(LO) with \herwig, using the CTEQ6L1 PDF~\cite{Pumplin:2002vw}.  The
$WZ$ and $ZZ$ cross-section values used are $20.3 \pm 0.8$~pb and $7.2
\pm 0.3$~pb respectively, as calculated at NLO with
MCFM~\cite{Campbell:1999mcfm,Campbell:2011bn} and the CT10 PDF.  The
$WW$ cross section is assumed to be $70.4 \pm 7$~pb, derived by
scaling the MCFM value of 58.7~pb by a factor of $1.20 \pm 0.12$.
This scale factor and its uncertainty correspond to an approximate
mean of the two scale factors for $WW$ production with zero and one
extra jet, as discussed in ref.~\cite{HIGG-2013-13}.
They  
are consistent with the recent ATLAS measurement of the $WW$ cross section at 
$\sqrt{s}=8$~\TeV, which yields a value of $71.1 \pm 1.1\,$(stat)$\,^{+5.7}_{-5.0}\,$(sys)$\,\pm 1.4\,$pb\cite{Aad:2016wpd}.

All MC samples used in the analysis include the effects of FSR, multiple interactions per bunch crossing
(``pile-up'') , and detector simulation. FSR is simulated using
\photos~\cite{fsr_ref}, except for samples hadronised by \herwig++
which includes a native FSR simulation. The effects of pile-up are
accounted for by overlaying simulated minimum-bias
events~\cite{SOFT-2010-01}.
The interactions of particles with the detector are
modelled using a full \mbox{ATLAS} detector
simulation~\cite{SOFT-2010-01} based on \geant4~\cite{geant}.  Finally,
several corrections are applied to the simulated samples, accounting
for differences between data and simulation in the lepton trigger,
reconstruction, identification, and isolation efficiencies as well as
lepton resolution and muon momentum scale.

An overview of the simulated event samples is given in table~\ref{tab:mc}. 
\begin{table}[h]
\centering
\small
\begin{tabular}{c|llll}
\hline
\hline
Process          & Generator    & Parton shower 	& Generator PDF 	& Model parameters (``Tune'')	  \\
\hline
Drell--Yan	& \powheg   	& \pythia~8.162		& CT10			& AU2~\cite{ATL-PHYS-PUB-2012-003}\\
Drell--Yan	& \mcatnlo~4.09	& \herwig++~2.6.3	& CT10			& UE-EE-3~\cite{Bahr:2008pv}	  \\
PI		& \pythia~8.170	& \pythia~8.170		& MRST2004qed		& 4C~\cite{Corke:2010yf}	  \\
\ttbar		& \powheg   	&  \pythia~6.427.2	& CT10			& AUET2 \cite{ATL-PHYS-PUB-2011-008}\\
$\ttbar$	& \mcatnlo~4.06 & \herwig~6.520		& CT10			& AUET2	\\
$Wt$		& \mcatnlo~4.06 & \herwig~6.520		& CT10			& AUET2	\\
Diboson		& \herwig~6.520	& \herwig~6.520  	& CTEQ6L1		& AUET2 \\
\hline
\hline
\end{tabular}
\caption{Overview of simulated event samples used.}
\label{tab:mc}
\end{table}

%-------------------------------------------------------------------------------
\section{Event selection}
\label{sec:selection}
%-------------------------------------------------------------------------------

Events are required to be recorded during stable beam condition periods
and must pass detector and data-quality requirements. 
Due to differences in the detector response to electrons and
muons the selection is optimised separately for
each channel and is described in the following.

%%%%%%%%%%%%%%%%%%%%%%%%%%%%%%%%%%%%%%%%%%%%%%%%%%%%%%%%%%%%%%%%%%%%%%%%%%%%%%%%%%%%%%%
\subsection{Electron channel}
\label{sec:sel_elec}
The electron data are collected by a trigger which uses
calorimetric information to identify two compact electromagnetic energy
depositions. Identification algorithms use calorimeter shower
shape information to find candidate electron pairs with a minimum
transverse energy 
of $35$~\GeV\ and $25$~\GeV\ for the leading and subleading
electron. 
The candidate electron pairs are not matched to inner
detector tracks in the trigger
allowing
the same trigger to be used for the multijet and $W$+jets data-driven background 
estimation studies, where a background-enriched sample is required.

Electrons are reconstructed by clustering energy deposits in the
electromagnetic calorimeter using a sliding-window algorithm. These
clusters are then matched to tracks reconstructed in the inner
detector.  The calorimeter provides the energy measurement and the
track is used to determine the angular information of the electron
trajectory.  An energy scale correction determined from $Z\to e^+e^-$,
$W\to e \nu$, and $J/\psi\to e^+e^-$ decays~\cite{PERF-2013-05} is
applied to data.  Candidates are required to have a pseudorapidity within the inner
detector tracking region, $\left|\eta^e\right|<2.47$, excluding a
region, $1.37<\left|\eta^e\right|<1.52$, where the transition between
the barrel and endcap electromagnetic calorimeters is not well
modelled in the simulation. Each candidate is required to satisfy the
``medium'' electron
identification~\cite{PERF-2013-03,ATLAS-CONF-2014-032} criteria based
on calorimetric shower shapes and track parameters.

Leptons produced in the Drell--Yan process are expected to be
well isolated from energy depositions not associated with the lepton. The degree of
isolation for electrons is defined as the scalar sum of transverse
energy, $\sum E_{\rm T}$, of additional energy contained in a cone
of size $\Delta R=\sqrt{(\Delta\phi)^2+(\Delta\eta)^2}$ around the
electron, omitting the electron transverse energy $E^e_{\rm
  T}$. This calorimetric isolation is required to satisfy
$\sum E_{\rm T}(\Delta R = 0.4)<0.007\cdot E_{\rm T}^e + 5$~\GeV\ for
the leading electron, and $\sum E_{\rm T}(\Delta R = 0.4)<0.022\cdot
E_{\rm T}^e + 6$~\GeV\ for the subleading electron, in order to 
retain a high efficiency of approximately 99\% per electron over a large 
range in $E_{\rm T}^e$.

Candidate events are required to have at least two electrons with
$\et^e>30$~\GeV\ and at least one of the electrons satisfying
$\et^e>40~$\GeV\ to ensure the selected electron is on the efficiency
plateau of the trigger. The invariant mass of the pair is required to
be in the range $116\leq m_{ee}\leq 1500$~\GeV. The absolute difference
in pseudorapidity between the two electrons, $|\Delta\eta_{ee}|$, is
restricted to be less than $3.5$ in order to suppress the multijet
background which is dominated by $t$--channel processes. No charge
requirements are placed on the lepton pair.

%%%%%%%%%%%%%%%%%%%%%%%%%%%%%%%%%%%%%%%%%%%%%%%%%%%%%%%%%%%%%%%%%%%%%%%%%%%%%%%%%%%%%%%
\subsection{Muon channel}
\label{sec:sel_muon}
Candidate events in the muon channel are collected using two triggers,
each requiring a single muon, but with different transverse momentum
thresholds as measured in the higher-level trigger system. A
high-threshold trigger demands that the muon transverse momentum be
above $36$~\GeV\ and collects most of the data sample. A supplementary
low-threshold trigger requires an isolated muon with transverse
momentum above $24$~\GeV. The isolation for muons is defined using the
scalar sum of transverse momenta, $\sum p_{\rm T}$, of additional
tracks divided by $p^{\mu}_{\rm T}$, the transverse momentum of the
muon. This provides a good discriminant against the multijet
background arising from the semileptonic decays of heavy quarks. This
isolation definition is implemented in the low-threshold trigger in
which the candidate muons are required to satisfy $\sum p_{\rm
  T}(\Delta R=0.2)/p^{\mu}_{\rm T}<0.12$.

Muons are identified by tracks reconstructed in the muon spectrometer
matched to tracks reconstructed in the inner detector and must satisfy
$|\eta^{\mu}| < 2.4$. In addition they must pass the ``medium''
identification criteria~\cite{PERF-2014-05}, based on requirements on
the number of hits in the different inner detector and muon
spectrometer subsystems, as well as the significance of the charge /
momentum ratio imbalance between the ID and MS
measurements. Background from multijet events is efficiently
suppressed by imposing the isolation condition $\sum p_{\rm T}(\Delta
R=0.2)/p^{\mu}_{\rm T}<0.1$. A small contribution of cosmic-ray muons is
removed by requiring the magnitude of the longitudinal impact parameter
to the primary interaction vertex, $z_0$, to be less than
$10$~mm. The primary interaction vertex is taken to be the one with
the largest sum of squared transverse momenta of all associated tracks.

Events are selected if they contain at least two oppositely
charged muons with $p^{\mu}_{\rm T} > 30$~\GeV\ and at least one of the
muons satisfies $p^{\mu}_{\rm T} > 40$~\GeV\ in order to have the same
phase space as in the electron channel measurement. Finally the
dilepton invariant mass is required to be in the range $116\leq
m_{\mu\mu} \leq 1500$~\GeV. No requirement is placed on
$|\Delta\eta_{\mu\mu}|$.

%-------------------------------------------------------------------------------
\section{Background estimate}
\label{sec:bg}
%-------------------------------------------------------------------------------

The background from processes with two or more isolated final-state leptons of
the same flavour is derived from MC simulation. The processes with
non-negligible contributions are \ttbar, $Wt$ (hereafter termed the top-quark
background) and diboson ($WW$, $WZ$
and $ZZ$) production, see table~\ref{tab:mc}. The background arising
from the $Z/\gamma^* \rightarrow \tau\tau$ process is predicted  
using MC simulation and found to be negligible.

Background contributions from events where at least one final-state
jet or photon passes the electron or muon selection criteria are
determined using data.  This includes contributions from light- and
heavy-flavour multijet processes, and $\gamma$ + jet production, referred to
hereafter as the multijet background. Additional contributions are due
to \wpjet\ processes and \ttbar\ and $Wt$ production with less than
two isolated final-state leptons,
referred to hereafter as \wpjet\ background.  The data-driven
estimates are described in detail below.

\begin{figure}[t]
\begin{center}
\includegraphics[width=0.495\textwidth]{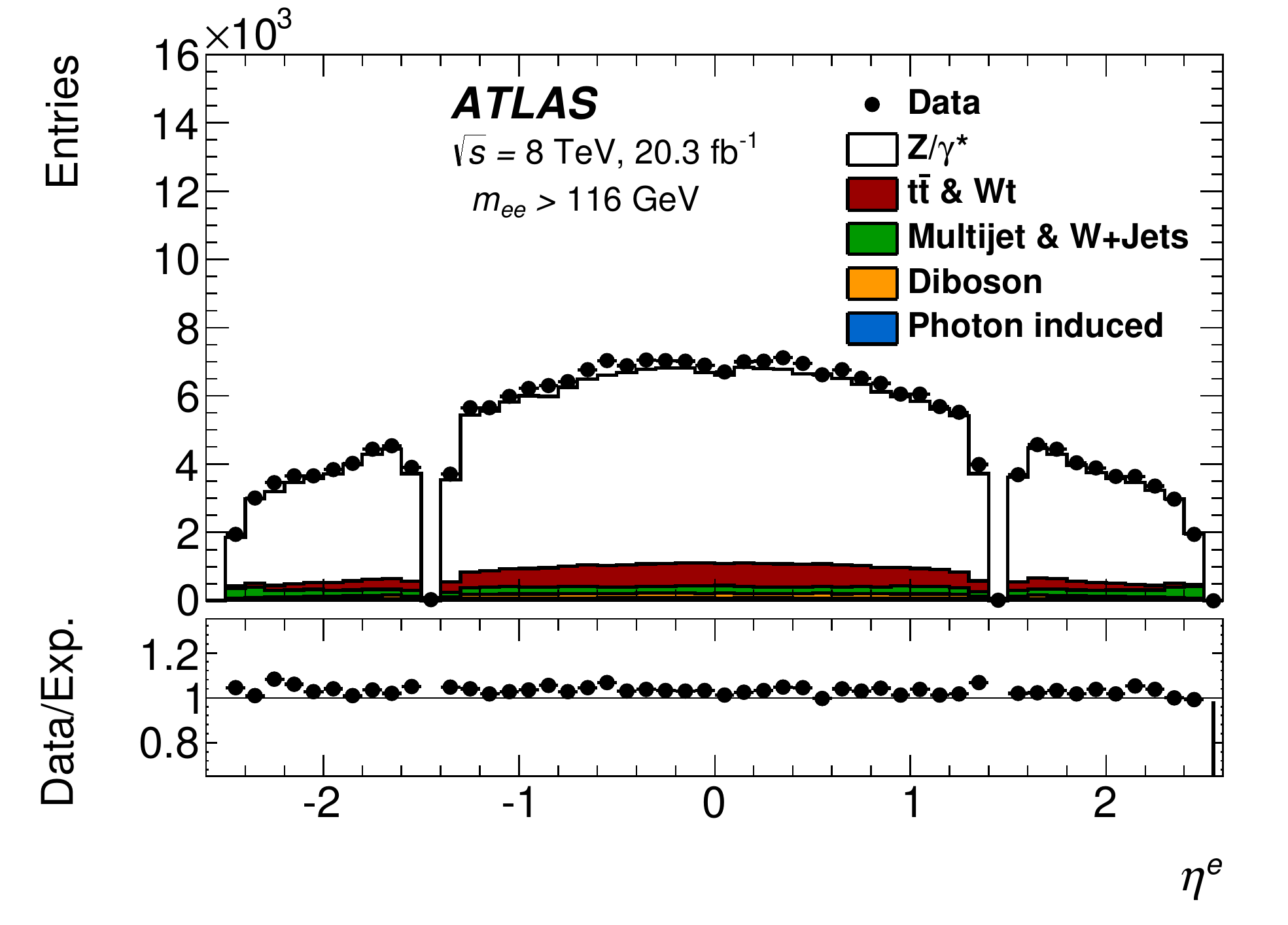}
\includegraphics[width=0.495\textwidth]{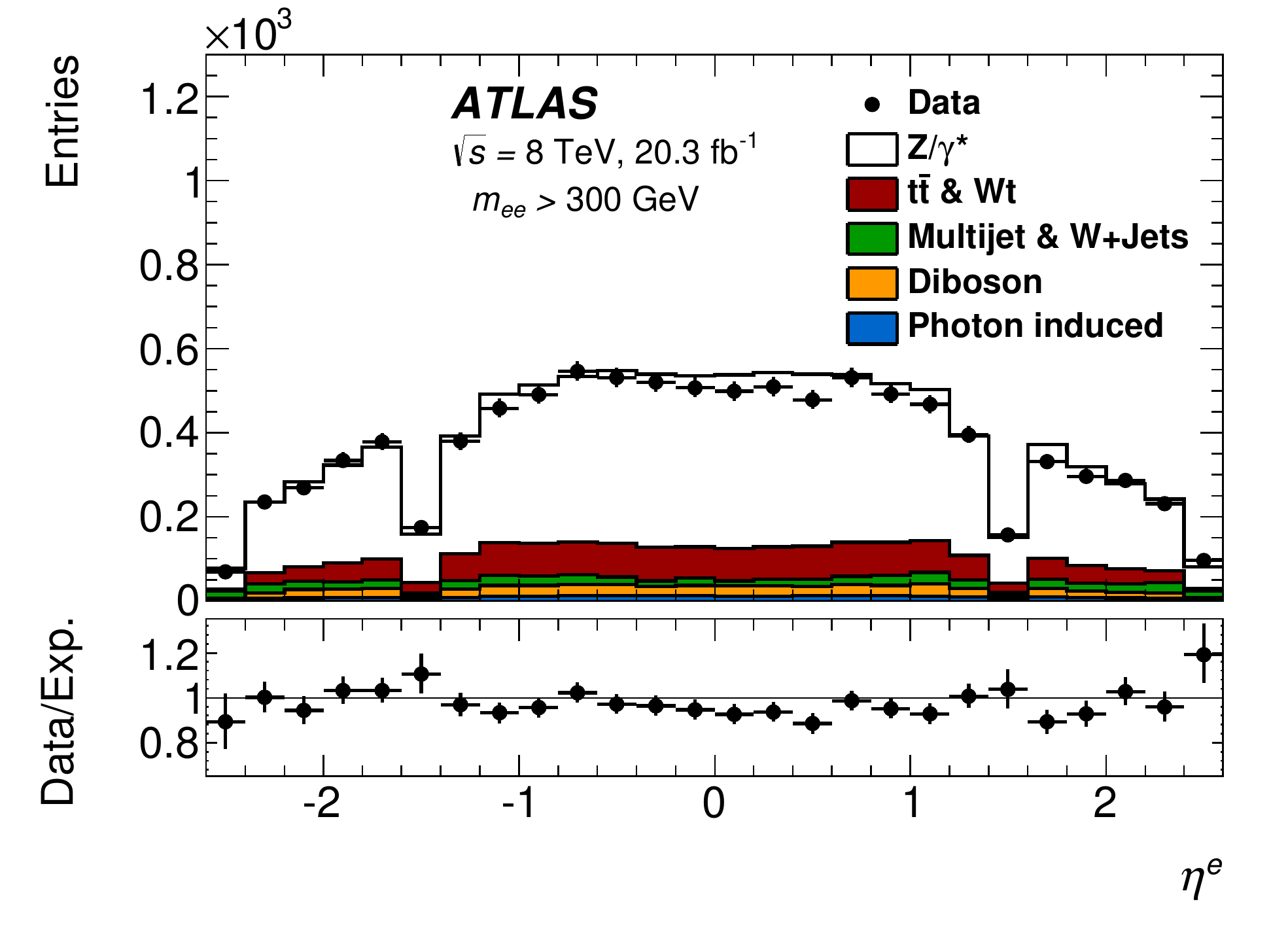}
\includegraphics[width=0.495\textwidth]{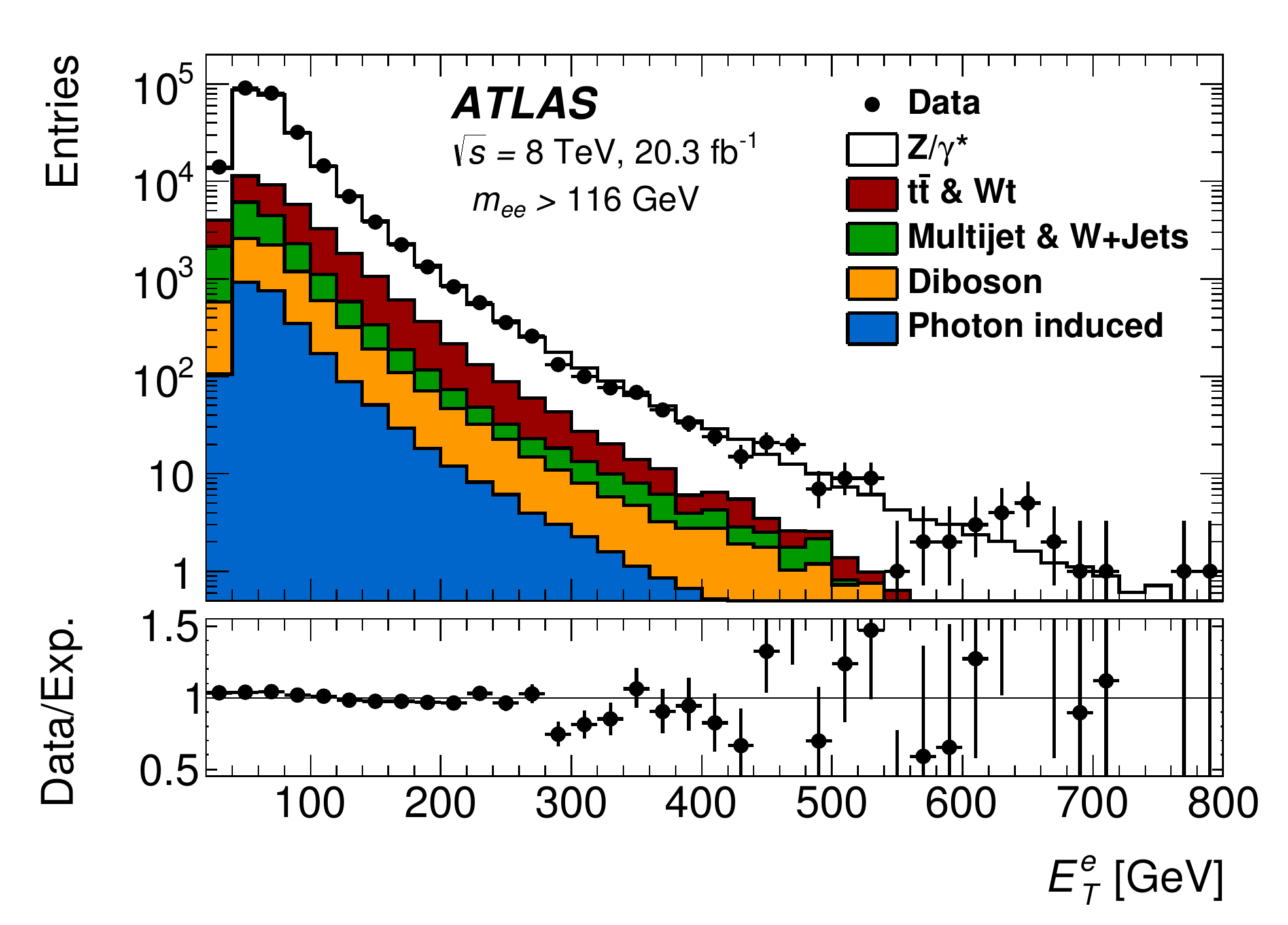}
\includegraphics[width=0.495\textwidth]{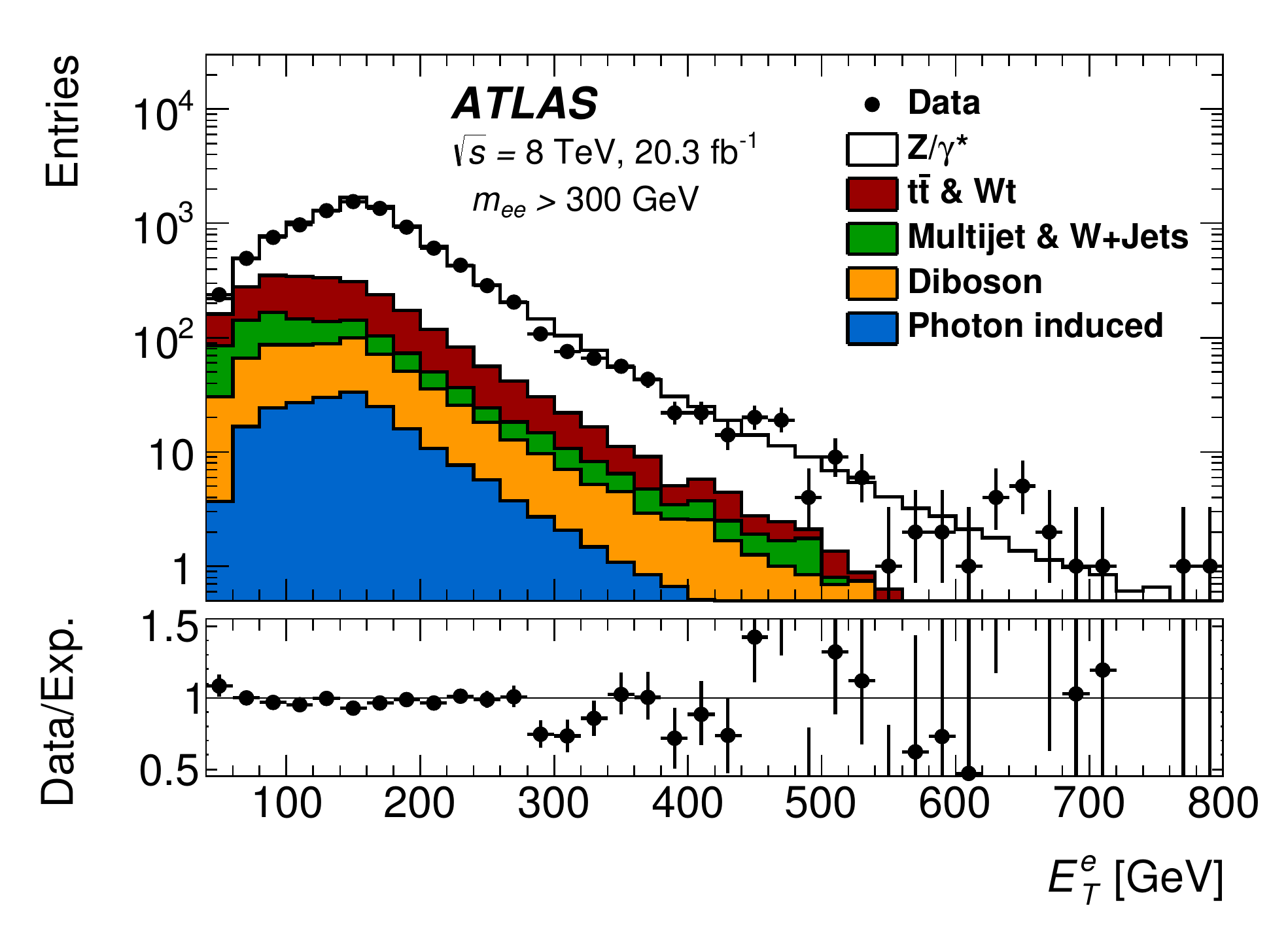}
\end{center}
\caption{Distribution of electron pseudorapidity $\eta^e$ (upper
  plots) and transverse energy $E_{\rm T}^e$ (lower
  plots) for invariant masses $m_{ee}>116$~\GeV\ (left plots), and $m_{ee}>300$~\GeV\
  (right plots), shown for data (solid points) and expectation (stacked
  histogram) after the complete selection. The lower
  panels show the ratio of data with its statistical uncertainty to the expectation.}
\label{fig:elec_controlPlots1}
\end{figure}

\begin{figure}[t]
\begin{center}
\includegraphics[width=0.495\textwidth]{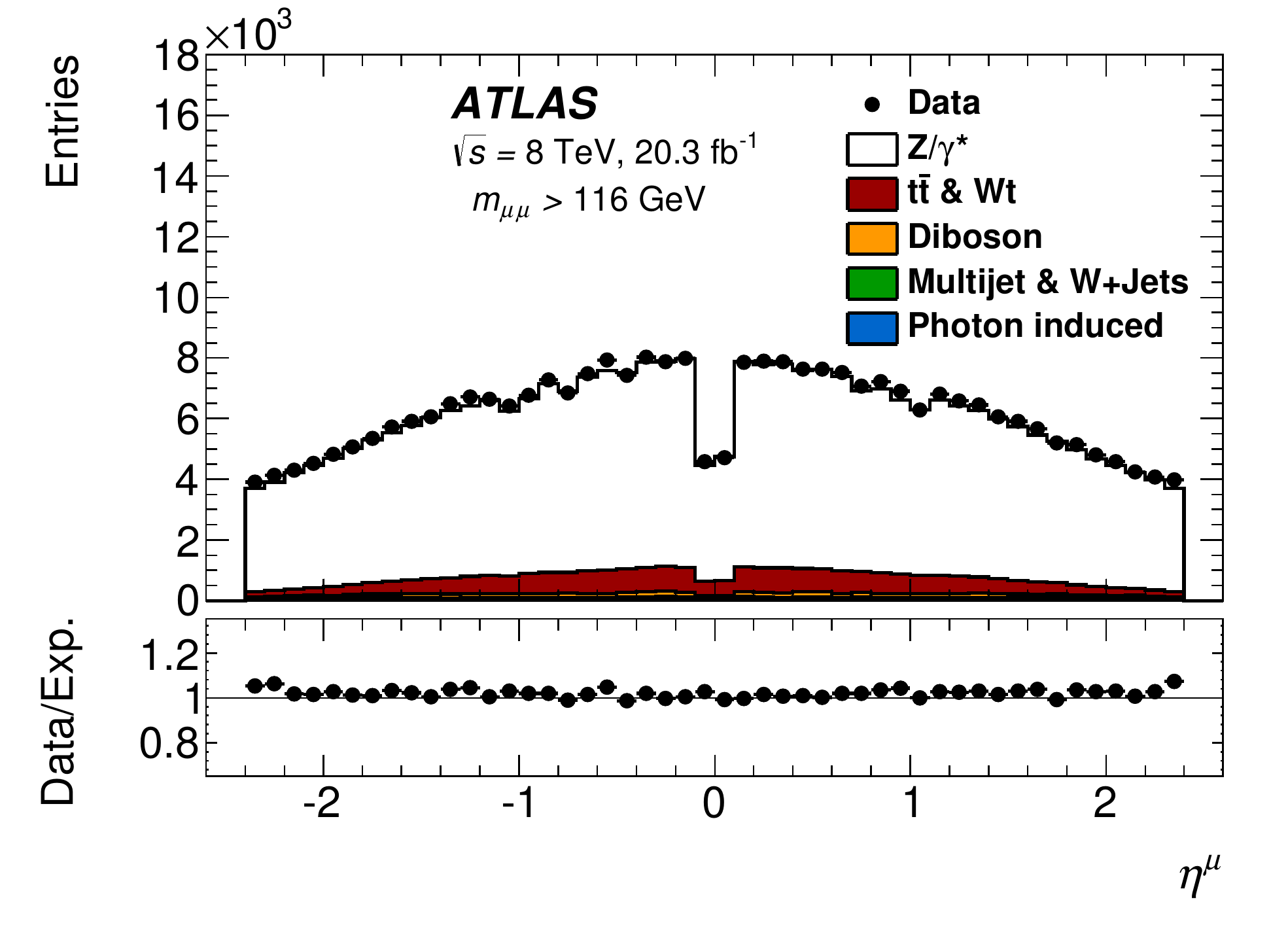}
\includegraphics[width=0.495\textwidth]{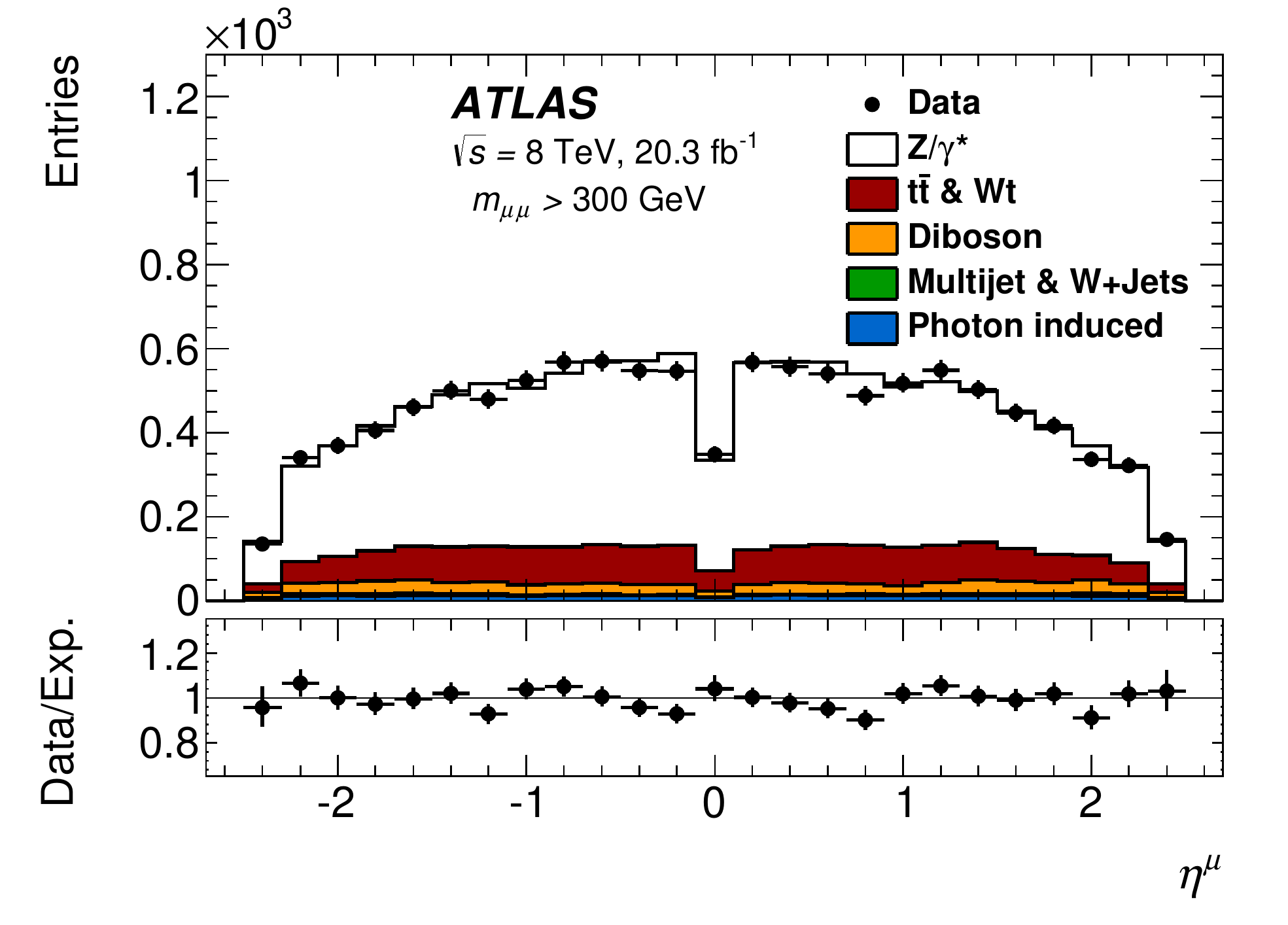}
\includegraphics[width=0.495\textwidth]{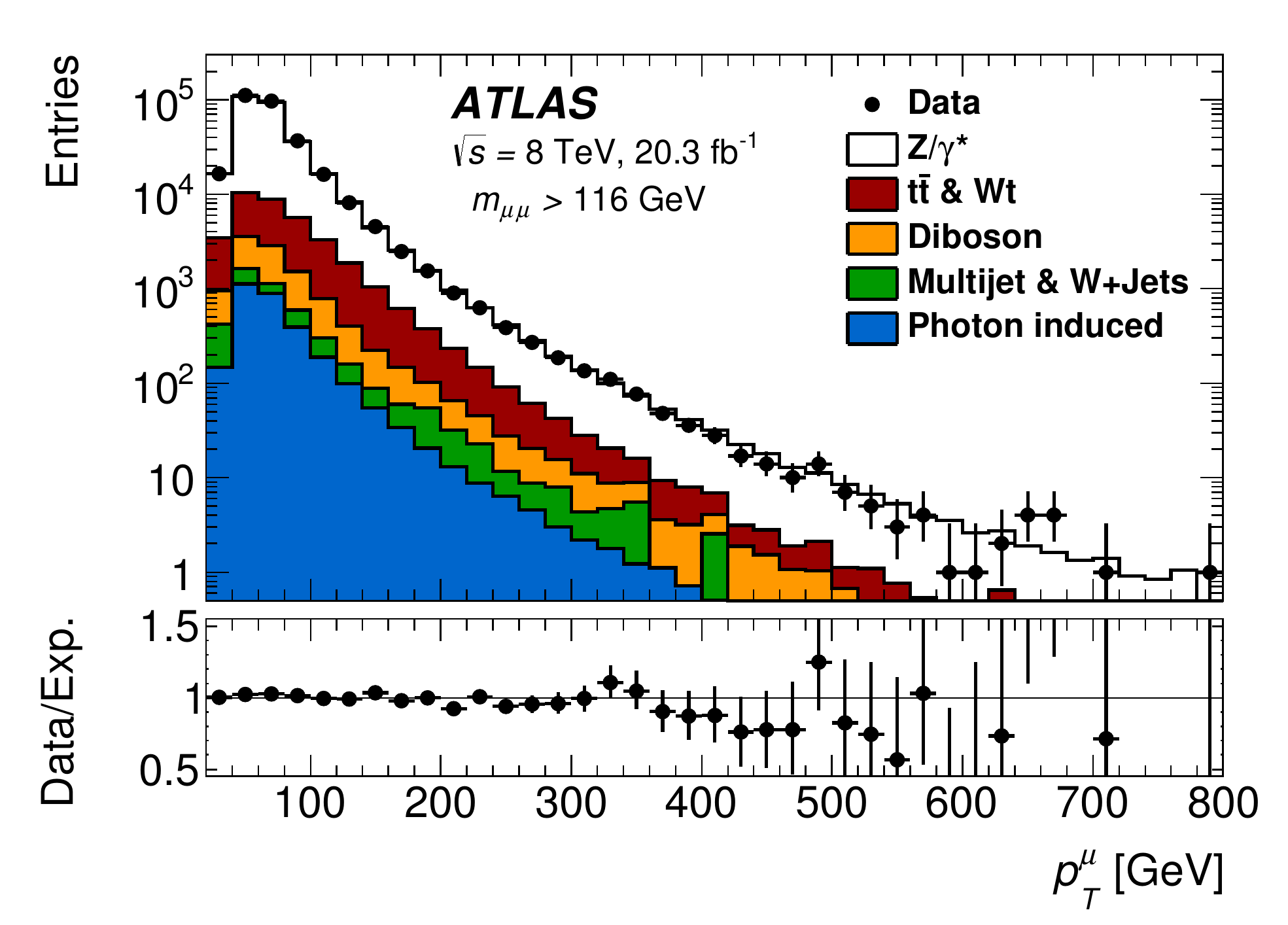}
\includegraphics[width=0.495\textwidth]{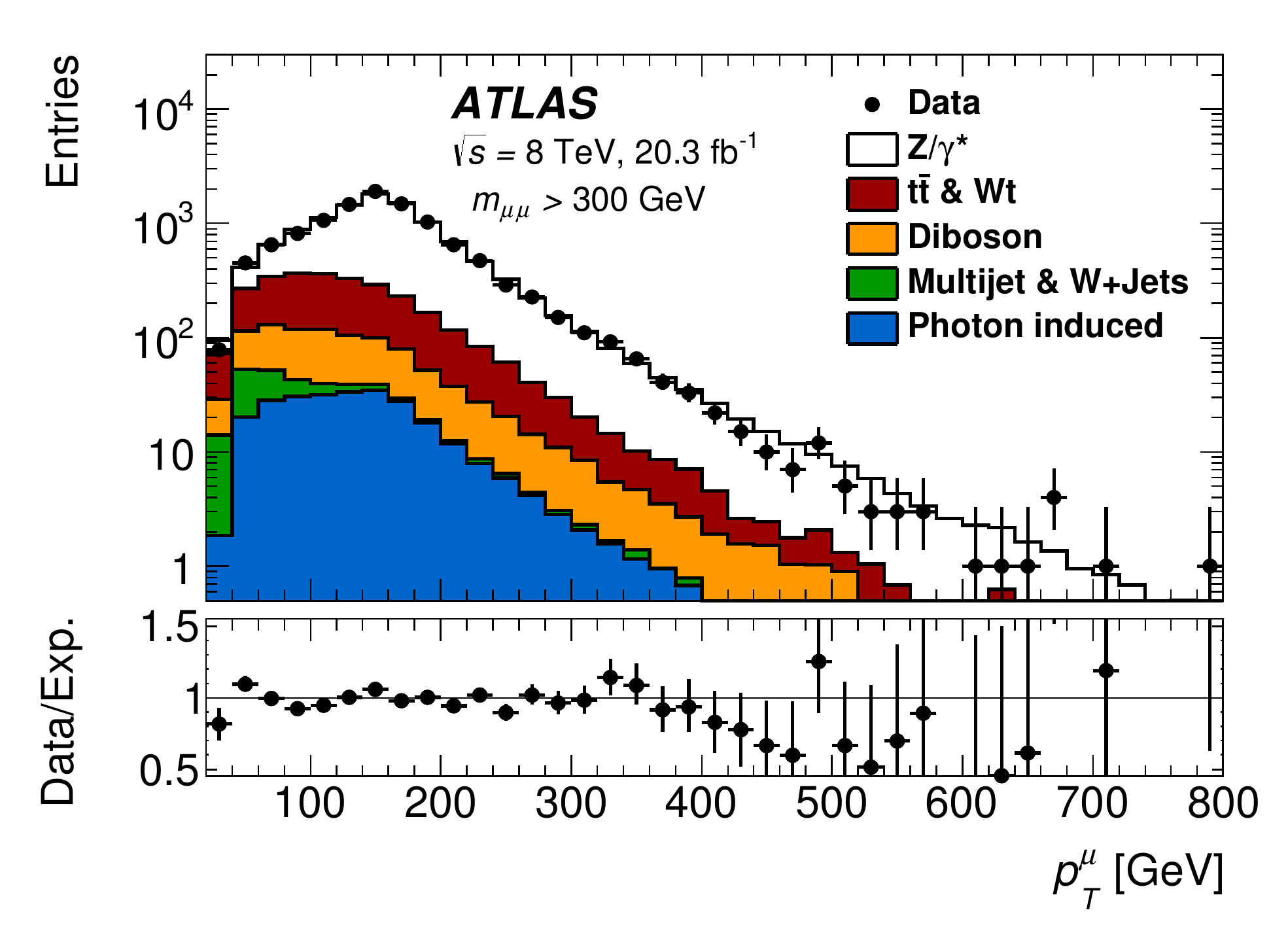}
\end{center}
\caption{Distribution of muon pseudorapidity $\eta^{\mu}$ (upper
  plots) and transverse momentum $p_{\rm T}^{\mu}$
  (lower plots) for invariant masses $m_{\mu\mu}>116$~\GeV\ (left plots), and
  $m_{\mu\mu}>300$~\GeV\ (right plots), shown for data (solid points)
  and expectation (stacked histogram) after the complete
  selection. The lower panels show the ratio of data with its statistical uncertainty to the
  expectation.}
\label{fig:muon_controlPlots1}
\end{figure}

\begin{figure}[t]
\begin{center}
\includegraphics[width=0.495\textwidth]{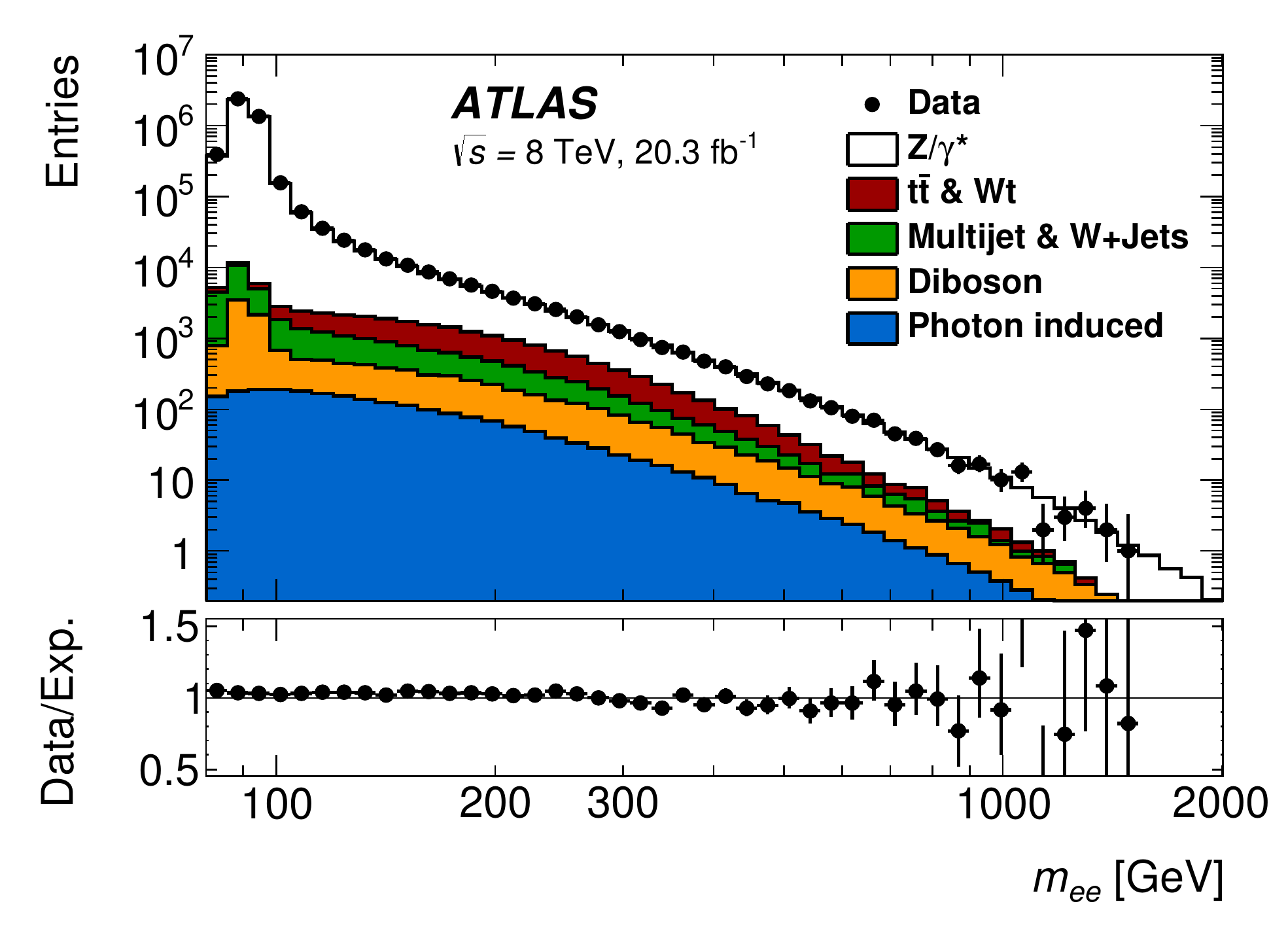}
\includegraphics[width=0.495\textwidth]{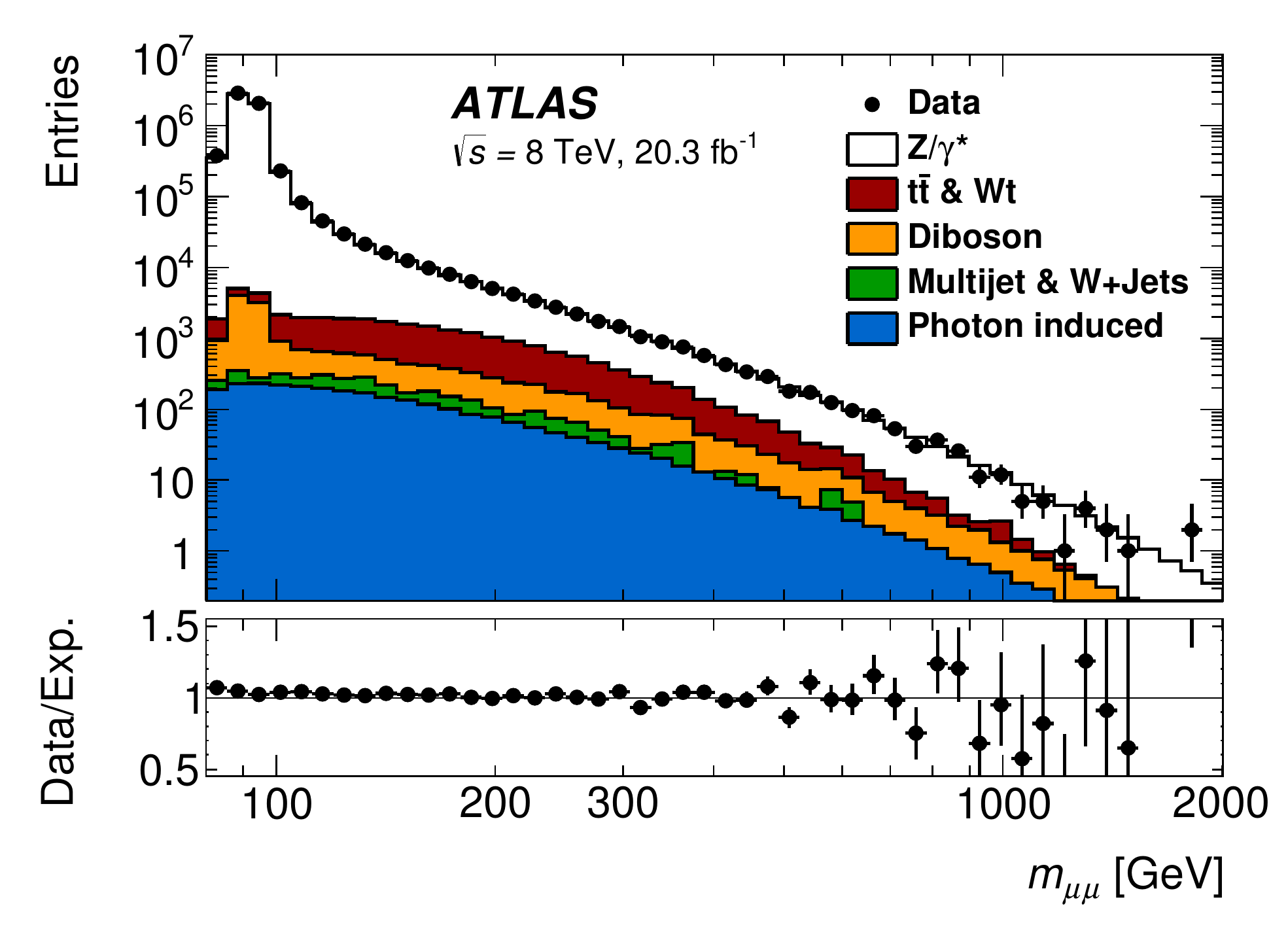}
\end{center}
\caption{The invariant mass (\mll ) distribution after event selection
  for the electron selection (left) and muon selection (right), shown
  for data (solid points) compared to the expectation (stacked
  histogram). The lower panels show the ratio of data with its statistical uncertainty to the
  expectation.}
\label{fig:invmass_controlPlots}
\end{figure}

\begin{figure}[t]
\begin{center}
\includegraphics[width=0.495\textwidth]{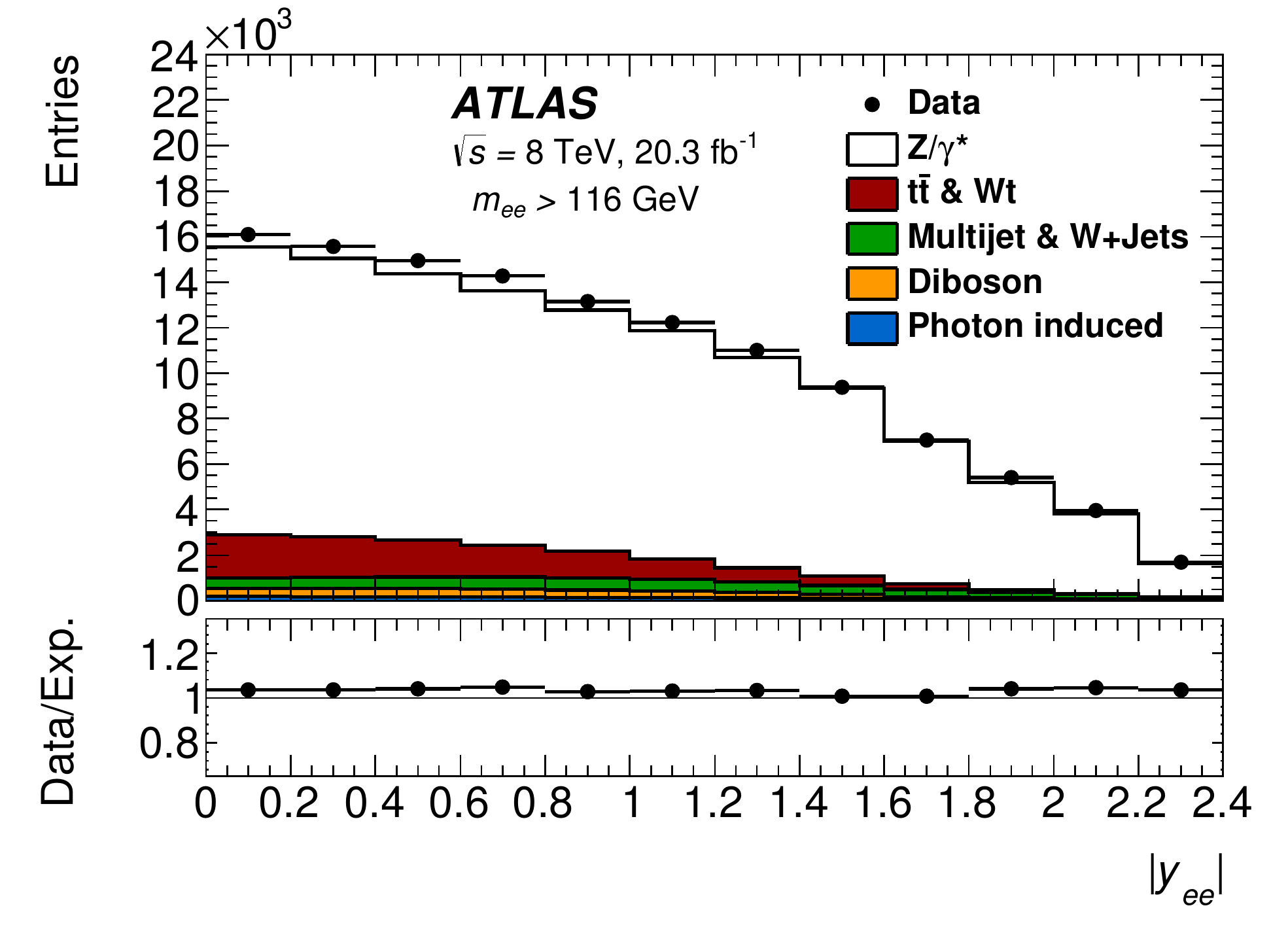}
\includegraphics[width=0.495\textwidth]{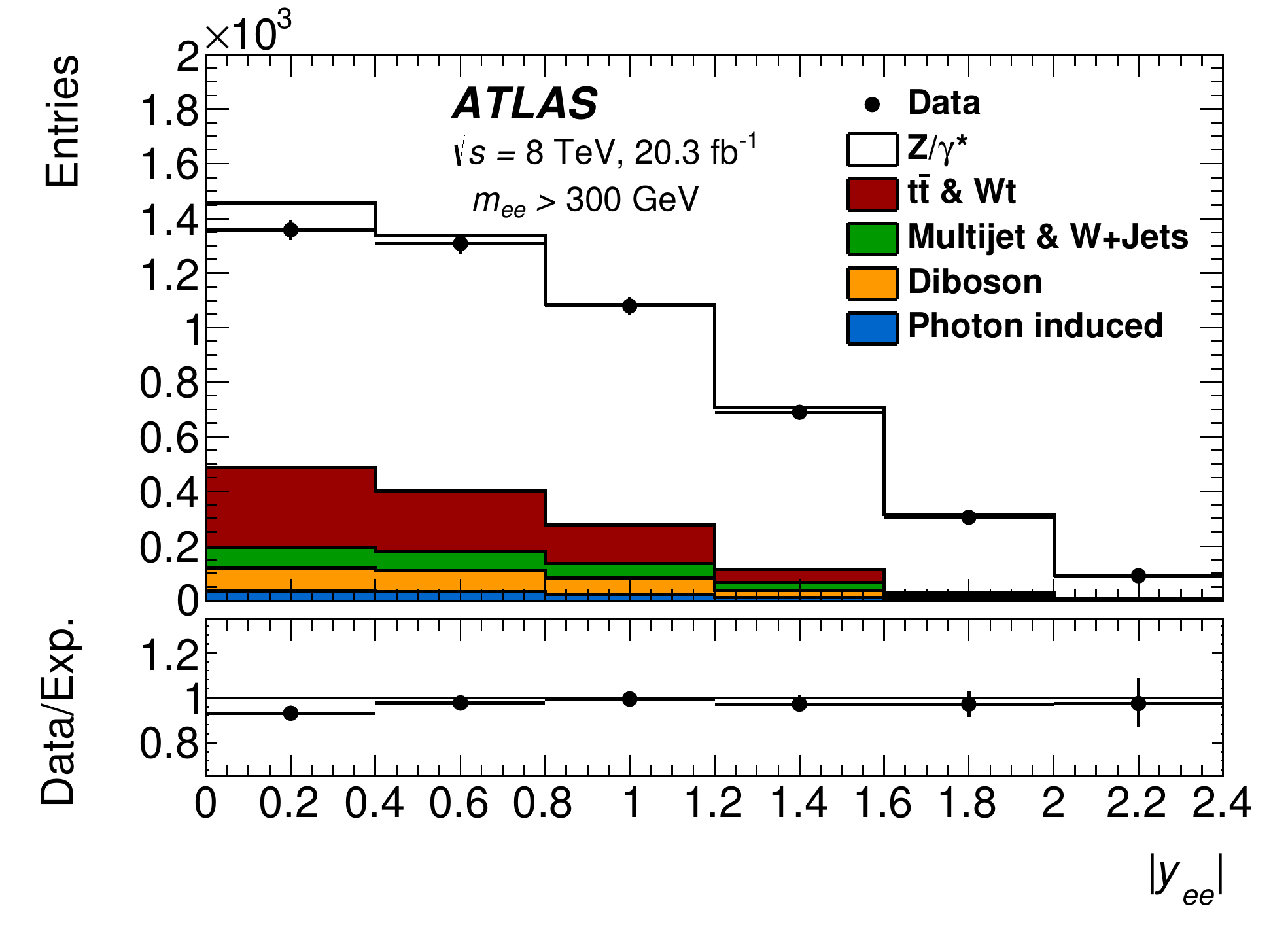}
\includegraphics[width=0.495\textwidth]{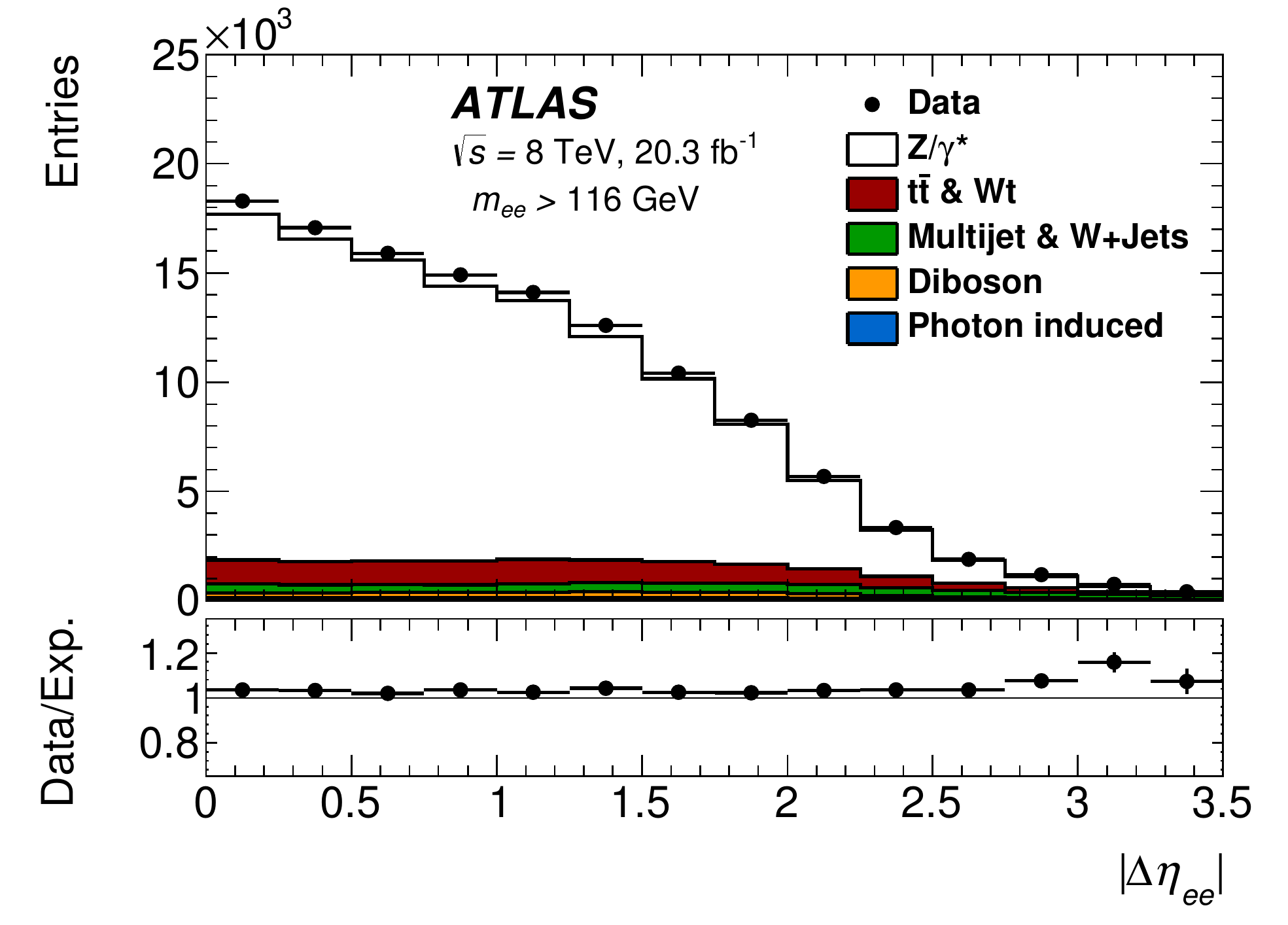}
\includegraphics[width=0.495\textwidth]{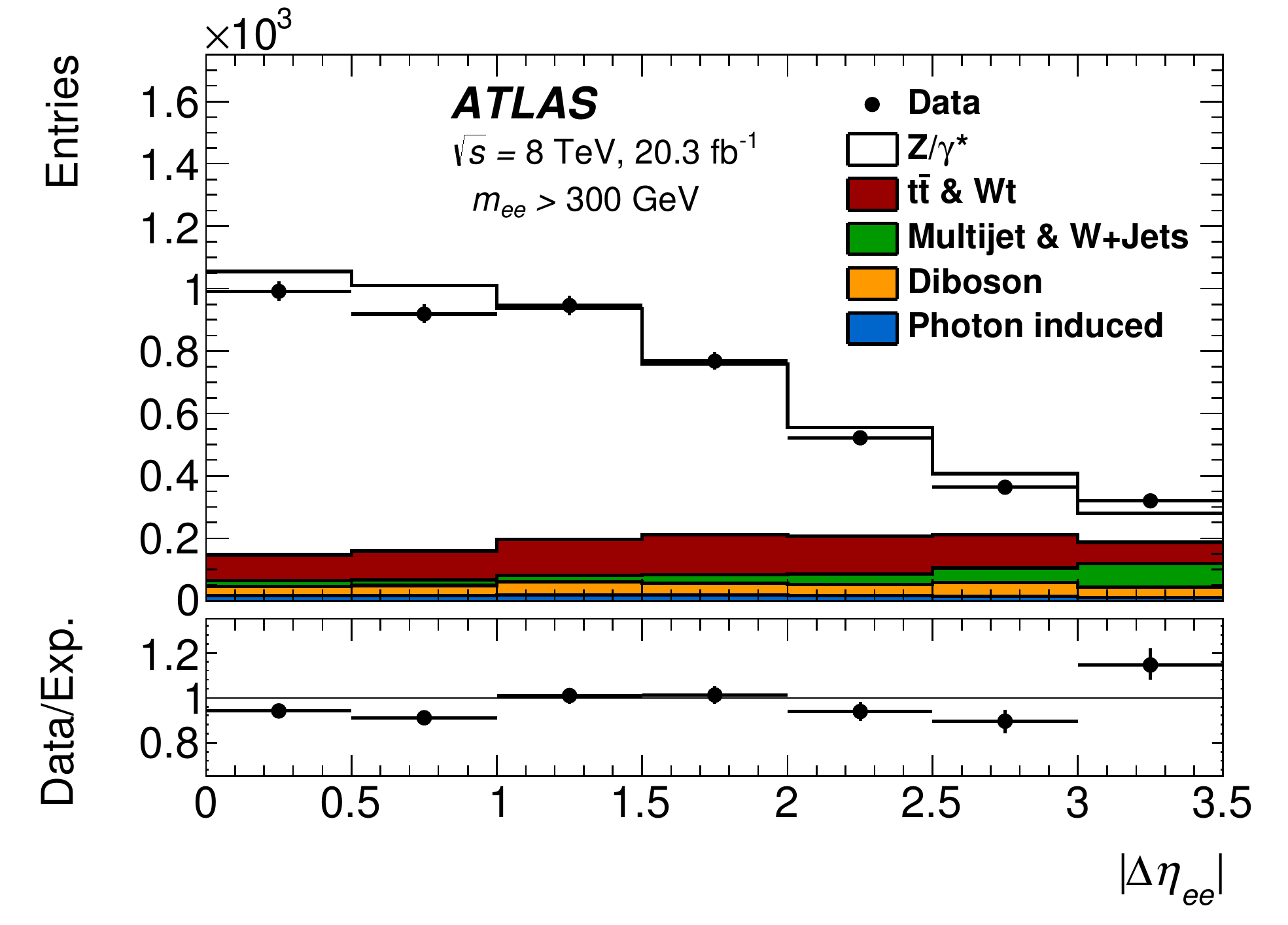}
\end{center}
\caption{Distribution of absolute dielectron rapidity $|y_{ee}|$ (upper plots) and
  absolute dielectron pseudorapidity separation $|\Delta\eta_{ee}|$
  (lower plots) for invariant mass $m_{ee}>116$~\GeV\ (left plots),
  and $m_{ee}>300$~\GeV\ (right plots), shown for data (solid points)
  and expectation (stacked histogram) after the complete 
  selection. The lower panels show the ratio of data with its statistical uncertainty to the
  expectation.}
\label{fig:elec_controlPlots2}
\end{figure}

\begin{figure}[t]
\begin{center}
\includegraphics[width=0.495\textwidth]{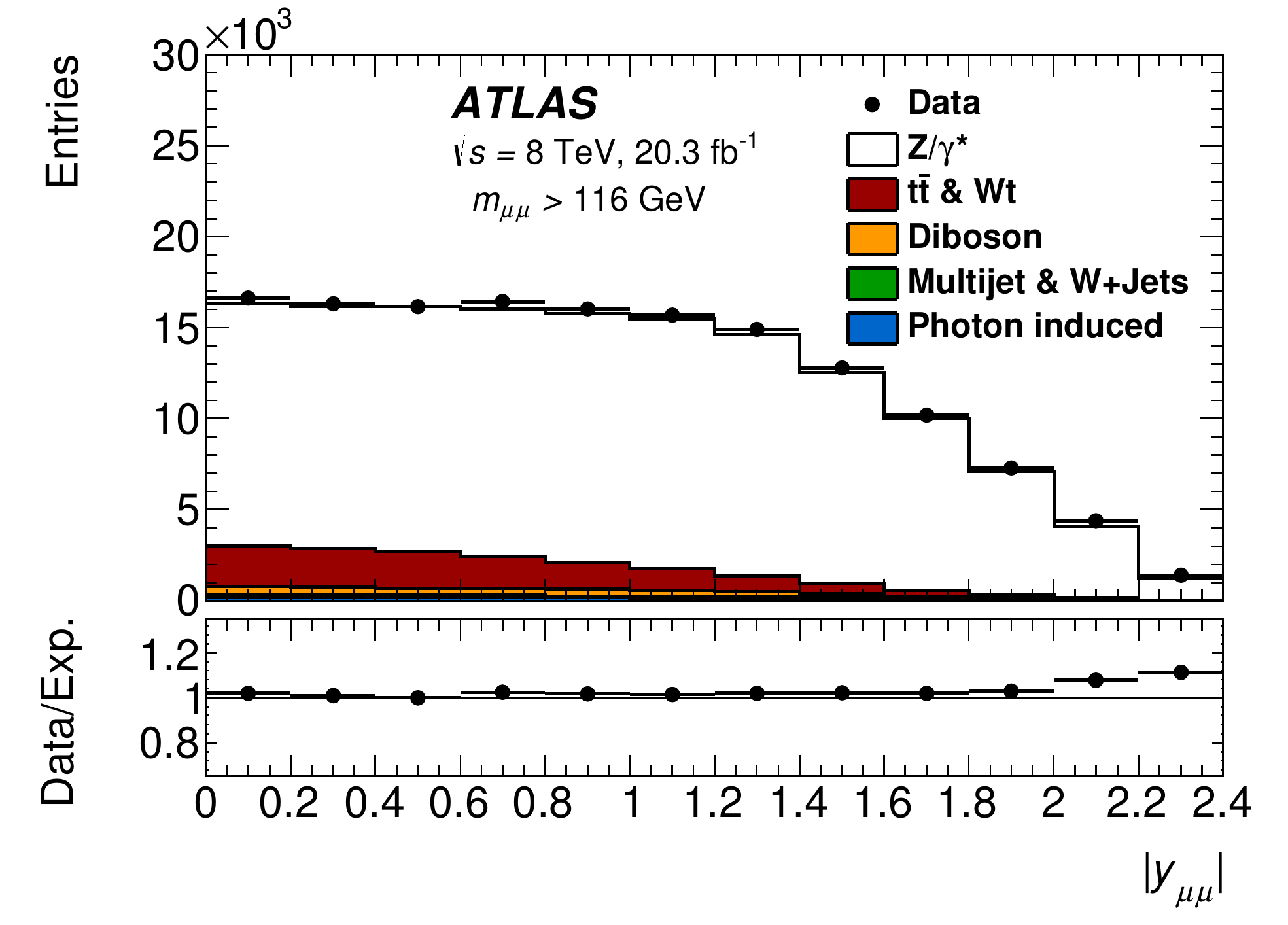}
\includegraphics[width=0.495\textwidth]{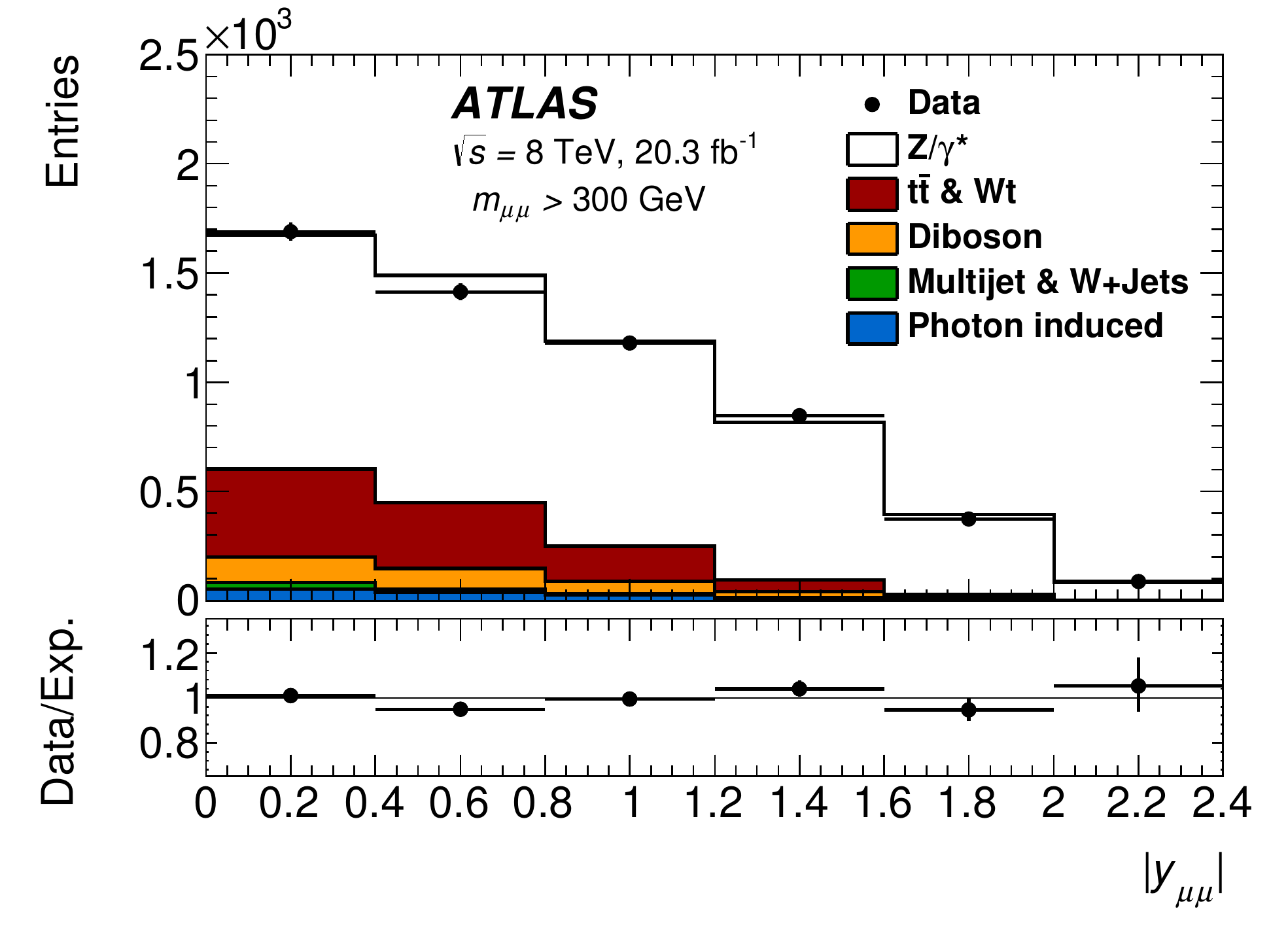}
\includegraphics[width=0.495\textwidth]{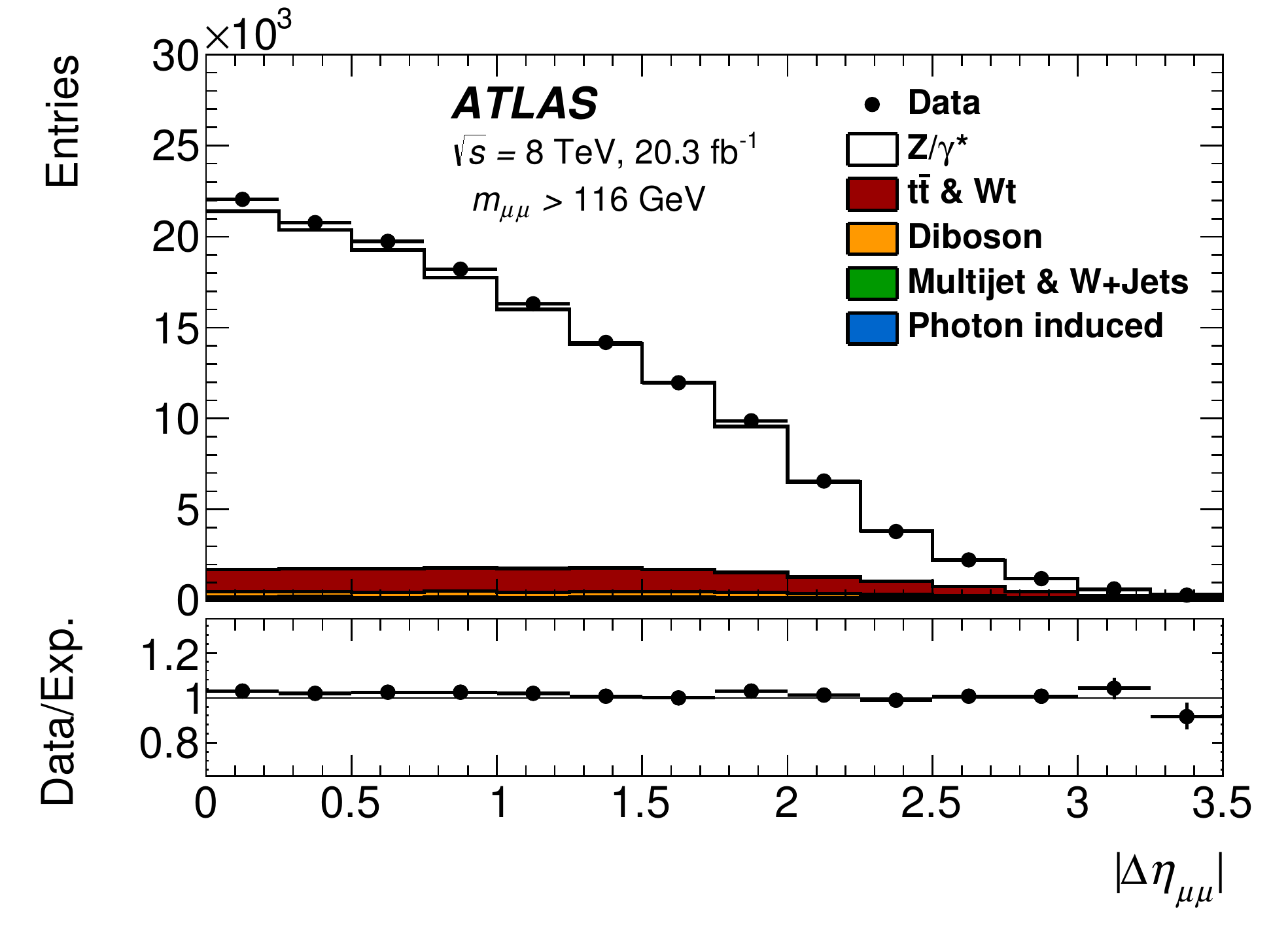}
\includegraphics[width=0.495\textwidth]{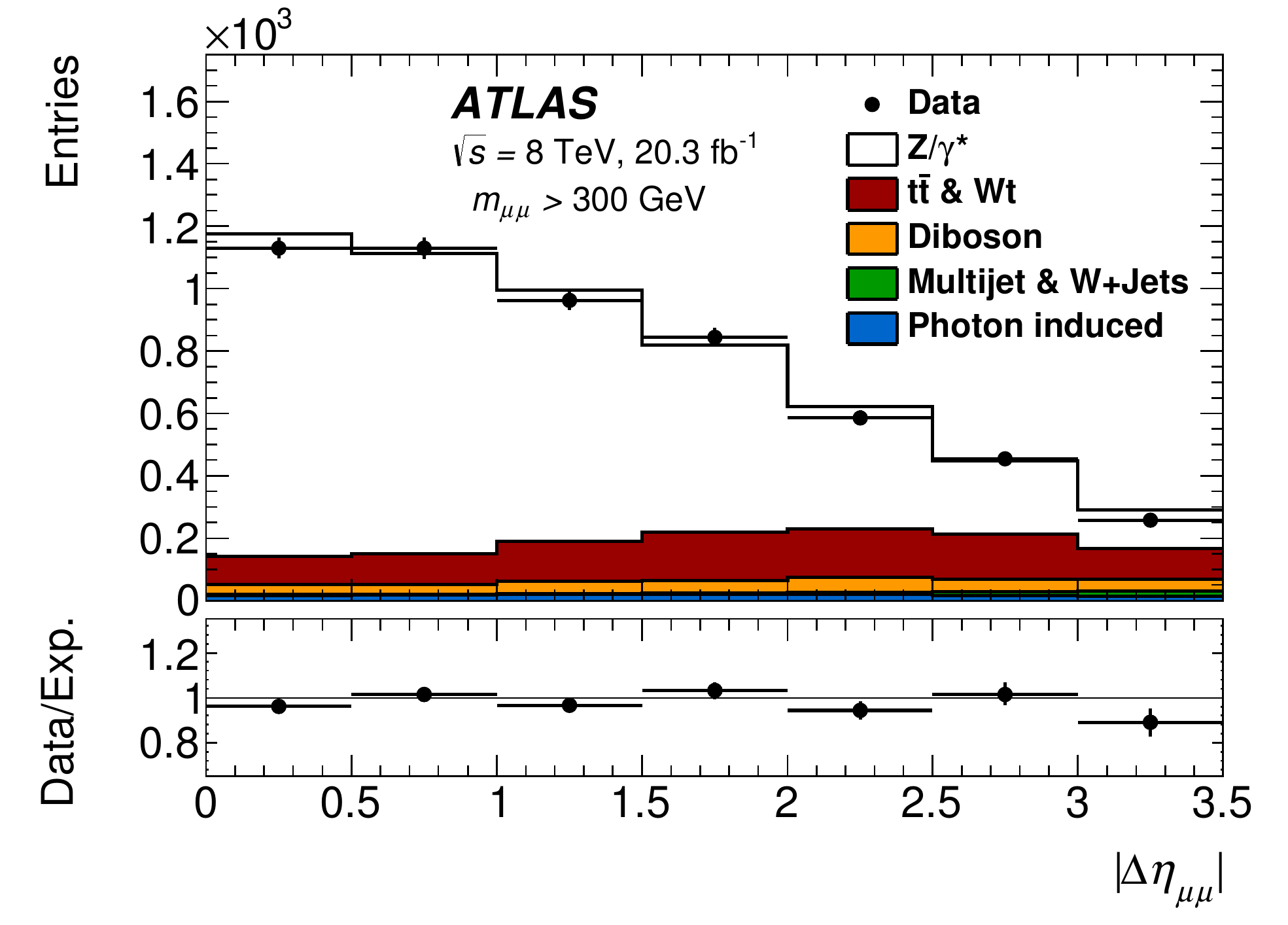}
\end{center}
\caption{ Distribution of absolute dimuon rapidity $|y_{\mu\mu}|$ (upper plots) and
  absolute dimuon pseudorapidity separation $|\Delta\eta_{\mu\mu}|$
  (lower plots) for invariant mass $m_{\mu\mu}>116$~\GeV\ (left
  plots), and $m_{\mu\mu}>300$~\GeV\ (right plots), shown for data
  (solid points) and expectation (stacked histogram) after the complete
  selection. The lower panels show the ratio of data with its statistical uncertainty to
  the expectation.}
\label{fig:muon_controlPlots2}
\end{figure}

The number of expected events is calculated as the sum of the
data-driven and simulated background estimates, and the expected event
yield predicted by the DY and PI MC simulations. As can be seen in
figures~\ref{fig:elec_controlPlots1}--\ref{fig:muon_controlPlots2},
good agreement is found in both the $ee$ and $\mu\mu$ channels
comparing data and expectation for the $\eta^{\ell}$ and $p_{\rm
  T}^{\ell}$ distributions of the leptons, as well as for the
distributions in invariant mass, rapidity and $\Delta
\eta_{\ell\ell}$. The background contributions are stacked in order of
increasing importance. In the electron channel the top-quark, multijet and
diboson contributions to the expectation are found to be approximately
$9\%$, $4\%$ and $2\%$ respectively in the phase space of the
measurement. In the muon channel the top-quark and diboson backgrounds
constitute about $9\%$ and $2\%$ of the total expectation, whereas the
multijet contribution is below $1\%$ everywhere. The predicted PI
contribution is $1\%$ for both channels but can reach as much as
$16\%$ in the bin at highest $m_{\ell\ell}$ and largest $\Delta\eta_{\ell\ell}$.

\subsection{Multijet and $W$+jets background estimate in the electron channel}
\label{sec:bg_elec_data_driven}

The probability that a jet is misidentified as an electron (the ``fake
rate'') is determined as a function of transverse energy, \et\, and
pseudorapidity, $\eta$, of the electron candidate using
background-enriched data samples.  These samples are recorded using
a set of single-jet triggers with \et\ thresholds in the range
25--360~\GeV.  In each of these samples, the fake rate $f_1$ ($f_2$) is
calculated as the fraction of leading (subleading) electron candidates
that pass the nominal electron identification and leading (subleading)
electron isolation requirements, 
with respect to the entire
sample of ``loose'' electron candidates.  The loose candidates satisfy
only a subset of the nominal electron identification criteria. 
To reject prompt-electron contributions from $W$ decays or
the DY process, events are vetoed in the following cases: if
the missing transverse momentum \cite{ATLAS-CONF-2013-082} is larger than 25~\GeV, if they contain
two identified electrons satisfying strict criteria or if they contain
two electrons satisfying less strict criteria but with an invariant
mass between 71~\GeV\ and 111~\GeV.  A weighted average of the fake rates
obtained from the jet samples is then calculated.

In addition to the fake rate, the probability $r_1$ ($r_2$) that a
prompt electron in
this loose selection satisfies the nominal electron identification and
leading (subleading) isolation requirements is used in evaluating this
background.  This probability is taken from the MC simulation as a
function of \et\ and $\eta$.  Potential differences between data and
simulated samples in lepton identification and isolation efficiencies
are accounted for by applying scale factors \cite{ATLAS-CONF-2014-032}
to the simulation, which are generally close to unity.

A system of equations is used to solve for the unknown 
contribution to the background from events with one or more fake
electrons in the sample triggered with the default analysis trigger.
The relation between the number of true paired objects $N_{ab}$, with
$E^a_{\rm T} > E^b_{\rm T}$ and $a,b \in \{R,F\}$, and the number of
measured pairs
$N_{xy}$ , with $x,y \in \{T,L\}$, can be written as:
\begin{equation}
\label{ff_matrix_full}
\begin{pmatrix} \NTT \\ \NTL \\ \NLT \\ \NLL \end{pmatrix}
=
  \begin{pmatrix}
  r_1 r_2 & r_1 f_2 & f_1 r_2 & f_1 f_2 \\
  r_1 (1-r_2) & r_1 (1-f_2) & f_1(1-r_2) & f_1 (1-f_2) \\
  (1-r_1)r_2 & (1-r_1)f_2 & (1-f_1) r_2 & (1-f_1)f_2 \\
  (1-r_1)(1-r_2) & (1-r_1)(1-f_2) & (1-f_1)(1-r_2) & (1-f_1)(1-f_2)
  \end{pmatrix}
  \begin{pmatrix} \NRR \\ \NRF \\ \NFR \\ \NFF \end{pmatrix}.
\end{equation}

The subscripts $R$ and $F$ refer to prompt electrons and fake electrons (jets)
respectively. The subscript $T$ refers to electrons that pass the nominal
selection. The subscript $L$ corresponds to
electrons that pass the loose requirements described above but
fail the nominal requirements.

The background originating from pairs of objects with at least one
fake electron ($N_{TT}^{\rm{Multijet} \& \wpjet}$) in the total number
of pairs, where both objects are reconstructed as signal-like
(i.e. contribute to $\NTT$) is given by:
\begin{equation}
\label{bkg_true_quantities}
N_{TT}^{\rm{Multijet} \& \wpjet\ }  = r_1 f_2 N_{RF} + f_1 r_2 N_{FR} + f_1 f_2 N_{FF}.
\end{equation}
The number of true paired objects on the right-hand side of
equation~(\ref{bkg_true_quantities}) can be expressed in terms of
measurable quantities (\NTT, \NTL, \NLT, \NLL) by inverting the matrix
in equation~(\ref{ff_matrix_full}). 
The normalisation and shape of the background in each variable of interest
are automatically derived by using the measurable quantities
as a function of that same variable. The estimated multijet background
over the full invariant mass range is found to be about $3\%$.

\subsection{Multijet and $W$+jets background estimate in the muon channel}
%%%%%%%%%%%%%%%%%%%%%%%%%%%%%%%%%%%%%%%%%%%%%%%%%

The multijet background remaining after the complete event selection
in the muon channel is largely due to heavy flavour $b$- and $c$-quark
decays, and is estimated using a data-driven technique in two
s-eps-converted-to.pdf. This method also accounts for any potential $W$+jets background,
however, the contribution of this component is expected to be
negligible. First the normalisation of the multijet background in each $m_{\mu\mu}$
bin is determined, and then the shape in the $|y_{\mu\mu}|$ and in
$|\Delta \eta_{\mu\mu}|$ variables is estimated.

The background in each invariant mass region is determined using three
orthogonal control regions with inverted muon isolation requirements,
and/or inverted muon-pair charge requirements. The two variables are
largely uncorrelated for the multijet background. In each control
region the contamination from signal, top-quark, and diboson background is
subtracted using simulation. The yield of multijet events in the
signal region is predicted using the constraint that the yield ratio
of opposite-charge to same-charge muon pairs is identical in the isolated
and non-isolated regions.  A comparison of the isolation distribution
for muons in events with either same-charge and opposite-charge muon pairs shows
a small linear deviation of up to $10\%$ when extrapolated into the
isolated signal region. This is found to be independent of
$m_{\mu\mu}$, and is corrected for. For the region
$m_{\mu\mu}>500$~\GeV\ there are insufficient same-charge isolated muon
pairs to give a reliable estimate. Therefore, the background yield in
the region $m_{\mu\mu}<500$~\GeV\ is fitted to two alternative
functional forms and extrapolated to larger $m_{\mu\mu}$ where the
averaged prediction is taken as the estimate of the background yield.
The $|y_{\mu\mu}|$ and $|\Delta \eta_{\mu\mu}|$ dependence of the
background in each $m_{\mu\mu}$ region is obtained from a
multijet-enriched data control region in which pairs of same-charge
and opposite-charge muons satisfy $\sum p_{\rm T}(\Delta
R=0.2)/p^{\mu}_{\rm T}>0.1$. Signal, top-quark and diboson contamination in
this control region is subtracted using MC simulation. The resulting
$|y_{\mu\mu}|$ and $|\Delta \eta_{\mu\mu}|$ spectra in each
$m_{\mu\mu}$ region are normalised to the yield obtained in the first
step. For $m_{\mu\mu}>500$~\GeV\ the $|y_{\mu\mu}|$ or $|\Delta
\eta_{\mu\mu}|$ shape is taken from the region
$300<m_{\mu\mu}<500$~\GeV.  Overall the total multijet background
varies from $1\%$ to $0.1\%$ over the complete invariant mass range.

%-------------------------------------------------------------------------------
\section{Cross-section measurement}
\label{sec:methodology}
%-------------------------------------------------------------------------------

The Drell--Yan cross section, including the irreducible contribution
from the PI process, is measured differentially in $12$ bins of
$m_{\ell\ell}$ from $116$~GeV to $1500$~GeV, as well as
double-differentially in five bins of $m_{\ell\ell}$ as a function of
$|y_{\ell\ell}|$ and $|\Delta \eta_{\ell\ell}|$.  The results are
presented in the fiducial region of the measurement, in which the
leading (subleading) lepton has a $p_{\rm T}^{\ell} > 40$~GeV ($p_{\rm
  T}^{\ell} > 30$~GeV) and both leptons are within $|\eta^{\ell}| <
2.5$. The kinematic variables are defined by the leptons before FSR, i.e. the results are
given at the Born-level in QED. Results at the ``dressed'' level, where leptons
after FSR are recombined with radiated photons within a cone of
$\Delta R = 0.1$, are obtained by multiplying the Born-level results
with the dressed correction factors $k_{\rm dressed}$, provided in tables
~\ref{table:elecXsec_Mass}--\ref{table:muonXsec_dMdeta} in the appendix. These
correction factors are obtained from the {\sc Powheg} and {\sc
  Pythia}~8 MC samples for the DY and PI processes, respectively.

The double-differential cross section as a function of invariant mass and rapidity is 
calculated as
\begin{equation} 
 \frac{{\rm{d^2}} \sigma} {{\rm{d}} m_{\ell\ell} \, {{\rm{d}}
     |y_{\ell\ell}|}} = \frac{N_{\rm data} - N_{\rm bkg} }{C_{\rm
     {DY}}\,\mathcal{L}_{\rm{int}}} \, \frac{1}
{\Delta_{m_{\ell\ell}} \, 2  \Delta_{|y_{\ell\ell}|}},
\label{eq:xsec}
\end{equation}
where $N_{\rm data}$ is the number of candidate events observed in a
given bin of $m_{\ell\ell}$ and $|y_{\ell\ell}|$ of width
$\Delta_{m_{\ell\ell}}$ and $\Delta_{|y_{\ell\ell}|}$ respectively.
The total background in that bin is denoted as $N_{\rm bkg}$ and
$\mathcal{L}_{\rm{int}}$ is the integrated luminosity. The factor of
two in the denominator accounts for the modulus in the rapidity bin
width. The double-differential
cross section as a function of mass and $|\Delta \eta_{\ell\ell}|$ and the
single-differential measurement as a function of invariant mass are defined
accordingly.

The factor, $C_{\rm {DY}}$, takes into account the
efficiency of the signal selection and bin migration effects. It is
defined as the number of MC generated events that pass the signal
selection in a certain measurement bin calculated from the reconstructed
lepton kinematics divided by the total number of generated events
within the fiducial region, in the corresponding bin, calculated from
Born-level or dressed-level lepton kinematics.  It is obtained from the
Drell--Yan and PI MC samples after correction for differences in the
reconstruction, identification, trigger, and isolation efficiencies
between data and simulation, as well as for momentum scale and
resolution mismodelling effects. In general the $C_{\rm {DY}}$ factors
are found to be in the range 60--80\% across the measured kinematic
range.

The $C_{\rm {DY}}$ factor also includes extrapolations
over the small regions that are excluded for reconstructed electron
($1.37<|\eta^e|<1.52$ and $2.47<|\eta^e|<2.5$) or muon
($2.4<|\eta^{\mu}|<2.5$) candidates.  In the electron channel, the
fiducial cross section measurements as a function of $m_{ee}$ and $|y_{ee}|$, and the
single-differential measurement, are
extrapolated over the unmeasured region $|\Delta\eta_{ee}|>3.5$. The
extrapolation correction is included in the $C_{\rm {DY}}$
factor. No such extrapolation is required for the 
double-differential measurement as function of mass and
$|\Delta \eta_{ee}|$ which only extends to $|\Delta \eta_{ee}|=3$.

The Born-level bin purity is defined as the fraction of reconstructed
MC signal events in a given bin which were also generated in the same
bin using Born-level lepton kinematics. An analagous definition is
used for the dressed-level bin purity. The bin purities are found to
be typically above $85\%$, and above $75\%$ everywhere. This ensures
that the bin migration effects are small, and the corrections applied
to account for bin migrations have small uncertainties.

\FloatBarrier

%-------------------------------------------------------------------------------
\section{Systematic uncertainties}
\label{sec:sys}
%-------------------------------------------------------------------------------

The systematic uncertainties on the measurements are discussed
separately for those sources which arise only in the electron channel,
those which arise only in the muon channel, and those which are common
to both measurements. In each section the sources are discussed in
order of importance, with the largest sources of uncertainty listed
first. Each source is classified as being correlated or uncorrelated
between measurement bins in a single channel.  The uncorrelated
sources are propagated using the pseudo-experiment method in which
the correction factors used to improve the
modelling of data by the simulation are randomly shifted in an
ensemble of pseudo-experiments according to the mean and standard
deviation of the correction factor.  The resulting uncertainty on the
measured cross section is determined from the variance of the
measurements for the ensemble. The correlated contributions are
propagated by the offset method in which the values
from each source are coherently shifted upwards and downwards by one
standard deviation and the magnitude of the change in the measurement
is computed. The sign of the uncertainty corresponds to a one standard
deviation upward shift of the uncertainty source.

\subsection{Electron channel}
The systematic uncertainties on the cross section that are unique to
the electron channel are dominated by the uncertainties in the
determination of the multijet and $W$+jets background described in
section~\ref{sec:bg_elec_data_driven}, and in the electron energy
scale. In addition, a large contribution to the uncertainty also
arises from the top-quark and diboson background subtraction, and is
discussed in section~\ref{sec:sysCommon}.

All correlated and uncorrelated contributions to the systematic
uncertainties are given in each bin of the measurement in
tables~\ref{table:elecXsec_Mass}, \ref{table:elecXsec_dMdy},
and~\ref{table:elecXsec_dMdeta} of the appendix.

\subsubsection{Multijet and $W$+jets background}

In order to derive the uncertainty on the data-driven background
estimate described in section~\ref{sec:bg_elec_data_driven}, the
default ``matrix method'' is altered by assuming $r_1 = r_2 = 1$.
This second matrix method leads to a simplification of the matrix in
equation~(\ref{ff_matrix_full}), but also necessitates the use of MC
corrections to account for the identification and isolation inefficiencies of
real electrons. Large MC corrections can be avoided in a third matrix
method where the contamination from real electrons is reduced.
The subscript $L$ in equation~(\ref{ff_matrix_full}) now corresponds to electrons that pass the loose requirements
but fail the requirement on the matching between track and cluster, instead of failing the full
identification and isolation requirements.

In addition, two alternative background-enriched samples 
are obtained using 
a tag-and-probe technique on the jet-triggered sample and on the
sample triggered by the default analysis triggers, 
requiring the tag to fail certain aspects of the electron identification 
depending on the trigger.
Furthermore, the event should have a missing transverse momentum
smaller than 25 GeV, the probe needs to have the same charge as the
tag and the invariant mass of the tag-and-probe pair needs to be
outside the $Z$ mass window from 71 to 111~GeV.

The default and the two additional matrix methods are each used in
conjunction with the default and the two alternative
background-enriched samples, leading to a default and eight
alternative background estimates.
Out of the eight alternative background estimates those two are
identified that in general, i.e. in almost all bins except for
fluctuations,
yield the largest and smallest background contribution.  In each bin,
the average absolute difference between those two and the default
background estimate is used as a systematic uncertainty on the method.

Another systematic uncertainty can arise if fake rates are different
for the various processes contributing to this background, and if the
relative contributions of these processes differ between the data
samples from which the fake rates are measured and the data sample to
which the fake rates are applied.  For example, jets originating from
bottom quarks have a higher fake rate than light-quark jets, but the
effect of this is negligible as the number of $b$-jets is small and
similar in both samples.  However, as an additional check the
background is recalculated using all nine methods discussed above, but
with separate fake rates for different background processes. As the
mean of these nine methods is in agreement with the default background
estimate
no additional systematic uncertainty is applied.

The uncertainty on the default fake-rate calculation is derived by
varying the requirements used to suppress real electron contamination
in the data sample used to measure the fake rate.  The largest
deviation of about 5\% on the background occurs when the value of the
missing transverse energy requirement is changed.  It is added in
quadrature to the systematic uncertainty on the method to obtain the
full systematic uncertainty ($\delta^{\rm mult.}_{\rm cor}$) on the
cross section that is correlated between bins. The value of
$\delta^{\rm mult.}_{\rm cor}$ is found to be around $1\%$, rising to
almost $4\%$ at large $|\Delta\eta_{ee}|$.

The uncorrelated part consists of the statistical uncertainty on the
fake rates, which results in an uncertainty on the background of at most
5\%, and of the statistical uncertainty from the sample to which the
fake rates are applied. These two sources are added in quadrature and
yield the uncertainty ($\delta^{\rm mult.}_{\rm unc}$) on the cross section that is
uncorrelated between bins and is typically less than $0.5\%$,
increasing to $3\%$ at large $|\Delta\eta_{ee}|$.

\subsubsection{Energy scale and resolution}
The electron energy scale and resolution as well as the corresponding
uncertainties are determined using $Z\to e^+e^-$, $W\to e \nu$, and
$J/\psi\to e^+e^-$ decays~\cite{PERF-2013-05}.  The uncertainty on the
energy scale is separated into 14 uncorrelated systematic sources as
well as one statistical component. The statistical uncertainty on the
energy scale is found to be negligible. Adding the effects of the 14
sources of uncertainty on the energy scale
in quadrature after propagating to the measured cross
sections, the combined uncertainty is
denoted as $\delta^{\rm Escale}_{\rm cor}$, and is 1--4\% for
$m_{ee}>200$~\GeV, but is better than $0.5\%$ at lower $m_{ee}$ and
central rapidity.

The uncertainty on the energy resolution is separated into seven
uncorrelated systematic sources which are propagated to the cross-section measurements individually and then quadratically summed. This
combined uncertainty is denoted as $\delta^{\rm Eres}_{\rm cor}$ and is
typically 0.1--0.2\% everywhere except at large $|y_{ee}|$ or large
$|\Delta\eta_{ee}|$.

\subsubsection{Reconstruction, identification and isolation efficiency}
The reconstruction and identification efficiencies of electrons are
determined from data for electrons with $E_{\rm T}^e$ up to
about 100~\GeV, using various tag-and-probe methods in $Z$ and $J/\psi$
decays, following the prescription of
ref.~\cite{PERF-2013-03}  
with certain improvements and adjustments for the
2012 conditions~\cite{ATLAS-CONF-2014-032}.  In order to extend the
measurement range of the identification efficiency, the tag-and-probe
method using the isolation distribution of the probe for the
discrimination between signal and background in $Z\to e^+e^-$
decays~\cite{ATLAS-CONF-2014-032} is carried out up to about 500~\GeV\ in
$E_{\rm T}$. Within statistical uncertainties, the identification efficiencies are
found to be stable and consistent with the one derived in the last bin
($E_{\rm T}^e$ > 80~\GeV) in ref.~\cite{ATLAS-CONF-2014-032}.

The differences between the measured reconstruction and identification
efficiencies and their values in MC simulation are taken as $\eta$-
and $E_{\rm T}$-dependent scale factors with which the 
$C_{\rm {DY}}$ factor derived from simulation is corrected. Similarly, scale factors for the
isolation requirements on the leading and subleading electron are
derived using a tag-and-probe method in $Z\to e^+e^-$ decays. They
are applied as a function of $E_{\rm T}^e$ only, as the $\eta$ dependence is negligible.

The uncertainties on the cross section due to the systematic
uncertainties on the scale factors for the electron reconstruction,
identification and isolation as well as the statistical uncertainty on
the isolation are denoted as $\delta^{\rm reco}_{\rm cor}$, $\delta^{\rm
  id}_{\rm cor}$, $\delta^{\rm iso}_{\rm cor}$ and $\delta^{\rm iso}_{\rm unc}$
respectively. Of these, the largest component is $\delta^{\rm
  id}_{\rm cor}$, which is found to be 0.5--1\% everywhere. The
uncertainty $\delta^{\rm reco}_{\rm cor}$ is generally below $0.3\%$ and
better than $1\%$ everywhere. Both components of the isolation
efficiency uncertainty are found to be $0.2\%$ or better for
$m_{ee}<300$~\GeV.

\subsubsection{Trigger efficiency}
The trigger efficiency is measured in data and in the MC simulation
using a tag-and-probe method in $Z\to e^+e^-$ decays. The differences
as a function of $E_{\rm T}^e$ are found to be smaller than 1\% 
everywhere, with no dependence on $\eta$.  Therefore, $E_{\rm
  T}^e$-dependent scale factors are used to correct $C_{\rm {DY}}$.  The
uncertainty on the cross section due to the statistical ($\delta^{\rm
  trig}_{\rm unc}$) and systematic ($\delta^{\rm trig}_{\rm cor}$)
uncertainties on the trigger efficiency are each found to be $0.1\%$
or better for $m_{ee}<300$~\GeV.

%%%%%%%%%%%%%%%%%%%%%%%%%%%%%%%%%%%%%%%%%%%%%
\subsection{Muon channel}
%%%%%%%%%%%%%%%%%%%%%%%%%%%%%%%%%%%%%%%%%%%%%
Uncertainties related to the muon trigger, reconstruction, isolation
and impact parameter efficiencies, as well as the muon momentum scale
and resolution are all studied using the $Z\rightarrow\mu^+\mu^-$
process and a tag-and-probe method. Of these, the largest contribution
to the measurement precision arises from the reconstruction efficiency
modelling, and the muon momentum scale calibration. However, the
top-quark and diboson background subtraction is also a dominant source of
uncertainty, and is discussed in section~\ref{sec:sysCommon}.  
A detailed breakdown of the uncertainties is provided in
tables~\ref{table:muonXsec_Mass},~\ref{table:muonXsec_dMdy}
and~\ref{table:muonXsec_dMdeta} in the appendix.

\subsubsection{Reconstruction efficiency}
This is the dominant source of muon-related correlated systematic
uncertainty and is dominated at large $p_{\rm T}^{\mu}$ by the uncertainty
in contributions from catastrophic muon energy loss via
bremsstrahlung~\cite{PERF-2014-05}. When propagated to the cross
section this source is found to be typically $0.5\%$, rising to $1\%$
at the highest $p_{\rm T}^{\mu}$. This contribution is denoted as
$\delta^{\rm reco}_{\rm cor}$.

\subsubsection{Momentum scale and resolution}
The corrections on muon momentum scale and resolution are obtained from
fits to the $Z\rightarrow\mu^+\mu^-$ line-shape with
scale and resolution parameters in local $\eta^{\mu}$ and $\phi^{\mu}$ regions, separately
for muon tracks reconstructed in the ID and the MS~\cite{PERF-2014-05}. Uncertainties
arising from the methodology result in a correlated systematic
uncertainty on the measured cross sections of typically $0.4\%$. These contributions are listed as
$\delta^{\rm pT}_{\rm cor}$ for the momentum scale, and
$\delta^{\rm MSres}_{\rm cor}$ and $\delta^{\rm IDres}_{\rm cor}$ for the MS
and ID resolution uncertainties respectively.

\subsubsection{Isolation and impact parameter efficiency}
Modelling of the muon isolation and impact parameter selection
efficiencies can give rise to additional systematic uncertainties and
are estimated together. The sources considered include the remaining
background contamination, 
the residual variation in $\eta^{\mu}$, 
and a possible bias from the event topology 
estimated by varying the azimuthal opening angle between the two muons 
used in the tag-and-probe method.
The resulting correlated cross-section
uncertainty is found to be typically $0.1\%$, rising to $0.5\%$ at
large $m_{\mu\mu}$. This contribution is labelled as $\delta^{\rm
  iso}_{\rm cor}$.

\subsubsection{Multijet and $W$+jets background}
The uncertainty on the multijet background estimation consists of
several sources. The statistical uncertainty of the control regions is
propagated in the appropriate way and can be significant, in
particular from the isolated same-sign control sample. The subtracted
top-quark and diboson contamination in the control regions is varied within
the theoretical cross section uncertainties given in
section~\ref{sec:MC}.  
The subtracted signal contamination is varied by $\pm5\%$. 
The extrapolation uncertainty in the multijet
estimate at large invariant mass is estimated taking half the
difference between the two extrapolation functions and is the largest
contribution to the uncertainty in this region. The uncertainty in the
shape of the $|y_{\mu\mu}|$ and $|\Delta\eta_{\mu\mu}|$ spectra is
determined from the RMS of these distributions in regions of small,
moderate and large non-isolation of the muon pairs obtained from the
control regions. This component typically dominates the uncertainty at
large $|y_{\mu\mu}|$ and small $|\Delta\eta_{\mu\mu}|$. These
sources are combined into correlated and uncorrelated components,
where the latter is due to the statistical uncertainty of the
method, and are denoted as $\delta^{\rm mult.}_{\rm cor}$ and
$\delta^{\rm mult.}_{\rm unc}$. The combined
uncertainty in the background estimation ranges from $20\%$ at low
$m_{\mu\mu}$ to above $100\%$ at the highest $m_{\mu\mu}$, however
when propagated onto the cross section measurements it is below $1\%$
in all double differential measurement bins.

\subsubsection{Trigger efficiency}
The trigger efficiency corrections obtained using the tag-and-probe
method as described in ref.~\cite{TRIG-2012-03} are parameterised in terms of muon
pseudorapidity $\eta^{\mu}$, azimuthal angle $\phi^{\mu}$, and
electric charge. The uncertainty sources considered are the background
contamination, a possible residual dependence on muon $p_{\rm T}^{\mu}$, and
an uncertainty based on the event topology. 
This results in a correlated uncertainty on the measured cross
sections of $0.1\%$ and is denoted as
$\delta^{\rm trig}_{\rm cor}$.

%%%%%%%%%%%%%%%%%%%%%%%%%%%%%%%%%%%%%%%%%%%%%
\subsection{Systematic uncertainties common to both channels}
%%%%%%%%%%%%%%%%%%%%%%%%%%%%%%%%%%%%%%%%%%%%%
\label{sec:sysCommon}
The systematic uncertainties common to both channels are derived using identical
methods and are assumed to be fully correlated between the
channels. They are dominated by the uncertainty on the
\ttbar~background and the luminosity. The statistical uncertainties of
the MC samples used are uncorrelated between the measurement channels.

\subsubsection{Top and diboson background}
In most of the phase space the largest background in both the electron
and the muon channel is due to top-quark production. The normalisation
uncertainty on this background is taken to be 6\% as already described
in section~\ref{sec:MC}. The top-quark background is dominated by
\ttbar~production, which is modelled using two different MC generators, see
table~\ref{tab:mc}.  No systematic differences have been found when
comparing the distributions in dilepton invariant mass, rapidity and $\Delta
\eta$.
Another important background is due to diboson ($WW$, $WZ$ and $ZZ$)
production. The normalisation uncertainties are about $10\%$ (see
section~\ref{sec:MC}).

To further validate the normalisation and shape of the \ttbar\ and diboson simulations
used in this analysis, a data control region is defined by an
opposite-charge electron--muon pair selection that is chosen to match the
nominal muon and electron pair selections as closely as possible. This
selection strongly suppresses the Drell--Yan process and leads to a
sample with about an 80\% contribution from top-quark production at lower
invariant masses, decreasing to about 50\% for the highest masses. The
remaining events are mainly due to diboson production.  As no
systematic differences are found comparing dilepton invariant mass,
rapidity and $\Delta \eta$ distributions in data and simulation no further
uncertainties are assigned. The normalisation uncertainties of the top
($\delta^{\rm top}_{\rm cor}$) and diboson ($\delta^{\rm diboson}_{\rm cor}$)
backgrounds as well as the statistical uncertainty of both simulations
combined ($\delta^{\rm bgMC}_{\rm unc}$) are given
in the appendix.

\subsubsection{Luminosity}
The uncertainty on the integrated luminosity is 1.9\%, which is
derived by following the methodology detailed in
ref.~\cite{DAPR-2011-01}.

\subsubsection{MC statistics and MC modelling}
The bin-by-bin correction factor $C_{\rm DY}$ is calculated from the
Drell--Yan and PI MC samples as already described in
section~\ref{sec:methodology}. In order to check for a potential
model dependence, two different Drell--Yan samples are generated, see
table~\ref{tab:mc}, and no systematic differences have been found for
$C_{\rm DY}$ as a function of invariant mass, rapidity and $\Delta \eta$.
Similarly, a negligible influence on these distributions is found when
assigning a 40\% uncertainty to the cross section of the PI MC
sample. This value is derived by calculating the PI contribution in a
constituent and a current quark mass scheme~\cite{MRST2004QED}, and taking the magnitude
of the difference between either scheme and their
average~\cite{STDM-2012-10}.  The statistical uncertainty due to the
finite number of signal MC events used in the calculation of $C_{\rm DY}$
is denoted as $\delta^{\rm MC}_{\rm unc}$.

\subsubsection{Bin-by-bin correction}

The bin-by-bin correction used in the calculation of the cross section 
is compared to an iterative Bayesian unfolding
technique~\cite{D'Agostini:1994zf} as implemented in the program
RooUnfold~\cite{Adye:2011gm}.  The differences between these two
approaches are found to be negligible.

\subsubsection{PDF uncertainty}
As discussed in section \ref{sec:methodology}, the $C_{\rm DY}$ correction
factor also includes a small extrapolation from the measured region to
the fiducial region. This acceptance correction is about 13\% for
electrons and 5\% for muons, but
can be larger in certain bins of the two-dimensional cross-section
measurement. The PDF uncertainties due to the acceptance correction
are estimated using the CT10 PDF eigenvector set at 68\% confidence
level (CL). 
They are found to be negligible, 
with uncertainties of the order of 0.1\% or below 
for most cross-section measurement bins, and below 0.2\% everywhere.

\FloatBarrier
%-------------------------------------------------------------------------------
\section{Results}
\label{sec:results}
%-------------------------------------------------------------------------------

The measured Born-level fiducial cross sections for the electron and
muon channel analyses are in good agreement with one another and each
achieve a precision of $1\%$ at low $m_{\ell\ell}$ where they are
dominated by the experimental systematic uncertainties. For
$m_{\ell\ell}\gtrsim 400$~\GeV\ the statistical uncertainty of the data
samples dominates the measurement precision. The measurements are
presented in
tables~\ref{table:elecXsec_Mass}--\ref{table:muonXsec_dMdeta} of the
appendix, including the systematic uncertainties (excluding the
luminosity measurement uncertainty) separated into those which are
point-to-point correlated and those which are uncorrelated.

The two measurement channels are defined with a common fiducial region
given in section~\ref{sec:methodology}, therefore, they can be
combined to further reduce the statistical uncertainty. A further
reduction in the total uncertainty is also achieved by taking into
account systematic uncertainties which are not correlated between the two
measurement channels. A $\chi^2$ minimisation technique is used to
perform the combination of the cross
sections~\cite{Glazov:2005rn,Aaron:2009bp,Aaron:2009aa}. This method
introduces a free nuisance parameter for each correlated systematic
error source which contributes to the total $\chi^2$, and therefore
gives results that are different from a simple weighted average. The
combination is performed separately for each differential
cross-section measurement. The sources of uncertainty considered are
discussed in section~\ref{sec:sys}, some of which consist of several
contributions.  However, for ease of presentation only the major
sources are reported in
tables~\ref{table:elecXsec_Mass}--\ref{table:muonXsec_dMdeta} giving a
total of $35$ nuisance parameters. Only the normalisation
uncertainties of the \ttbar~and diboson backgrounds are correlated
between the channels, using two common nuisance parameters.  After the
minimisation, no nuisance parameter is shifted by more than one
standard deviation. The $\chi^2$ per degree of freedom, $\chi^2/dof$,
is found to be $14.2/12=1.19$ for the single-differential cross
section, $53.1/48=1.11$ for the cross sections differential in
$m_{\ell\ell}$ and $|y_{\ell\ell}|$, and $59.3/47=1.26$ for the cross
sections differential in $m_{\ell\ell}$ and $|\Delta
\eta_{\ell\ell}|$.

The combined fiducial cross sections are presented in tables~\ref{table:combXsec_Mass},
~\ref{table:combXsec_dMdy} and~\ref{table:combXsec_dMdeta}. The
combination procedure yields orthogonal systematic uncertainty sources,
as listed in the tables, which are formed of linear combinations of the
original uncertainties. The combined cross sections have an improved
precision across the kinematic range, where at low $m_{\ell\ell}$ an
accuracy of better than $1\%$ is attained. The double-differential
cross sections have an accuracy of between $1\%$ and $7\%$ throughout the
kinematic range of the measurements.

\begin{sidewaystable}[!htb]
\begin{center}
\resizebox{\textwidth}{!}{
\begin{tabular}{r|rrrr|rrrrrrrrrrrrrrrrrrrrrrrrrrrrrrrrrrrr}
\hline\hline
$m_{\ell\ell}$ & $\frac{\text{d}\sigma}{\text{d}m_{\ell\ell}}$  & 
$\delta^{\rm stat}$           &  $\delta^{\rm sys}$          & $\delta^{\rm tot}$ &
$\delta^{\rm unc}$          &  
$\delta^{\rm 1}_{\rm cor}$     &  $\delta^{\rm 2}_{\rm cor}$   &
$\delta^{\rm 3}_{\rm cor}$     &  $\delta^{\rm 4}_{\rm cor}$   &
$\delta^{\rm 5}_{\rm cor}$     &  $\delta^{\rm 6}_{\rm cor}$   &
$\delta^{\rm 7}_{\rm cor}$     &  $\delta^{\rm 8}_{\rm cor}$   &
$\delta^{\rm 9}_{\rm cor}$     &  $\delta^{\rm 10}_{\rm cor}$   &
$\delta^{\rm 11}_{\rm cor}$     &  $\delta^{\rm 12}_{\rm cor}$   &
$\delta^{\rm 13}_{\rm cor}$     &  $\delta^{\rm 14}_{\rm cor}$   &
$\delta^{\rm 15}_{\rm cor}$     &  $\delta^{\rm 16}_{\rm cor}$   &
$\delta^{\rm 17}_{\rm cor}$     &  $\delta^{\rm 18}_{\rm cor}$   &
$\delta^{\rm 19}_{\rm cor}$     &  $\delta^{\rm 20}_{\rm cor}$   &
$\delta^{\rm 21}_{\rm cor}$     &  $\delta^{\rm 22}_{\rm cor}$   &
$\delta^{\rm 23}_{\rm cor}$     &  $\delta^{\rm 24}_{\rm cor}$   &
$\delta^{\rm 25}_{\rm cor}$     &  $\delta^{\rm 26}_{\rm cor}$   &
$\delta^{\rm 27}_{\rm cor}$     &  $\delta^{\rm 28}_{\rm cor}$   &
$\delta^{\rm 29}_{\rm cor}$     &  $\delta^{\rm 30}_{\rm cor}$   &
$\delta^{\rm 31}_{\rm cor}$     &  $\delta^{\rm 32}_{\rm cor}$   &
$\delta^{\rm 33}_{\rm cor}$     &  $\delta^{\rm 34}_{\rm cor}$     &  $\delta^{\rm 35}_{\rm cor}$\\
{[\GeV]}  & [pb/\GeV] & [\%] & [\%] & [\%] & [\%] & [\%]& [\%] & [\%]& [\%] & [\%]& [\%] & [\%]& [\%] & [\%]& [\%] & [\%] & [\%] & [\%] & [\%] & [\%] & [\%] & [\%] & [\%] & [\%] & [\%] & [\%] & [\%] & [\%] & [\%] & [\%] & [\%] & [\%] & [\%] & [\%] & [\%] & [\%] & [\%] & [\%] & [\%] & [\%] \\
\hline
116--130 & $2.28 \times 10^{-1}$ & 0.34 & 0.53 & 0.63 & 0.12 & 0.24 & -0.01 & 0.08 & -0.03 & -0.03 & 0.00 & 0.01 & 0.01 & 0.00 & 0.01 & 0.00 & -0.00 & 0.00 & 0.00 & -0.01 & 0.01 & -0.00 & 0.01 & -0.02 & 0.01 & -0.02 & -0.05 & -0.31 & 0.15 & 0.18 & 0.03 & -0.04 & -0.01 & -0.02 & 0.11 & 0.04 & 0.07 & 0.07 & 0.09 & -0.05 \\ 
130--150 & $1.04 \times 10^{-1}$ & 0.44 & 0.67 & 0.80 & 0.13 & 0.38 & -0.00 & 0.03 & -0.05 & -0.02 & 0.03 & -0.01 & 0.00 & 0.00 & 0.01 & -0.00 & -0.00 & 0.00 & 0.00 & -0.01 & -0.01 & -0.01 & 0.02 & -0.01 & 0.01 & 0.08 & -0.08 & -0.38 & 0.10 & 0.15 & 0.04 & -0.08 & 0.01 & 0.04 & 0.22 & 0.03 & 0.08 & 0.14 & 0.08 & -0.07 \\ 
150--175 & $4.98 \times 10^{-2}$ & 0.57 & 0.91 & 1.08 & 0.18 & 0.56 & 0.01 & -0.02 & -0.05 & -0.01 & 0.05 & -0.03 & 0.00 & 0.00 & 0.01 & -0.00 & 0.00 & 0.00 & 0.00 & -0.02 & -0.02 & -0.01 & 0.03 & 0.00 & 0.00 & 0.15 & -0.10 & -0.47 & 0.06 & 0.15 & 0.04 & -0.12 & -0.00 & 0.09 & 0.35 & 0.04 & 0.10 & 0.16 & 0.10 & -0.09 \\ 
175--200 & $2.54 \times 10^{-2}$ & 0.81 & 1.18 & 1.43 & 0.25 & 0.74 & -0.00 & -0.06 & -0.07 & -0.01 & 0.06 & -0.05 & -0.01 & -0.00 & 0.01 & -0.00 & 0.00 & 0.01 & 0.00 & -0.02 & -0.03 & -0.01 & 0.04 & 0.01 & 0.00 & 0.23 & -0.11 & -0.58 & 0.02 & 0.14 & 0.07 & -0.12 & -0.00 & 0.13 & 0.47 & 0.03 & 0.12 & 0.17 & 0.10 & -0.12 \\ 
200--230 & $1.37 \times 10^{-2}$ & 1.02 & 1.42 & 1.75 & 0.32 & 0.89 & 0.02 & -0.09 & -0.06 & -0.01 & 0.08 & -0.05 & -0.00 & 0.00 & 0.00 & -0.00 & 0.00 & 0.00 & -0.00 & -0.02 & -0.04 & -0.00 & 0.04 & 0.02 & -0.00 & 0.29 & -0.12 & -0.67 & -0.01 & 0.17 & 0.05 & -0.16 & 0.02 & 0.16 & 0.58 & 0.05 & 0.16 & 0.21 & 0.16 & -0.15 \\ 
230--260 & $7.89 \times 10^{-3}$ & 1.36 & 1.59 & 2.09 & 0.43 & 0.99 & -0.01 & -0.12 & -0.08 & 0.00 & 0.07 & -0.06 & -0.01 & 0.00 & 0.01 & -0.00 & 0.00 & 0.01 & -0.00 & -0.02 & -0.04 & 0.00 & 0.05 & 0.03 & -0.01 & 0.28 & -0.11 & -0.74 & 0.04 & 0.19 & 0.09 & -0.14 & -0.00 & 0.23 & 0.65 & 0.06 & 0.23 & 0.10 & 0.22 & -0.18 \\ 
260--300 & $4.43 \times 10^{-3}$ & 1.58 & 1.67 & 2.30 & 0.46 & 1.06 & 0.02 & -0.11 & -0.05 & 0.01 & 0.12 & -0.06 & -0.00 & 0.01 & 0.00 & -0.00 & -0.00 & -0.00 & -0.00 & -0.04 & -0.06 & 0.01 & 0.07 & 0.04 & 0.00 & 0.35 & -0.19 & -0.73 & 0.00 & 0.17 & 0.05 & -0.15 & 0.04 & 0.17 & 0.68 & 0.08 & 0.22 & 0.18 & 0.22 & -0.19 \\ 
300--380 & $1.87 \times 10^{-3}$ & 1.73 & 1.80 & 2.50 & 0.56 & 1.12 & -0.02 & -0.11 & -0.09 & 0.01 & 0.09 & -0.07 & -0.01 & 0.00 & 0.01 & -0.01 & 0.00 & 0.01 & -0.00 & -0.03 & -0.06 & -0.00 & 0.06 & 0.01 & -0.01 & 0.29 & -0.18 & -0.79 & 0.03 & 0.15 & 0.08 & -0.13 & -0.00 & 0.20 & 0.76 & 0.06 & 0.30 & 0.03 & 0.29 & -0.20 \\ 
380--500 & $6.20 \times 10^{-4}$ & 2.42 & 1.71 & 2.96 & 0.63 & 1.03 & 0.00 & -0.08 & -0.10 & 0.04 & 0.14 & -0.07 & -0.00 & 0.01 & 0.01 & -0.00 & -0.00 & -0.00 & 0.00 & -0.06 & -0.08 & 0.01 & 0.08 & 0.01 & 0.02 & 0.30 & -0.26 & -0.69 & 0.09 & 0.20 & 0.05 & -0.13 & 0.05 & 0.16 & 0.59 & 0.06 & 0.36 & 0.03 & 0.39 & -0.25 \\ 
500--700 & $1.53 \times 10^{-4}$ & 3.65 & 1.68 & 4.02 & 0.57 & 0.87 & -0.08 & -0.06 & -0.14 & 0.04 & 0.15 & 0.01 & 0.02 & 0.01 & -0.00 & -0.01 & 0.01 & 0.03 & 0.00 & -0.05 & -0.05 & 0.03 & 0.17 & 0.04 & 0.12 & 0.02 & -0.21 & -0.56 & 0.10 & 0.03 & 0.01 & 0.06 & -0.15 & 0.06 & 0.38 & 0.13 & 0.96 & -0.09 & 0.35 & -0.18 \\ 
700--1000 & $2.66 \times 10^{-5}$ & 6.98 & 1.85 & 7.22 & 1.02 & 0.73 & -0.09 & 0.04 & -0.13 & 0.07 & 0.17 & 0.03 & 0.03 & 0.03 & 0.00 & -0.01 & -0.01 & 0.02 & 0.01 & -0.07 & -0.07 & 0.05 & 0.17 & -0.01 & 0.14 & -0.15 & -0.26 & -0.44 & 0.23 & 0.08 & 0.06 & 0.13 & -0.09 & 0.02 & 0.17 & 0.19 & 1.00 & -0.17 & 0.50 & -0.17 \\ 
1000--1500 & $2.66 \times 10^{-6}$ & 17.05 & 2.95 & 17.31 & 2.26 & 0.71 & 0.04 & 0.16 & -0.01 & 0.06 & 0.33 & 0.10 & 0.08 & 0.05 & -0.04 & -0.00 & -0.00 & -0.01 & -0.01 & -0.15 & -0.14 & 0.02 & 0.22 & -0.08 & 0.16 & -0.10 & -0.49 & -0.32 & 0.21 & 0.23 & 0.08 & -0.17 & 0.01 & -0.34 & 0.28 & 0.32 & 1.21 & -0.03 & 0.69 & -0.35 \\ 
\hline
\end{tabular} }%end of resizebox
\caption{The combined Born-level single-differential cross section $\frac{\text{d}\sigma}{\text{d}m_{\ell\ell}}$. 
The measurements are listed together with the statistical ($\delta^{\rm stat}$), systematic ($\delta^{\rm sys}$) and total ($\delta^{\rm tot}$) uncertainties.
In addition the contributions from the individual correlated ($\delta^{\rm 1}_{\rm cor}$-$\delta^{\rm 35}_{\rm cor}$) and uncorrelated ($\delta^{\rm unc}$) systematic error sources are also provided.
The luminosity uncertainty of 1.9\% is not shown and not included in the overall systematic and total uncertainties.}
\label{table:combXsec_Mass}
\end{center}
\end{sidewaystable}

\begin{sidewaystable}[!htb]
\begin{center}
\resizebox{\textwidth}{!}{
\begin{tabular}{rc|rrrr|rrrrrrrrrrrrrrrrrrrrrrrrrrrrrrrrrrrr}
\hline\hline
$m_{\ell\ell}$ & $|y_{\ell\ell}|$ & $\frac{\text{d}^2\sigma}{\text{d}m_{\ell\ell}\text{d}|y_{\ell\ell}|}$  & 
$\delta^{\rm stat}$           &  $\delta^{\rm sys}$          & $\delta^{\rm tot}$ &
$\delta^{\rm unc}$          &  
$\delta^{\rm 1}_{\rm cor}$     &  $\delta^{\rm 2}_{\rm cor}$   &
$\delta^{\rm 3}_{\rm cor}$     &  $\delta^{\rm 4}_{\rm cor}$   &
$\delta^{\rm 5}_{\rm cor}$     &  $\delta^{\rm 6}_{\rm cor}$   &
$\delta^{\rm 7}_{\rm cor}$     &  $\delta^{\rm 8}_{\rm cor}$   &
$\delta^{\rm 9}_{\rm cor}$     &  $\delta^{\rm 10}_{\rm cor}$   &
$\delta^{\rm 11}_{\rm cor}$     &  $\delta^{\rm 12}_{\rm cor}$   &
$\delta^{\rm 13}_{\rm cor}$     &  $\delta^{\rm 14}_{\rm cor}$   &
$\delta^{\rm 15}_{\rm cor}$     &  $\delta^{\rm 16}_{\rm cor}$   &
$\delta^{\rm 17}_{\rm cor}$     &  $\delta^{\rm 18}_{\rm cor}$   &
$\delta^{\rm 19}_{\rm cor}$     &  $\delta^{\rm 20}_{\rm cor}$   &
$\delta^{\rm 21}_{\rm cor}$     &  $\delta^{\rm 22}_{\rm cor}$   &
$\delta^{\rm 23}_{\rm cor}$     &  $\delta^{\rm 24}_{\rm cor}$   &
$\delta^{\rm 25}_{\rm cor}$     &  $\delta^{\rm 26}_{\rm cor}$   &
$\delta^{\rm 27}_{\rm cor}$     &  $\delta^{\rm 28}_{\rm cor}$   &
$\delta^{\rm 29}_{\rm cor}$     &  $\delta^{\rm 30}_{\rm cor}$   &
$\delta^{\rm 31}_{\rm cor}$     &  $\delta^{\rm 32}_{\rm cor}$   &
$\delta^{\rm 33}_{\rm cor}$     &  $\delta^{\rm 34}_{\rm cor}$     &  $\delta^{\rm 35}_{\rm cor}$\\
{[GeV]} &  & [pb/GeV] & [\%] & [\%] & [\%] & [\%] & [\%]& [\%] & [\%]& [\%] & [\%]& [\%] & [\%]& [\%] & [\%]& [\%] & [\%] & [\%] & [\%] & [\%] & [\%] & [\%] & [\%] & [\%] & [\%] & [\%] & [\%] & [\%] & [\%] & [\%] & [\%] & [\%] & [\%] & [\%] & [\%] & [\%] & [\%] & [\%] & [\%] & [\%] & [\%] \\
\hline
116--150 & 0.0--0.2 & $4.10 \times 10^{-2}$ & 0.81 & 0.62 & 1.02 & 0.20 & 0.00 & 0.02 & 0.00 & -0.03 & 0.01 & 0.06 & 0.07 & -0.05 & -0.07 & -0.01 & 0.02 & 0.02 & 0.01 & 0.15 & -0.15 & -0.02 & -0.02 & -0.08 & 0.09 & -0.19 & -0.08 & 0.05 & -0.14 & 0.09 & -0.21 & 0.13 & -0.19 & 0.09 & -0.15 & 0.04 & 0.01 & 0.25 & -0.03 & -0.02 & 0.03  \\ 
116--150 & 0.2--0.4 & $4.04 \times 10^{-2}$ & 0.82 & 0.63 & 1.03 & 0.21 & 0.00 & 0.01 & 0.01 & -0.02 & 0.01 & 0.06 & 0.08 & -0.05 & -0.05 & 0.00 & 0.02 & 0.04 & 0.02 & 0.14 & -0.16 & -0.01 & -0.01 & -0.10 & 0.07 & -0.20 & -0.08 & 0.06 & -0.19 & 0.05 & -0.20 & 0.10 & -0.19 & 0.10 & -0.16 & 0.01 & -0.02 & 0.24 & -0.02 & -0.06 & 0.03  \\ 
116--150 & 0.4--0.6 & $4.01 \times 10^{-2}$ & 0.83 & 0.62 & 1.04 & 0.21 & 0.01 & 0.01 & 0.02 & -0.02 & -0.01 & 0.06 & 0.07 & -0.05 & -0.05 & -0.00 & 0.02 & 0.03 & 0.03 & 0.13 & -0.17 & -0.03 & -0.03 & -0.08 & 0.09 & -0.17 & -0.07 & 0.06 & -0.19 & 0.03 & -0.18 & 0.13 & -0.19 & 0.14 & -0.12 & -0.01 & -0.01 & 0.26 & -0.01 & -0.06 & 0.04  \\ 
116--150 & 0.6--0.8 & $4.04 \times 10^{-2}$ & 0.83 & 0.59 & 1.02 & 0.21 & 0.01 & 0.01 & 0.03 & -0.02 & -0.02 & 0.06 & 0.07 & -0.04 & -0.06 & -0.01 & 0.02 & 0.01 & 0.01 & 0.13 & -0.16 & -0.03 & -0.03 & -0.04 & 0.09 & -0.16 & -0.09 & 0.05 & -0.18 & 0.06 & -0.18 & 0.05 & -0.15 & 0.08 & -0.19 & 0.01 & 0.05 & 0.23 & 0.01 & -0.05 & 0.04  \\ 
116--150 & 0.8--1.0 & $3.97 \times 10^{-2}$ & 0.84 & 0.59 & 1.03 & 0.22 & 0.02 & 0.01 & 0.03 & -0.02 & -0.02 & 0.06 & 0.07 & -0.04 & -0.05 & -0.01 & 0.01 & 0.01 & 0.03 & 0.11 & -0.14 & -0.04 & -0.04 & -0.06 & 0.04 & -0.17 & -0.06 & 0.09 & -0.19 & -0.01 & -0.16 & 0.06 & -0.12 & 0.15 & -0.18 & -0.00 & 0.06 & 0.25 & 0.03 & 0.02 & 0.03  \\ 
116--150 & 1.0--1.2 & $3.95 \times 10^{-2}$ & 0.84 & 0.56 & 1.01 & 0.22 & 0.02 & 0.00 & 0.04 & -0.01 & -0.03 & 0.06 & 0.07 & -0.04 & -0.03 & 0.01 & -0.01 & 0.01 & 0.03 & 0.09 & -0.12 & -0.04 & -0.04 & -0.05 & 0.05 & -0.12 & -0.04 & 0.06 & -0.19 & -0.05 & -0.10 & 0.05 & -0.13 & 0.12 & -0.20 & -0.04 & 0.07 & 0.27 & 0.00 & -0.06 & 0.04  \\ 
116--150 & 1.2--1.4 & $3.84 \times 10^{-2}$ & 0.84 & 0.58 & 1.02 & 0.23 & 0.02 & -0.00 & 0.03 & -0.01 & -0.03 & 0.05 & 0.07 & -0.04 & -0.02 & 0.01 & -0.02 & 0.01 & 0.02 & 0.07 & -0.11 & -0.05 & -0.05 & -0.03 & 0.01 & -0.12 & -0.05 & 0.05 & -0.19 & -0.09 & -0.06 & -0.00 & -0.13 & 0.18 & -0.15 & -0.02 & 0.08 & 0.29 & 0.01 & -0.09 & 0.06  \\ 
116--150 & 1.4--1.6 & $3.48 \times 10^{-2}$ & 0.89 & 0.58 & 1.06 & 0.24 & 0.02 & -0.01 & 0.02 & -0.01 & -0.01 & 0.06 & 0.07 & -0.04 & -0.01 & 0.01 & -0.04 & 0.02 & 0.03 & 0.04 & -0.11 & -0.05 & -0.05 & -0.04 & -0.02 & -0.10 & 0.01 & 0.08 & -0.17 & -0.14 & -0.08 & -0.03 & -0.14 & 0.07 & -0.13 & -0.08 & 0.00 & 0.31 & -0.03 & -0.15 & 0.08  \\ 
116--150 & 1.6--1.8 & $2.92 \times 10^{-2}$ & 1.00 & 0.64 & 1.18 & 0.28 & 0.02 & -0.00 & 0.01 & -0.01 & -0.01 & 0.06 & 0.08 & -0.04 & -0.01 & 0.00 & -0.04 & 0.01 & 0.03 & 0.04 & -0.11 & -0.05 & -0.05 & -0.01 & -0.04 & -0.10 & -0.00 & 0.07 & -0.19 & -0.22 & -0.06 & -0.02 & -0.09 & 0.11 & -0.13 & -0.04 & -0.00 & 0.35 & -0.04 & -0.12 & 0.10  \\ 
116--150 & 1.8--2.0 & $2.29 \times 10^{-2}$ & 1.14 & 0.70 & 1.33 & 0.31 & 0.02 & -0.01 & 0.01 & -0.01 & -0.01 & 0.07 & 0.09 & -0.05 & -0.01 & 0.01 & -0.06 & 0.01 & 0.03 & 0.02 & -0.12 & -0.07 & -0.07 & -0.05 & -0.04 & -0.06 & -0.01 & 0.07 & -0.21 & -0.28 & -0.07 & -0.09 & -0.09 & 0.12 & -0.13 & -0.07 & -0.04 & 0.34 & -0.05 & -0.14 & 0.11  \\ 
116--150 & 2.0--2.2 & $1.54 \times 10^{-2}$ & 1.40 & 0.88 & 1.65 & 0.40 & 0.02 & 0.00 & 0.00 & -0.00 & -0.02 & 0.06 & 0.09 & -0.07 & -0.02 & -0.01 & -0.05 & 0.00 & 0.01 & 0.07 & -0.12 & -0.07 & -0.07 & 0.03 & -0.03 & -0.04 & -0.01 & 0.07 & -0.31 & -0.15 & 0.01 & -0.15 & -0.02 & 0.25 & -0.15 & 0.18 & 0.12 & 0.45 & -0.24 & -0.01 & 0.17  \\ 
116--150 & 2.2--2.4 & $7.31 \times 10^{-3}$ & 2.33 & 1.46 & 2.75 & 0.72 & 0.03 & -0.00 & -0.03 & 0.03 & -0.01 & 0.05 & 0.13 & -0.08 & 0.01 & 0.02 & -0.08 & 0.02 & 0.01 & 0.04 & -0.15 & -0.07 & -0.07 & 0.03 & -0.07 & -0.12 & -0.00 & 0.10 & -0.48 & -0.27 & 0.13 & -0.14 & -0.00 & 0.18 & -0.34 & 0.26 & 0.26 & 0.74 & -0.46 & 0.13 & 0.33  \\ 
\hline
150--200 & 0.0--0.2 & $1.04 \times 10^{-2}$ & 1.41 & 1.31 & 1.93 & 0.32 & -0.00 & 0.04 & 0.00 & -0.06 & 0.01 & 0.07 & 0.10 & -0.07 & -0.12 & -0.00 & 0.12 & 0.06 & 0.04 & 0.31 & -0.30 & -0.02 & -0.02 & -0.22 & 0.29 & -0.43 & -0.22 & 0.03 & -0.26 & 0.32 & -0.42 & 0.37 & -0.50 & 0.11 & -0.27 & -0.06 & -0.08 & 0.42 & -0.00 & -0.03 & 0.07  \\ 
150--200 & 0.2--0.4 & $1.04 \times 10^{-2}$ & 1.41 & 1.28 & 1.91 & 0.33 & -0.00 & 0.04 & 0.01 & -0.05 & 0.00 & 0.07 & 0.10 & -0.06 & -0.13 & -0.00 & 0.11 & 0.05 & 0.03 & 0.30 & -0.30 & -0.01 & -0.01 & -0.19 & 0.27 & -0.42 & -0.21 & 0.05 & -0.26 & 0.32 & -0.42 & 0.36 & -0.46 & 0.15 & -0.26 & -0.03 & -0.05 & 0.42 & 0.01 & -0.05 & 0.07  \\ 
150--200 & 0.4--0.6 & $1.03 \times 10^{-2}$ & 1.42 & 1.22 & 1.87 & 0.32 & -0.00 & 0.04 & 0.03 & -0.05 & 0.00 & 0.07 & 0.10 & -0.06 & -0.12 & -0.01 & 0.10 & 0.05 & 0.02 & 0.29 & -0.29 & -0.03 & -0.03 & -0.17 & 0.26 & -0.40 & -0.19 & 0.05 & -0.23 & 0.29 & -0.38 & 0.33 & -0.45 & 0.13 & -0.28 & -0.03 & -0.01 & 0.42 & 0.01 & -0.03 & 0.07  \\ 
150--200 & 0.6--0.8 & $1.05 \times 10^{-2}$ & 1.39 & 1.11 & 1.78 & 0.32 & 0.01 & 0.03 & 0.02 & -0.04 & -0.02 & 0.06 & 0.09 & -0.06 & -0.09 & -0.01 & 0.08 & 0.05 & 0.04 & 0.25 & -0.25 & -0.04 & -0.04 & -0.16 & 0.23 & -0.33 & -0.19 & 0.04 & -0.30 & 0.23 & -0.29 & 0.33 & -0.38 & 0.13 & -0.29 & -0.05 & 0.01 & 0.40 & 0.00 & -0.03 & 0.07  \\ 
150--200 & 0.8--1.0 & $1.02 \times 10^{-2}$ & 1.42 & 1.02 & 1.75 & 0.33 & 0.01 & 0.02 & 0.03 & -0.04 & -0.02 & 0.07 & 0.08 & -0.06 & -0.10 & -0.01 & 0.07 & 0.03 & 0.02 & 0.22 & -0.24 & -0.05 & -0.05 & -0.15 & 0.19 & -0.29 & -0.16 & 0.06 & -0.20 & 0.20 & -0.26 & 0.25 & -0.39 & 0.10 & -0.23 & -0.11 & 0.01 & 0.41 & -0.02 & -0.01 & 0.08  \\ 
150--200 & 1.0--1.2 & $9.68 \times 10^{-3}$ & 1.44 & 0.94 & 1.72 & 0.32 & 0.02 & 0.01 & 0.02 & -0.03 & -0.03 & 0.07 & 0.08 & -0.05 & -0.07 & 0.01 & 0.05 & 0.03 & 0.03 & 0.16 & -0.20 & -0.05 & -0.05 & -0.10 & 0.15 & -0.23 & -0.16 & 0.04 & -0.26 & 0.14 & -0.19 & 0.19 & -0.34 & 0.14 & -0.19 & -0.10 & 0.04 & 0.44 & -0.08 & -0.09 & 0.12  \\ 
150--200 & 1.2--1.4 & $9.12 \times 10^{-3}$ & 1.47 & 0.80 & 1.67 & 0.31 & 0.03 & 0.01 & 0.02 & -0.03 & -0.03 & 0.06 & 0.07 & -0.06 & -0.07 & -0.02 & 0.02 & 0.01 & 0.02 & 0.13 & -0.18 & -0.08 & -0.08 & -0.05 & 0.10 & -0.20 & -0.12 & 0.05 & -0.19 & 0.07 & -0.19 & 0.14 & -0.27 & 0.09 & -0.20 & -0.03 & 0.05 & 0.38 & -0.04 & -0.15 & 0.10  \\ 
150--200 & 1.4--1.6 & $7.79 \times 10^{-3}$ & 1.59 & 0.78 & 1.77 & 0.32 & 0.03 & 0.01 & 0.01 & -0.02 & -0.02 & 0.07 & 0.06 & -0.06 & -0.05 & -0.02 & -0.01 & 0.01 & 0.03 & 0.10 & -0.18 & -0.09 & -0.09 & -0.06 & 0.06 & -0.13 & -0.10 & 0.04 & -0.21 & -0.03 & -0.17 & 0.09 & -0.21 & 0.09 & -0.17 & -0.11 & 0.00 & 0.41 & -0.04 & -0.18 & 0.14  \\ 
150--200 & 1.6--1.8 & $6.64 \times 10^{-3}$ & 1.73 & 0.75 & 1.88 & 0.33 & 0.03 & -0.00 & 0.01 & -0.01 & -0.03 & 0.07 & 0.07 & -0.07 & -0.04 & -0.00 & -0.02 & 0.02 & 0.03 & 0.07 & -0.17 & -0.05 & -0.05 & -0.07 & 0.01 & -0.06 & -0.10 & 0.06 & -0.24 & -0.12 & -0.05 & 0.03 & -0.20 & 0.07 & -0.16 & -0.04 & -0.04 & 0.39 & -0.05 & -0.21 & 0.15  \\ 
150--200 & 1.8--2.0 & $4.83 \times 10^{-3}$ & 2.03 & 0.75 & 2.16 & 0.38 & 0.03 & -0.00 & 0.01 & -0.02 & -0.03 & 0.08 & 0.07 & -0.07 & -0.04 & -0.01 & -0.04 & 0.01 & 0.01 & 0.08 & -0.17 & -0.09 & -0.09 & 0.00 & -0.02 & -0.03 & -0.08 & 0.10 & -0.21 & -0.15 & -0.10 & -0.02 & -0.14 & 0.17 & -0.12 & 0.05 & 0.01 & 0.37 & -0.12 & -0.12 & 0.14  \\ 
150--200 & 2.0--2.2 & $3.35 \times 10^{-3}$ & 2.42 & 0.88 & 2.57 & 0.46 & 0.03 & -0.00 & 0.00 & -0.02 & -0.02 & 0.08 & 0.07 & -0.07 & -0.05 & -0.03 & -0.04 & 0.00 & -0.01 & 0.07 & -0.15 & -0.10 & -0.10 & 0.03 & -0.03 & 0.05 & -0.09 & 0.10 & -0.22 & -0.18 & -0.07 & -0.02 & -0.09 & 0.16 & -0.18 & 0.10 & 0.06 & 0.41 & -0.32 & 0.11 & 0.17  \\ 
150--200 & 2.2--2.4 & $1.63 \times 10^{-3}$ & 3.91 & 1.45 & 4.17 & 0.81 & 0.02 & -0.01 & -0.05 & -0.00 & -0.02 & 0.08 & 0.11 & -0.11 & -0.01 & 0.00 & -0.07 & 0.06 & 0.02 & 0.06 & -0.14 & -0.10 & -0.10 & 0.02 & -0.02 & 0.03 & -0.12 & 0.23 & -0.42 & -0.31 & 0.09 & -0.15 & -0.08 & 0.15 & -0.35 & 0.20 & 0.06 & 0.68 & -0.39 & 0.21 & 0.38  \\ 
\hline
200--300 & 0.0--0.2 & $2.42 \times 10^{-3}$ & 2.20 & 2.14 & 3.07 & 0.55 & -0.00 & 0.07 & 0.02 & -0.08 & 0.02 & 0.08 & 0.14 & -0.08 & -0.20 & 0.00 & 0.19 & 0.09 & 0.07 & 0.48 & -0.45 & 0.00 & 0.00 & -0.31 & 0.49 & -0.73 & -0.31 & 0.08 & -0.39 & 0.61 & -0.69 & 0.62 & -0.76 & 0.23 & -0.45 & -0.02 & -0.03 & 0.67 & 0.03 & -0.12 & 0.12  \\ 
200--300 & 0.2--0.4 & $2.39 \times 10^{-3}$ & 2.21 & 2.06 & 3.03 & 0.56 & -0.00 & 0.06 & 0.02 & -0.08 & 0.01 & 0.08 & 0.14 & -0.09 & -0.19 & -0.01 & 0.18 & 0.08 & 0.07 & 0.47 & -0.44 & -0.01 & -0.01 & -0.30 & 0.47 & -0.71 & -0.31 & 0.06 & -0.37 & 0.57 & -0.65 & 0.62 & -0.75 & 0.18 & -0.45 & -0.06 & -0.05 & 0.65 & 0.00 & -0.09 & 0.11  \\ 
200--300 & 0.4--0.6 & $2.43 \times 10^{-3}$ & 2.18 & 1.93 & 2.91 & 0.56 & 0.01 & 0.05 & 0.01 & -0.08 & -0.02 & 0.09 & 0.13 & -0.09 & -0.17 & -0.01 & 0.14 & 0.05 & 0.06 & 0.37 & -0.40 & -0.06 & -0.06 & -0.26 & 0.42 & -0.63 & -0.30 & 0.06 & -0.38 & 0.54 & -0.55 & 0.57 & -0.74 & 0.21 & -0.33 & -0.13 & -0.07 & 0.68 & -0.05 & -0.08 & 0.15  \\ 
200--300 & 0.6--0.8 & $2.48 \times 10^{-3}$ & 2.12 & 1.67 & 2.70 & 0.52 & 0.02 & 0.04 & 0.01 & -0.06 & -0.01 & 0.08 & 0.12 & -0.09 & -0.15 & 0.01 & 0.12 & 0.06 & 0.06 & 0.34 & -0.35 & -0.07 & -0.07 & -0.23 & 0.34 & -0.51 & -0.29 & 0.04 & -0.35 & 0.42 & -0.46 & 0.44 & -0.64 & 0.15 & -0.31 & -0.16 & -0.07 & 0.61 & -0.02 & -0.15 & 0.15  \\ 
200--300 & 0.8--1.0 & $2.29 \times 10^{-3}$ & 2.19 & 1.51 & 2.65 & 0.49 & 0.02 & 0.04 & 0.02 & -0.05 & -0.02 & 0.07 & 0.10 & -0.09 & -0.14 & -0.00 & 0.08 & 0.04 & 0.07 & 0.31 & -0.33 & -0.07 & -0.07 & -0.18 & 0.28 & -0.47 & -0.23 & 0.07 & -0.34 & 0.32 & -0.44 & 0.38 & -0.52 & 0.16 & -0.32 & -0.12 & -0.02 & 0.56 & -0.03 & -0.23 & 0.15  \\ 
200--300 & 1.0--1.2 & $2.16 \times 10^{-3}$ & 2.20 & 1.29 & 2.55 & 0.46 & 0.02 & 0.02 & 0.02 & -0.05 & -0.02 & 0.08 & 0.09 & -0.08 & -0.13 & -0.02 & 0.04 & 0.04 & 0.03 & 0.23 & -0.28 & -0.10 & -0.10 & -0.12 & 0.19 & -0.32 & -0.21 & 0.06 & -0.29 & 0.26 & -0.34 & 0.29 & -0.44 & 0.08 & -0.24 & -0.13 & -0.04 & 0.56 & -0.06 & -0.30 & 0.20  \\ 
200--300 & 1.2--1.4 & $1.88 \times 10^{-3}$ & 2.32 & 1.14 & 2.59 & 0.45 & 0.03 & 0.01 & 0.02 & -0.04 & -0.02 & 0.08 & 0.08 & -0.07 & -0.08 & -0.00 & 0.01 & 0.04 & 0.04 & 0.17 & -0.26 & -0.08 & -0.08 & -0.12 & 0.13 & -0.23 & -0.18 & 0.07 & -0.24 & 0.11 & -0.25 & 0.27 & -0.39 & 0.09 & -0.15 & -0.18 & -0.05 & 0.53 & -0.11 & -0.30 & 0.20  \\ 
200--300 & 1.4--1.6 & $1.66 \times 10^{-3}$ & 2.41 & 0.95 & 2.59 & 0.44 & 0.04 & 0.01 & 0.01 & -0.02 & -0.03 & 0.08 & 0.08 & -0.08 & -0.08 & -0.02 & -0.00 & 0.04 & 0.02 & 0.16 & -0.22 & -0.08 & -0.08 & -0.04 & 0.07 & -0.15 & -0.15 & 0.08 & -0.22 & 0.03 & -0.16 & 0.12 & -0.24 & 0.12 & -0.25 & -0.04 & 0.02 & 0.46 & -0.08 & -0.26 & 0.18  \\ 
200--300 & 1.6--1.8 & $1.34 \times 10^{-3}$ & 2.71 & 0.90 & 2.85 & 0.47 & 0.04 & 0.01 & 0.00 & -0.02 & -0.02 & 0.08 & 0.07 & -0.07 & -0.08 & -0.05 & -0.04 & 0.03 & 0.02 & 0.11 & -0.20 & -0.10 & -0.10 & -0.02 & 0.02 & -0.07 & -0.15 & 0.09 & -0.19 & -0.05 & -0.14 & 0.01 & -0.21 & 0.13 & -0.15 & -0.04 & -0.06 & 0.42 & -0.12 & -0.30 & 0.18  \\ 
200--300 & 1.8--2.0 & $1.00 \times 10^{-3}$ & 3.13 & 0.94 & 3.27 & 0.52 & 0.04 & -0.00 & 0.01 & -0.01 & -0.01 & 0.10 & 0.08 & -0.09 & -0.05 & -0.05 & -0.07 & 0.01 & 0.01 & 0.05 & -0.17 & -0.08 & -0.08 & 0.00 & -0.03 & 0.01 & -0.14 & 0.08 & -0.17 & -0.20 & -0.13 & 0.02 & -0.11 & 0.08 & -0.21 & -0.03 & -0.12 & 0.42 & -0.20 & -0.29 & 0.20  \\ 
200--300 & 2.0--2.2 & $6.04 \times 10^{-4}$ & 3.97 & 1.09 & 4.12 & 0.62 & 0.03 & -0.01 & -0.00 & -0.02 & -0.00 & 0.10 & 0.09 & -0.08 & -0.04 & -0.04 & -0.10 & -0.00 & 0.01 & 0.09 & -0.20 & -0.08 & -0.08 & 0.02 & -0.06 & 0.03 & -0.15 & 0.13 & -0.17 & -0.18 & -0.06 & -0.12 & -0.06 & 0.24 & -0.30 & 0.06 & -0.07 & 0.43 & -0.43 & 0.07 & 0.21  \\ 
200--300 & 2.2--2.4 & $2.43 \times 10^{-4}$ & 6.94 & 1.70 & 7.15 & 1.18 & 0.02 & 0.01 & -0.01 & -0.03 & 0.02 & 0.09 & 0.12 & -0.12 & -0.07 & -0.09 & -0.07 & 0.04 & 0.05 & 0.04 & -0.23 & -0.12 & -0.12 & 0.06 & -0.04 & 0.07 & -0.14 & 0.32 & -0.20 & -0.30 & -0.06 & -0.21 & 0.03 & 0.38 & -0.46 & 0.26 & -0.15 & 0.59 & -0.33 & 0.17 & 0.35  \\ 
\hline
300--500 & 0.0--0.4 & $3.55 \times 10^{-4}$ & 2.97 & 2.54 & 3.90 & 0.92 & 0.02 & 0.03 & -0.01 & -0.08 & -0.01 & 0.12 & 0.15 & -0.12 & -0.19 & 0.03 & 0.16 & 0.09 & 0.01 & 0.42 & -0.47 & -0.09 & -0.09 & -0.29 & 0.49 & -0.78 & -0.41 & 0.04 & -0.52 & 0.74 & -0.67 & 0.75 & -0.98 & 0.23 & -0.41 & -0.18 & -0.15 & 0.89 & -0.09 & -0.14 & 0.23  \\ 
300--500 & 0.4--0.8 & $3.36 \times 10^{-4}$ & 2.96 & 2.10 & 3.62 & 0.76 & 0.04 & 0.03 & 0.01 & -0.07 & -0.02 & 0.11 & 0.12 & -0.11 & -0.18 & 0.00 & 0.11 & 0.07 & 0.02 & 0.36 & -0.41 & -0.14 & -0.14 & -0.16 & 0.35 & -0.63 & -0.31 & 0.07 & -0.41 & 0.57 & -0.64 & 0.57 & -0.74 & 0.23 & -0.40 & -0.08 & -0.05 & 0.71 & 0.03 & -0.43 & 0.18  \\ 
300--500 & 0.8--1.2 & $3.16 \times 10^{-4}$ & 2.90 & 1.49 & 3.26 & 0.58 & 0.06 & 0.02 & 0.02 & -0.04 & -0.03 & 0.09 & 0.08 & -0.10 & -0.14 & -0.03 & 0.03 & 0.05 & 0.03 & 0.22 & -0.32 & -0.13 & -0.13 & -0.08 & 0.22 & -0.36 & -0.23 & 0.10 & -0.32 & 0.24 & -0.40 & 0.34 & -0.51 & 0.16 & -0.19 & -0.16 & -0.13 & 0.54 & 0.02 & -0.52 & 0.19  \\ 
300--500 & 1.2--1.6 & $2.45 \times 10^{-4}$ & 3.12 & 1.15 & 3.33 & 0.51 & 0.06 & -0.01 & -0.00 & -0.01 & -0.03 & 0.10 & 0.05 & -0.11 & -0.08 & -0.02 & -0.03 & 0.05 & 0.02 & 0.13 & -0.20 & -0.11 & -0.11 & -0.04 & 0.04 & -0.08 & -0.20 & 0.05 & -0.32 & 0.04 & -0.10 & 0.11 & -0.24 & 0.11 & -0.20 & -0.12 & -0.06 & 0.55 & -0.12 & -0.49 & 0.27  \\ 
300--500 & 1.6--2.0 & $1.31 \times 10^{-4}$ & 4.29 & 1.24 & 4.46 & 0.59 & 0.07 & 0.00 & -0.01 & -0.01 & -0.03 & 0.10 & 0.05 & -0.11 & -0.10 & -0.06 & -0.10 & 0.04 & -0.01 & 0.09 & -0.23 & -0.13 & -0.13 & 0.05 & -0.03 & 0.03 & -0.14 & 0.11 & -0.17 & -0.17 & -0.10 & -0.02 & -0.02 & 0.19 & -0.18 & 0.04 & -0.23 & 0.35 & 0.11 & -0.81 & 0.20  \\ 
300--500 & 2.0--2.4 & $3.81 \times 10^{-5}$ & 7.99 & 1.78 & 8.18 & 1.10 & 0.05 & -0.00 & -0.02 & -0.03 & 0.01 & 0.12 & 0.09 & -0.14 & -0.11 & -0.08 & -0.11 & 0.03 & 0.01 & 0.13 & -0.32 & -0.14 & -0.14 & 0.06 & -0.10 & 0.07 & -0.03 & 0.09 & -0.24 & -0.12 & 0.17 & -0.20 & 0.35 & 0.33 & -0.82 & 0.23 & -0.45 & 0.52 & -0.25 & 0.00 & 0.24  \\ 
\hline
500--1500 & 0.0--0.4 & $1.47 \times 10^{-5}$ & 6.15 & 1.89 & 6.43 & 0.99 & 0.12 & -0.04 & -0.02 & -0.02 & -0.05 & 0.17 & 0.08 & -0.16 & -0.16 & 0.02 & -0.02 & 0.06 & -0.07 & 0.20 & -0.28 & -0.23 & -0.23 & 0.04 & 0.14 & -0.34 & -0.33 & 0.01 & -0.47 & 0.44 & -0.51 & 0.41 & -0.65 & 0.18 & -0.37 & -0.06 & -0.16 & 0.55 & 0.04 & -0.45 & 0.16  \\ 
500--1500 & 0.4--0.8 & $1.38 \times 10^{-5}$ & 6.08 & 1.61 & 6.29 & 0.90 & 0.14 & -0.02 & 0.02 & -0.01 & -0.04 & 0.14 & 0.03 & -0.12 & -0.16 & -0.04 & -0.07 & 0.04 & -0.04 & 0.12 & -0.26 & -0.20 & -0.20 & 0.10 & 0.03 & -0.21 & -0.24 & 0.09 & -0.34 & 0.22 & -0.47 & 0.29 & -0.36 & 0.15 & -0.27 & -0.09 & -0.14 & 0.41 & 0.13 & -0.70 & 0.16  \\ 
500--1500 & 0.8--1.2 & $1.14 \times 10^{-5}$ & 6.46 & 1.47 & 6.62 & 0.85 & 0.14 & -0.02 & 0.02 & -0.00 & -0.02 & 0.11 & 0.01 & -0.13 & -0.13 & -0.05 & -0.11 & 0.07 & -0.01 & 0.12 & -0.23 & -0.18 & -0.18 & 0.09 & 0.00 & -0.08 & -0.18 & 0.08 & -0.28 & 0.07 & -0.26 & 0.11 & -0.16 & 0.11 & -0.15 & -0.09 & -0.05 & 0.52 & -0.07 & -0.77 & 0.25  \\ 
500--1500 & 1.2--1.6 & $8.05 \times 10^{-6}$ & 7.39 & 1.44 & 7.53 & 0.71 & 0.13 & -0.02 & 0.02 & 0.01 & -0.02 & 0.12 & -0.00 & -0.13 & -0.12 & -0.05 & -0.15 & 0.11 & -0.02 & 0.09 & -0.23 & -0.15 & -0.15 & 0.04 & -0.10 & 0.11 & -0.14 & 0.09 & -0.17 & -0.16 & -0.09 & -0.05 & 0.02 & 0.15 & -0.18 & -0.20 & -0.30 & 0.33 & 0.16 & -0.94 & 0.21  \\ 
500--1500 & 1.6--2.0 & $2.21 \times 10^{-6}$ & 14.30 & 1.95 & 14.43 & 1.04 & 0.13 & -0.01 & -0.01 & -0.02 & -0.00 & 0.12 & 0.02 & -0.16 & -0.20 & -0.16 & -0.09 & 0.08 & 0.06 & 0.13 & -0.38 & -0.16 & -0.16 & 0.14 & -0.18 & 0.11 & -0.11 & 0.02 & -0.08 & -0.04 & -0.00 & -0.21 & 0.51 & 0.43 & -0.54 & 0.22 & -0.31 & 0.56 & 0.16 & -1.00 & 0.22  \\ 
500--1500 & 2.0--2.4 & $2.87 \times 10^{-7}$ & 35.70 & 7.63 & 36.51 & 6.94 & 0.04 & 0.17 & 0.06 & -0.11 & 0.12 & 0.14 & 0.14 & -0.26 & -0.28 & -0.26 & -0.28 & 0.05 & -0.15 & 0.23 & -0.54 & -0.16 & -0.16 & 0.53 & -0.28 & -0.24 & 0.14 & 0.19 & 0.07 & -0.15 & -0.54 & -0.40 & 1.73 & 0.83 & -1.00 & 0.61 & -0.41 & 1.63 & -0.26 & -0.47 & 0.36  \\ 
\hline
\end{tabular} }%end of resizebox
\caption{The combined Born-level double-differential cross section $\frac{\text{d}^2\sigma}{\text{d}m_{\ell\ell}\text{d}|y_{\ell\ell}|}$. 
The measurements are listed together with the statistical ($\delta^{\rm stat}$), systematic ($\delta^{\rm sys}$) and total ($\delta^{\rm tot}$) uncertainties.
In addition the contributions from the individual correlated ($\delta^{\rm 1}_{\rm cor}$-$\delta^{\rm 35}_{\rm cor}$) and uncorrelated ($\delta^{\rm unc}$) systematic error sources are also provided.
The luminosity uncertainty of 1.9\% is not shown and not included in the overall systematic and total uncertainties.}
\label{table:combXsec_dMdy}
\end{center}
\end{sidewaystable}

\begin{sidewaystable}[!htb]
\begin{center}
\resizebox{\textwidth}{!}{
\begin{tabular}{rc|rrrr|rrrrrrrrrrrrrrrrrrrrrrrrrrrrrrrrrrrr}
\hline\hline
$m_{\ell\ell}$ & $|\Delta \eta_{\ell\ell}|$ & $\frac{\text{d}^2\sigma}{\text{d}m_{\ell\ell}\text{d}|\Delta \eta_{\ell\ell}|}$  & 
$\delta^{\rm stat}$           &  $\delta^{\rm sys}$          & $\delta^{\rm tot}$ &
$\delta^{\rm unc}$          &  
$\delta^{\rm 1}_{\rm cor}$     &  $\delta^{\rm 2}_{\rm cor}$   &
$\delta^{\rm 3}_{\rm cor}$     &  $\delta^{\rm 4}_{\rm cor}$   &
$\delta^{\rm 5}_{\rm cor}$     &  $\delta^{\rm 6}_{\rm cor}$   &
$\delta^{\rm 7}_{\rm cor}$     &  $\delta^{\rm 8}_{\rm cor}$   &
$\delta^{\rm 9}_{\rm cor}$     &  $\delta^{\rm 10}_{\rm cor}$   &
$\delta^{\rm 11}_{\rm cor}$     &  $\delta^{\rm 12}_{\rm cor}$   &
$\delta^{\rm 13}_{\rm cor}$     &  $\delta^{\rm 14}_{\rm cor}$   &
$\delta^{\rm 15}_{\rm cor}$     &  $\delta^{\rm 16}_{\rm cor}$   &
$\delta^{\rm 17}_{\rm cor}$     &  $\delta^{\rm 18}_{\rm cor}$   &
$\delta^{\rm 19}_{\rm cor}$     &  $\delta^{\rm 20}_{\rm cor}$   &
$\delta^{\rm 21}_{\rm cor}$     &  $\delta^{\rm 22}_{\rm cor}$   &
$\delta^{\rm 23}_{\rm cor}$     &  $\delta^{\rm 24}_{\rm cor}$   &
$\delta^{\rm 25}_{\rm cor}$     &  $\delta^{\rm 26}_{\rm cor}$   &
$\delta^{\rm 27}_{\rm cor}$     &  $\delta^{\rm 28}_{\rm cor}$   &
$\delta^{\rm 29}_{\rm cor}$     &  $\delta^{\rm 30}_{\rm cor}$   &
$\delta^{\rm 31}_{\rm cor}$     &  $\delta^{\rm 32}_{\rm cor}$   &
$\delta^{\rm 33}_{\rm cor}$     &  $\delta^{\rm 34}_{\rm cor}$     &  $\delta^{\rm 35}_{\rm cor}$\\
{[GeV]} &  & [pb/GeV] & [\%] & [\%] & [\%] & [\%] & [\%]& [\%] & [\%]& [\%] & [\%]& [\%] & [\%]& [\%] & [\%]& [\%] & [\%] & [\%] & [\%] & [\%] & [\%] & [\%] & [\%] & [\%] & [\%] & [\%] & [\%] & [\%] & [\%] & [\%] & [\%] & [\%] & [\%] & [\%] & [\%] & [\%] & [\%] & [\%] & [\%] & [\%] & [\%] \\
\hline
116--150 & 0.00--0.25 & $4.94 \times 10^{-2}$ & 0.66 & 0.56 & 0.86 & 0.16 & -0.00 & -0.00 & 0.00 & -0.00 & -0.03 & -0.04 & 0.03 & -0.02 & -0.02 & -0.06 & 0.01 & -0.01 & 0.02 & 0.05 & 0.02 & -0.05 & -0.05 & -0.06 & 0.03 & 0.01 & 0.04 & 0.00 & -0.11 & -0.12 & -0.18 & -0.03 & 0.06 & 0.22 & -0.21 & 0.25 & 0.12 & -0.02 & -0.14 & 0.08 & 0.07 \\ 
116--150 & 0.25--0.50 & $4.68 \times 10^{-2}$ & 0.68 & 0.58 & 0.89 & 0.18 & -0.00 & -0.01 & -0.00 & 0.00 & -0.03 & -0.04 & 0.03 & -0.02 & -0.03 & -0.06 & 0.02 & -0.01 & 0.01 & 0.06 & 0.03 & -0.06 & -0.06 & -0.06 & 0.03 & 0.01 & 0.06 & -0.02 & -0.11 & -0.12 & -0.20 & -0.01 & 0.06 & 0.20 & -0.22 & 0.26 & 0.12 & -0.06 & -0.11 & 0.11 & 0.07 \\ 
116--150 & 0.50--0.75 & $4.43 \times 10^{-2}$ & 0.70 & 0.58 & 0.91 & 0.19 & -0.00 & -0.00 & 0.00 & -0.01 & -0.04 & -0.04 & 0.03 & -0.03 & -0.03 & -0.06 & 0.01 & -0.01 & 0.01 & 0.06 & 0.03 & -0.05 & -0.05 & -0.06 & 0.03 & 0.01 & 0.04 & -0.02 & -0.10 & -0.12 & -0.20 & -0.02 & 0.02 & 0.21 & -0.22 & 0.25 & 0.15 & -0.04 & -0.12 & 0.09 & 0.07 \\ 
116--150 & 0.75--1.00 & $4.03 \times 10^{-2}$ & 0.75 & 0.58 & 0.94 & 0.20 & -0.00 & -0.01 & 0.00 & -0.01 & -0.04 & -0.04 & 0.02 & -0.03 & -0.03 & -0.06 & 0.01 & -0.01 & 0.01 & 0.07 & 0.04 & -0.05 & -0.05 & -0.06 & 0.03 & 0.02 & 0.03 & -0.03 & -0.09 & -0.12 & -0.21 & -0.03 & 0.05 & 0.17 & -0.24 & 0.24 & 0.17 & -0.01 & -0.12 & 0.05 & 0.04 \\ 
116--150 & 1.00--1.25 & $3.63 \times 10^{-2}$ & 0.78 & 0.58 & 0.97 & 0.21 & -0.00 & -0.00 & 0.00 & -0.00 & -0.04 & -0.03 & 0.02 & -0.04 & -0.03 & -0.06 & 0.02 & -0.03 & 0.01 & 0.06 & 0.03 & -0.05 & -0.05 & -0.06 & 0.03 & 0.01 & 0.04 & -0.03 & -0.06 & -0.11 & -0.20 & -0.03 & 0.02 & 0.17 & -0.25 & 0.27 & 0.18 & -0.02 & -0.09 & 0.03 & 0.05 \\ 
116--150 & 1.25--1.50 & $3.16 \times 10^{-2}$ & 0.85 & 0.61 & 1.04 & 0.23 & -0.00 & -0.01 & 0.00 & -0.00 & -0.04 & -0.04 & 0.03 & -0.03 & -0.03 & -0.06 & 0.02 & -0.03 & 0.00 & 0.07 & 0.03 & -0.05 & -0.05 & -0.06 & 0.03 & 0.02 & 0.03 & -0.05 & -0.08 & -0.11 & -0.21 & -0.02 & 0.04 & 0.16 & -0.29 & 0.24 & 0.23 & -0.01 & -0.07 & -0.03 & 0.04 \\ 
116--150 & 1.50--1.75 & $2.54 \times 10^{-2}$ & 0.95 & 0.68 & 1.17 & 0.26 & -0.00 & -0.01 & 0.01 & -0.00 & -0.04 & -0.04 & 0.02 & -0.03 & -0.03 & -0.06 & 0.01 & -0.03 & -0.00 & 0.07 & 0.04 & -0.05 & -0.05 & -0.06 & 0.03 & 0.03 & 0.03 & -0.04 & -0.09 & -0.09 & -0.24 & -0.08 & 0.06 & 0.18 & -0.29 & 0.29 & 0.26 & 0.01 & -0.09 & -0.06 & 0.04 \\ 
116--150 & 1.75--2.00 & $2.03 \times 10^{-2}$ & 1.08 & 0.72 & 1.30 & 0.30 & -0.00 & -0.00 & 0.01 & -0.00 & -0.04 & -0.04 & 0.02 & -0.04 & -0.03 & -0.05 & 0.02 & -0.02 & -0.00 & 0.08 & 0.02 & -0.05 & -0.05 & -0.04 & 0.04 & 0.03 & 0.03 & -0.03 & -0.09 & -0.10 & -0.26 & -0.03 & 0.05 & 0.19 & -0.27 & 0.31 & 0.28 & 0.06 & -0.12 & -0.10 & 0.03 \\ 
116--150 & 2.00--2.25 & $1.16 \times 10^{-2}$ & 1.47 & 0.89 & 1.72 & 0.40 & -0.00 & -0.00 & 0.01 & -0.00 & -0.04 & -0.04 & 0.01 & -0.04 & -0.03 & -0.05 & 0.02 & -0.02 & -0.01 & 0.08 & 0.04 & -0.05 & -0.05 & -0.04 & 0.04 & 0.04 & 0.03 & -0.04 & -0.09 & -0.07 & -0.29 & -0.02 & 0.10 & 0.17 & -0.37 & 0.34 & 0.40 & 0.08 & -0.15 & -0.14 & 0.04 \\ 
116--150 & 2.25--2.50 & $3.94 \times 10^{-3}$ & 2.61 & 1.29 & 2.91 & 0.67 & -0.01 & 0.00 & 0.01 & 0.01 & -0.06 & -0.03 & 0.01 & -0.05 & -0.03 & -0.04 & 0.02 & -0.06 & -0.02 & 0.08 & -0.01 & -0.08 & -0.08 & -0.02 & 0.05 & -0.01 & -0.03 & -0.07 & 0.01 & -0.04 & -0.29 & -0.03 & 0.10 & 0.25 & -0.54 & 0.49 & 0.64 & 0.11 & -0.14 & -0.21 & 0.09 \\ 
116--150 & 2.50--2.75 & $6.18 \times 10^{-4}$ & 7.05 & 2.70 & 7.54 & 1.78 & -0.01 & 0.00 & 0.01 & 0.02 & -0.04 & -0.03 & -0.01 & -0.04 & -0.04 & -0.06 & 0.02 & -0.06 & -0.01 & 0.12 & -0.04 & -0.06 & -0.06 & -0.00 & 0.04 & 0.07 & 0.04 & -0.16 & 0.09 & 0.04 & -0.32 & -0.06 & 0.24 & 0.28 & -0.88 & 1.07 & 1.20 & 0.10 & -0.47 & -0.37 & 0.24 \\ 
\hline
150--200 & 0.00--0.25 & $1.09 \times 10^{-2}$ & 1.16 & 0.76 & 1.39 & 0.21 & -0.01 & -0.00 & -0.01 & 0.00 & -0.05 & -0.03 & 0.04 & -0.03 & -0.03 & -0.08 & 0.03 & -0.02 & 0.02 & 0.05 & 0.02 & -0.07 & -0.07 & -0.05 & 0.05 & -0.02 & 0.07 & -0.07 & -0.01 & -0.17 & -0.17 & 0.02 & 0.06 & 0.21 & -0.40 & 0.27 & 0.30 & 0.00 & -0.16 & 0.14 & 0.12 \\ 
150--200 & 0.25--0.50 & $1.04 \times 10^{-2}$ & 1.21 & 0.80 & 1.45 & 0.23 & -0.01 & 0.00 & -0.01 & -0.01 & -0.05 & -0.03 & 0.04 & -0.03 & -0.03 & -0.07 & 0.02 & -0.02 & 0.02 & 0.05 & 0.01 & -0.07 & -0.07 & -0.07 & 0.05 & -0.00 & 0.04 & -0.06 & -0.01 & -0.17 & -0.18 & -0.00 & 0.06 & 0.23 & -0.39 & 0.31 & 0.34 & 0.03 & -0.15 & 0.15 & 0.10 \\ 
150--200 & 0.50--0.75 & $9.54 \times 10^{-3}$ & 1.28 & 0.85 & 1.54 & 0.25 & -0.01 & -0.00 & -0.01 & -0.01 & -0.06 & -0.03 & 0.04 & -0.04 & -0.04 & -0.07 & 0.02 & -0.03 & 0.02 & 0.05 & 0.01 & -0.07 & -0.07 & -0.06 & 0.04 & 0.02 & 0.04 & -0.11 & 0.01 & -0.18 & -0.17 & -0.01 & 0.03 & 0.22 & -0.43 & 0.32 & 0.42 & 0.03 & -0.14 & 0.11 & 0.10 \\ 
150--200 & 0.75--1.00 & $9.15 \times 10^{-3}$ & 1.32 & 0.89 & 1.59 & 0.26 & -0.01 & 0.00 & -0.01 & -0.01 & -0.06 & -0.03 & 0.04 & -0.04 & -0.03 & -0.08 & 0.02 & -0.02 & 0.02 & 0.06 & 0.00 & -0.07 & -0.07 & -0.06 & 0.05 & 0.01 & 0.02 & -0.08 & 0.03 & -0.17 & -0.18 & -0.05 & 0.06 & 0.24 & -0.45 & 0.38 & 0.43 & 0.04 & -0.11 & 0.11 & 0.10 \\ 
150--200 & 1.00--1.25 & $8.03 \times 10^{-3}$ & 1.41 & 0.99 & 1.72 & 0.28 & -0.01 & -0.00 & 0.00 & -0.00 & -0.06 & -0.03 & 0.03 & -0.04 & -0.03 & -0.08 & 0.02 & -0.03 & 0.02 & 0.07 & -0.00 & -0.06 & -0.06 & -0.06 & 0.05 & 0.01 & 0.01 & -0.10 & 0.06 & -0.17 & -0.23 & -0.05 & 0.08 & 0.24 & -0.50 & 0.44 & 0.48 & 0.04 & -0.12 & 0.09 & 0.10 \\ 
150--200 & 1.25--1.50 & $7.02 \times 10^{-3}$ & 1.53 & 1.09 & 1.88 & 0.32 & -0.01 & -0.00 & 0.00 & -0.00 & -0.06 & -0.04 & 0.03 & -0.04 & -0.04 & -0.08 & 0.02 & -0.04 & 0.01 & 0.07 & 0.01 & -0.08 & -0.08 & -0.05 & 0.04 & 0.03 & 0.03 & -0.12 & 0.07 & -0.18 & -0.24 & -0.05 & 0.08 & 0.18 & -0.56 & 0.47 & 0.58 & 0.05 & -0.11 & 0.04 & 0.09 \\ 
150--200 & 1.50--1.75 & $6.06 \times 10^{-3}$ & 1.68 & 1.22 & 2.07 & 0.37 & -0.01 & -0.01 & 0.01 & 0.00 & -0.06 & -0.04 & 0.03 & -0.05 & -0.03 & -0.08 & 0.02 & -0.04 & 0.00 & 0.08 & 0.00 & -0.07 & -0.07 & -0.05 & 0.05 & 0.03 & -0.01 & -0.14 & 0.10 & -0.17 & -0.24 & -0.08 & 0.08 & 0.21 & -0.59 & 0.54 & 0.67 & 0.07 & -0.12 & -0.02 & 0.09 \\ 
150--200 & 1.75--2.00 & $4.94 \times 10^{-3}$ & 1.91 & 1.42 & 2.38 & 0.45 & -0.01 & -0.00 & 0.01 & -0.00 & -0.08 & -0.03 & 0.02 & -0.06 & -0.03 & -0.08 & 0.01 & -0.06 & -0.01 & 0.09 & 0.00 & -0.07 & -0.07 & -0.06 & 0.05 & 0.03 & -0.04 & -0.17 & 0.11 & -0.16 & -0.26 & -0.12 & 0.05 & 0.23 & -0.69 & 0.61 & 0.82 & 0.05 & -0.14 & -0.08 & 0.11 \\ 
150--200 & 2.00--2.25 & $3.77 \times 10^{-3}$ & 2.26 & 1.63 & 2.78 & 0.58 & -0.01 & 0.00 & 0.02 & -0.00 & -0.08 & -0.03 & 0.02 & -0.06 & -0.04 & -0.07 & 0.03 & -0.06 & -0.01 & 0.11 & 0.01 & -0.09 & -0.09 & -0.04 & 0.04 & 0.06 & -0.04 & -0.19 & 0.13 & -0.16 & -0.32 & -0.08 & 0.08 & 0.19 & -0.75 & 0.72 & 0.94 & 0.09 & -0.15 & -0.10 & 0.10 \\ 
150--200 & 2.25--2.50 & $2.92 \times 10^{-3}$ & 2.60 & 1.74 & 3.13 & 0.73 & -0.02 & -0.01 & -0.01 & -0.00 & -0.08 & -0.05 & 0.02 & -0.06 & -0.02 & -0.06 & 0.02 & -0.06 & -0.04 & 0.10 & 0.03 & -0.07 & -0.07 & -0.03 & 0.03 & 0.04 & -0.03 & -0.13 & 0.06 & -0.09 & -0.28 & -0.13 & 0.02 & 0.17 & -0.81 & 0.72 & 1.02 & 0.07 & -0.18 & -0.18 & 0.12 \\ 
150--200 & 2.50--2.75 & $1.98 \times 10^{-3}$ & 3.16 & 1.81 & 3.64 & 0.92 & -0.02 & -0.01 & -0.01 & -0.00 & -0.07 & -0.04 & 0.02 & -0.06 & -0.02 & -0.06 & 0.01 & -0.05 & -0.05 & 0.09 & 0.04 & -0.09 & -0.09 & -0.03 & 0.05 & 0.04 & -0.04 & -0.12 & 0.01 & -0.04 & -0.27 & -0.12 & 0.01 & 0.14 & -0.83 & 0.68 & 1.00 & 0.08 & -0.18 & -0.18 & 0.11 \\ 
150--200 & 3.00--3.00 & $8.36 \times 10^{-4}$ & 5.13 & 2.68 & 5.79 & 1.59 & -0.02 & -0.03 & -0.02 & 0.01 & -0.07 & -0.05 & 0.02 & -0.07 & -0.01 & -0.05 & 0.01 & -0.05 & -0.06 & 0.11 & 0.03 & -0.09 & -0.09 & -0.02 & 0.02 & 0.08 & 0.02 & -0.02 & -0.06 & 0.03 & -0.33 & -0.23 & 0.09 & 0.20 & -1.02 & 1.02 & 1.38 & 0.08 & -0.38 & -0.41 & 0.23 \\ 
\hline
200--300 & 0.00--0.25 & $2.19 \times 10^{-3}$ & 1.84 & 0.97 & 2.08 & 0.32 & -0.00 & -0.00 & -0.00 & 0.01 & -0.05 & -0.04 & 0.05 & -0.03 & -0.03 & -0.09 & 0.02 & -0.01 & 0.03 & 0.06 & -0.00 & -0.08 & -0.08 & -0.07 & 0.05 & -0.03 & 0.06 & -0.09 & -0.01 & -0.18 & -0.19 & 0.02 & 0.09 & 0.28 & -0.41 & 0.35 & 0.42 & 0.01 & -0.24 & 0.26 & 0.15 \\ 
200--300 & 0.25--0.50 & $2.11 \times 10^{-3}$ & 1.90 & 1.02 & 2.16 & 0.34 & -0.01 & -0.00 & -0.01 & -0.01 & -0.06 & -0.04 & 0.04 & -0.04 & -0.04 & -0.09 & 0.01 & -0.02 & 0.03 & 0.05 & 0.02 & -0.07 & -0.07 & -0.07 & 0.04 & -0.01 & 0.04 & -0.10 & 0.01 & -0.19 & -0.21 & 0.04 & 0.07 & 0.29 & -0.47 & 0.32 & 0.48 & -0.01 & -0.24 & 0.28 & 0.15 \\ 
200--300 & 0.50--0.75 & $1.99 \times 10^{-3}$ & 1.98 & 1.11 & 2.27 & 0.35 & -0.01 & -0.00 & -0.00 & -0.00 & -0.06 & -0.04 & 0.04 & -0.04 & -0.04 & -0.09 & 0.01 & -0.02 & 0.03 & 0.06 & 0.01 & -0.08 & -0.08 & -0.09 & 0.04 & -0.00 & 0.03 & -0.11 & 0.04 & -0.19 & -0.23 & 0.03 & 0.07 & 0.29 & -0.52 & 0.39 & 0.54 & -0.01 & -0.23 & 0.27 & 0.16 \\ 
200--300 & 0.75--1.00 & $1.91 \times 10^{-3}$ & 2.05 & 1.16 & 2.36 & 0.38 & -0.01 & -0.00 & -0.01 & -0.01 & -0.07 & -0.04 & 0.04 & -0.04 & -0.05 & -0.09 & 0.02 & -0.02 & 0.01 & 0.07 & 0.02 & -0.08 & -0.08 & -0.08 & 0.04 & 0.01 & 0.05 & -0.12 & 0.06 & -0.21 & -0.22 & 0.03 & 0.04 & 0.23 & -0.60 & 0.40 & 0.60 & 0.02 & -0.21 & 0.21 & 0.15 \\ 
200--300 & 1.00--1.25 & $1.69 \times 10^{-3}$ & 2.18 & 1.31 & 2.55 & 0.41 & -0.01 & -0.00 & -0.00 & 0.00 & -0.07 & -0.03 & 0.04 & -0.04 & -0.04 & -0.10 & 0.01 & -0.04 & 0.02 & 0.07 & 0.00 & -0.09 & -0.09 & -0.07 & 0.04 & 0.01 & 0.03 & -0.16 & 0.13 & -0.23 & -0.23 & -0.03 & 0.10 & 0.28 & -0.63 & 0.51 & 0.71 & 0.11 & -0.19 & 0.20 & 0.14 \\ 
200--300 & 1.25--1.50 & $1.45 \times 10^{-3}$ & 2.42 & 1.58 & 2.89 & 0.48 & -0.01 & -0.00 & 0.01 & -0.00 & -0.08 & -0.03 & 0.04 & -0.06 & -0.03 & -0.09 & -0.01 & -0.04 & 0.03 & 0.09 & -0.00 & -0.09 & -0.09 & -0.07 & 0.07 & 0.01 & -0.01 & -0.21 & 0.18 & -0.20 & -0.26 & -0.08 & 0.08 & 0.35 & -0.76 & 0.65 & 0.88 & 0.09 & -0.22 & 0.13 & 0.19 \\ 
200--300 & 1.50--1.75 & $1.27 \times 10^{-3}$ & 2.65 & 1.77 & 3.18 & 0.54 & -0.01 & -0.00 & 0.00 & 0.00 & -0.08 & -0.03 & 0.04 & -0.05 & -0.04 & -0.09 & 0.01 & -0.05 & 0.01 & 0.10 & -0.01 & -0.09 & -0.09 & -0.05 & 0.05 & 0.04 & -0.02 & -0.22 & 0.24 & -0.23 & -0.31 & -0.05 & 0.13 & 0.29 & -0.87 & 0.72 & 1.02 & 0.12 & -0.15 & 0.12 & 0.16 \\ 
200--300 & 1.75--2.00 & $1.06 \times 10^{-3}$ & 3.02 & 2.11 & 3.69 & 0.67 & -0.01 & -0.01 & 0.01 & 0.01 & -0.09 & -0.04 & 0.02 & -0.06 & -0.05 & -0.09 & 0.01 & -0.05 & 0.00 & 0.11 & 0.00 & -0.09 & -0.09 & -0.04 & 0.05 & 0.08 & -0.03 & -0.22 & 0.24 & -0.26 & -0.36 & -0.13 & 0.20 & 0.28 & -0.94 & 0.98 & 1.26 & 0.17 & -0.16 & 0.10 & 0.14 \\ 
200--300 & 2.00--2.25 & $9.16 \times 10^{-4}$ & 3.34 & 2.34 & 4.08 & 0.78 & -0.02 & -0.00 & 0.01 & -0.00 & -0.10 & -0.04 & 0.02 & -0.07 & -0.04 & -0.09 & 0.01 & -0.07 & -0.02 & 0.11 & -0.00 & -0.09 & -0.09 & -0.04 & 0.06 & 0.08 & -0.06 & -0.27 & 0.30 & -0.24 & -0.42 & -0.14 & 0.17 & 0.30 & -1.07 & 1.04 & 1.39 & 0.14 & -0.17 & 0.01 & 0.17 \\ 
200--300 & 2.25--2.50 & $6.50 \times 10^{-4}$ & 4.17 & 3.07 & 5.18 & 1.09 & -0.02 & -0.01 & 0.03 & -0.00 & -0.12 & -0.04 & 0.01 & -0.09 & -0.04 & -0.07 & -0.02 & -0.08 & -0.03 & 0.14 & -0.01 & -0.10 & -0.10 & -0.03 & 0.07 & 0.10 & -0.14 & -0.34 & 0.34 & -0.15 & -0.49 & -0.25 & 0.13 & 0.36 & -1.39 & 1.38 & 1.86 & 0.19 & -0.20 & -0.06 & 0.20 \\ 
200--300 & 2.50--2.75 & $5.34 \times 10^{-4}$ & 4.72 & 3.32 & 5.77 & 1.37 & -0.03 & -0.03 & -0.01 & 0.01 & -0.10 & -0.06 & 0.01 & -0.08 & -0.03 & -0.08 & 0.01 & -0.09 & -0.05 & 0.13 & 0.01 & -0.12 & -0.12 & -0.02 & 0.03 & 0.13 & -0.05 & -0.24 & 0.18 & -0.16 & -0.46 & -0.30 & 0.20 & 0.13 & -1.43 & 1.50 & 2.04 & 0.25 & -0.12 & -0.08 & 0.15 \\ 
200--300 & 3.00--3.00 & $3.98 \times 10^{-4}$ & 5.65 & 3.76 & 6.79 & 1.81 & -0.03 & -0.03 & 0.00 & 0.01 & -0.11 & -0.06 & 0.02 & -0.10 & -0.01 & -0.07 & -0.02 & -0.10 & -0.06 & 0.14 & 0.03 & -0.12 & -0.12 & -0.02 & 0.07 & 0.07 & -0.14 & -0.29 & 0.12 & -0.08 & -0.43 & -0.34 & 0.04 & 0.22 & -1.54 & 1.65 & 2.21 & 0.15 & -0.36 & -0.30 & 0.28 \\ 
\hline
300--500 & 0.00--0.50 & $2.67 \times 10^{-4}$ & 2.64 & 1.11 & 2.86 & 0.44 & -0.01 & -0.01 & -0.03 & 0.01 & -0.06 & -0.04 & 0.08 & -0.02 & -0.02 & -0.10 & -0.01 & 0.01 & 0.05 & 0.01 & -0.02 & -0.10 & -0.10 & -0.07 & 0.08 & -0.09 & 0.10 & -0.03 & -0.10 & -0.10 & -0.15 & -0.00 & 0.03 & 0.35 & -0.39 & 0.36 & 0.44 & -0.12 & -0.28 & 0.45 & 0.20 \\ 
300--500 & 0.50--1.00 & $2.52 \times 10^{-4}$ & 2.76 & 1.16 & 2.99 & 0.45 & -0.01 & -0.00 & -0.03 & 0.00 & -0.06 & -0.04 & 0.07 & -0.03 & -0.04 & -0.10 & 0.01 & 0.01 & 0.06 & 0.03 & 0.00 & -0.09 & -0.09 & -0.09 & 0.06 & -0.03 & 0.07 & -0.05 & -0.04 & -0.17 & -0.17 & 0.03 & 0.07 & 0.32 & -0.43 & 0.39 & 0.53 & -0.09 & -0.33 & 0.38 & 0.20 \\ 
300--500 & 1.00--1.50 & $2.19 \times 10^{-4}$ & 3.05 & 1.41 & 3.36 & 0.55 & -0.01 & -0.01 & -0.02 & 0.01 & -0.06 & -0.04 & 0.07 & -0.04 & -0.04 & -0.10 & 0.01 & -0.01 & 0.04 & 0.05 & -0.00 & -0.10 & -0.10 & -0.07 & 0.06 & -0.03 & 0.04 & -0.10 & 0.03 & -0.19 & -0.21 & -0.02 & 0.08 & 0.31 & -0.61 & 0.52 & 0.75 & -0.02 & -0.27 & 0.33 & 0.19 \\ 
300--500 & 1.50--2.00 & $1.70 \times 10^{-4}$ & 3.63 & 1.94 & 4.12 & 0.71 & -0.02 & -0.03 & -0.01 & -0.01 & -0.09 & -0.05 & 0.07 & -0.06 & -0.03 & -0.10 & -0.01 & -0.03 & 0.00 & 0.08 & 0.02 & -0.11 & -0.11 & -0.06 & 0.03 & 0.01 & -0.04 & -0.24 & 0.12 & -0.18 & -0.34 & -0.07 & 0.05 & 0.33 & -0.99 & 0.68 & 1.11 & -0.01 & -0.28 & 0.26 & 0.22 \\ 
300--500 & 2.00--2.50 & $1.02 \times 10^{-4}$ & 5.17 & 3.12 & 6.04 & 1.23 & -0.03 & -0.02 & -0.00 & 0.01 & -0.10 & -0.06 & 0.04 & -0.08 & -0.03 & -0.09 & -0.02 & -0.05 & -0.03 & 0.12 & -0.01 & -0.12 & -0.12 & -0.06 & 0.05 & 0.07 & -0.11 & -0.34 & 0.26 & -0.16 & -0.46 & -0.22 & 0.21 & 0.41 & -1.42 & 1.30 & 1.87 & 0.24 & -0.24 & 0.18 & 0.22 \\ 
300--500 & 2.50--3.00 & $6.29 \times 10^{-5}$ & 7.23 & 4.70 & 8.62 & 2.13 & -0.03 & -0.03 & 0.01 & -0.01 & -0.12 & -0.07 & 0.02 & -0.09 & -0.04 & -0.09 & -0.04 & -0.08 & -0.07 & 0.17 & 0.03 & -0.11 & -0.11 & -0.04 & 0.03 & 0.17 & -0.15 & -0.46 & 0.38 & -0.23 & -0.71 & -0.36 & 0.26 & 0.28 & -2.07 & 1.95 & 2.81 & 0.18 & -0.33 & 0.06 & 0.28 \\ 
\hline
500--1500 & 0.00--0.50 & $8.08 \times 10^{-6}$ & 6.58 & 1.38 & 6.72 & 0.84 & -0.03 & -0.01 & -0.08 & -0.01 & -0.04 & -0.06 & 0.12 & 0.01 & -0.04 & -0.14 & 0.00 & 0.10 & 0.12 & -0.03 & 0.01 & -0.13 & -0.13 & -0.14 & 0.07 & -0.12 & 0.20 & 0.11 & -0.24 & -0.13 & -0.06 & 0.10 & 0.02 & 0.33 & -0.17 & 0.32 & 0.27 & -0.22 & -0.41 & 0.59 & 0.20 \\ 
500--1500 & 0.50--1.00 & $8.79 \times 10^{-6}$ & 6.34 & 1.33 & 6.48 & 0.78 & -0.03 & -0.01 & -0.08 & -0.01 & -0.05 & -0.05 & 0.10 & 0.01 & -0.05 & -0.12 & 0.01 & 0.08 & 0.12 & -0.03 & 0.03 & -0.12 & -0.12 & -0.14 & 0.06 & -0.12 & 0.21 & 0.07 & -0.22 & -0.18 & -0.10 & 0.13 & 0.06 & 0.22 & -0.25 & 0.24 & 0.28 & -0.27 & -0.31 & 0.65 & 0.18 \\ 
500--1500 & 1.00--1.50 & $7.98 \times 10^{-6}$ & 6.76 & 1.39 & 6.90 & 0.81 & -0.02 & -0.01 & -0.06 & -0.02 & -0.05 & -0.05 & 0.11 & 0.00 & -0.04 & -0.14 & -0.02 & 0.07 & 0.09 & -0.03 & -0.01 & -0.13 & -0.13 & -0.09 & 0.06 & -0.12 & 0.20 & 0.06 & -0.20 & -0.18 & -0.10 & 0.05 & 0.04 & 0.25 & -0.25 & 0.31 & 0.42 & -0.17 & -0.28 & 0.70 & 0.16 \\ 
500--1500 & 1.50--2.00 & $7.00 \times 10^{-6}$ & 7.44 & 1.46 & 7.58 & 0.83 & -0.02 & -0.01 & -0.05 & -0.01 & -0.04 & -0.05 & 0.12 & -0.01 & -0.04 & -0.12 & -0.02 & 0.04 & 0.06 & -0.00 & -0.01 & -0.12 & -0.12 & -0.07 & 0.08 & -0.12 & 0.18 & 0.05 & -0.17 & -0.13 & -0.16 & 0.01 & 0.01 & 0.34 & -0.37 & 0.50 & 0.54 & -0.16 & -0.44 & 0.41 & 0.25 \\ 
500--1500 & 2.00--2.50 & $4.16 \times 10^{-6}$ & 10.47 & 2.37 & 10.74 & 1.64 & -0.02 & -0.02 & -0.05 & 0.01 & -0.06 & -0.06 & 0.09 & -0.03 & -0.03 & -0.12 & 0.00 & 0.00 & 0.02 & 0.04 & 0.01 & -0.14 & -0.14 & -0.03 & 0.04 & -0.04 & 0.18 & -0.01 & -0.15 & -0.20 & -0.23 & -0.04 & 0.08 & 0.14 & -0.70 & 0.82 & 1.09 & 0.03 & -0.23 & 0.51 & 0.12 \\ 
500--1500 & 2.50--3.00 & $2.50 \times 10^{-6}$ & 14.59 & 3.74 & 15.06 & 2.55 & -0.03 & -0.05 & -0.04 & 0.02 & -0.07 & -0.07 & 0.09 & -0.06 & 0.00 & -0.09 & -0.04 & -0.02 & -0.01 & 0.08 & 0.00 & -0.13 & -0.13 & -0.04 & 0.05 & -0.01 & 0.04 & -0.06 & -0.16 & -0.09 & -0.30 & -0.30 & 0.04 & 0.23 & -1.21 & 1.32 & 1.89 & 0.24 & -0.19 & 0.46 & 0.13 \\ 
\hline
\end{tabular} }%end of resizebox
\caption{The combined Born-level double-differential cross section $\frac{\text{d}^2\sigma}{\text{d}m_{\ell\ell}\text{d}|\Delta \eta_{\ell\ell}|}$. 
The measurements are listed together with the statistical ($\delta^{\rm stat}$), systematic ($\delta^{\rm sys}$) and total ($\delta^{\rm tot}$) uncertainties.
In addition the contributions from the individual correlated ($\delta^{\rm 1}_{\rm cor}$-$\delta^{\rm 35}_{\rm cor}$) and uncorrelated ($\delta^{\rm unc}$) systematic error sources are also provided.
The luminosity uncertainty of 1.9\% is not shown and not included in the overall systematic and total uncertainties.}
\label{table:combXsec_dMdeta}
\end{center}
\end{sidewaystable}

\begin{figure}[p]
\begin{center}
\includegraphics[width=0.6\textwidth]{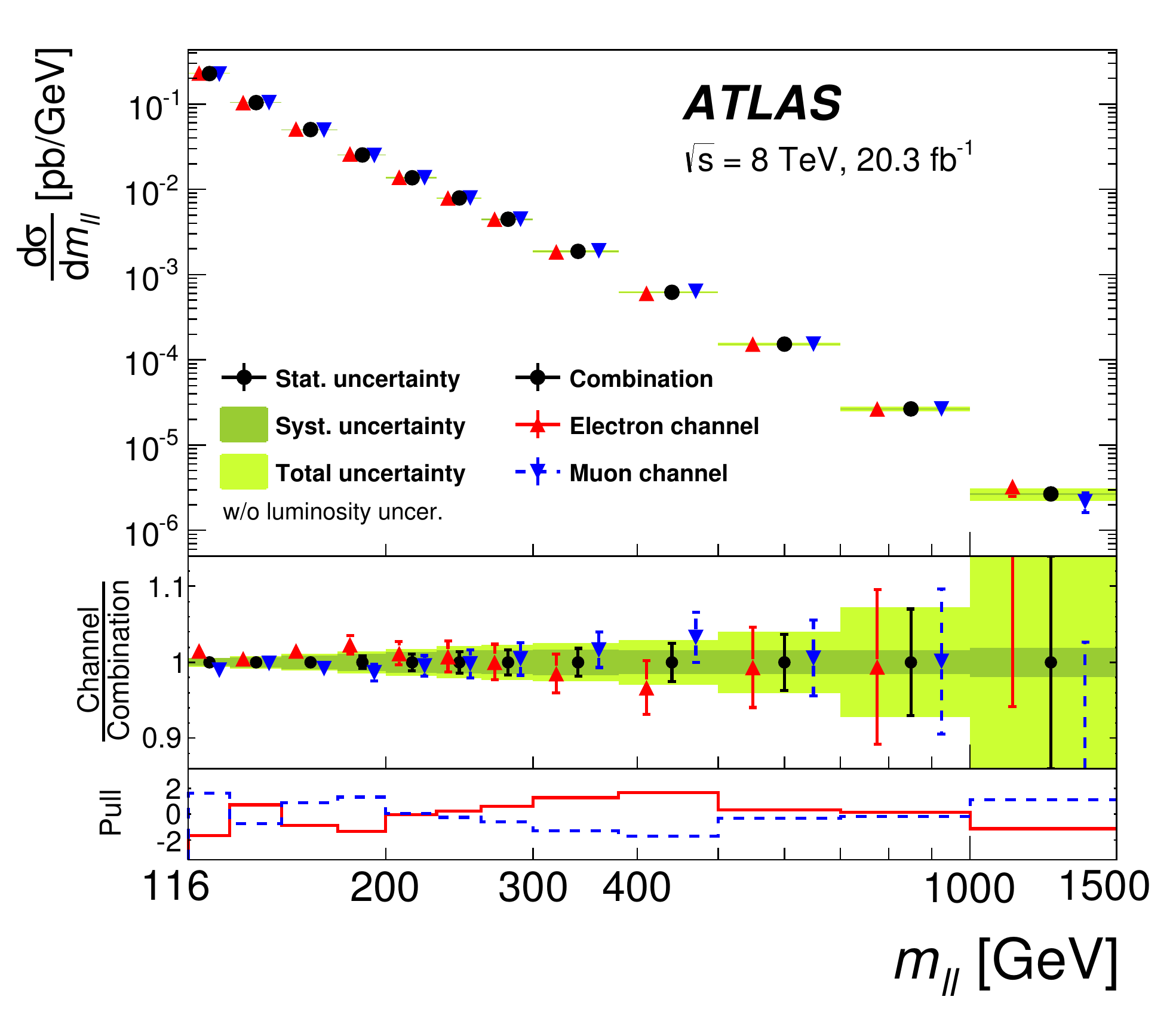}
\end{center}
\caption{Comparison of the electron (red points), muon (blue points)
  and combined (black points) single-differential fiducial Born-level
  cross sections as a function of invariant mass \mll. The error bars represent the statistical
  uncertainty. The inner shaded band represents the systematic
  uncertainty on the combined cross sections, and the outer shaded
  band represents the total measurement uncertainty (excluding the
  luminosity uncertainty). The central panel shows the ratio of each
  measurement channel to the combined data, and the lower panel shows
  the pull of the electron (red) and muon (blue) channel measurements
  with respect to the combined data.}
\label{fig:combinedMass}
\end{figure}

\begin{figure}[p]
\begin{center}
\includegraphics[width=0.49\textwidth]{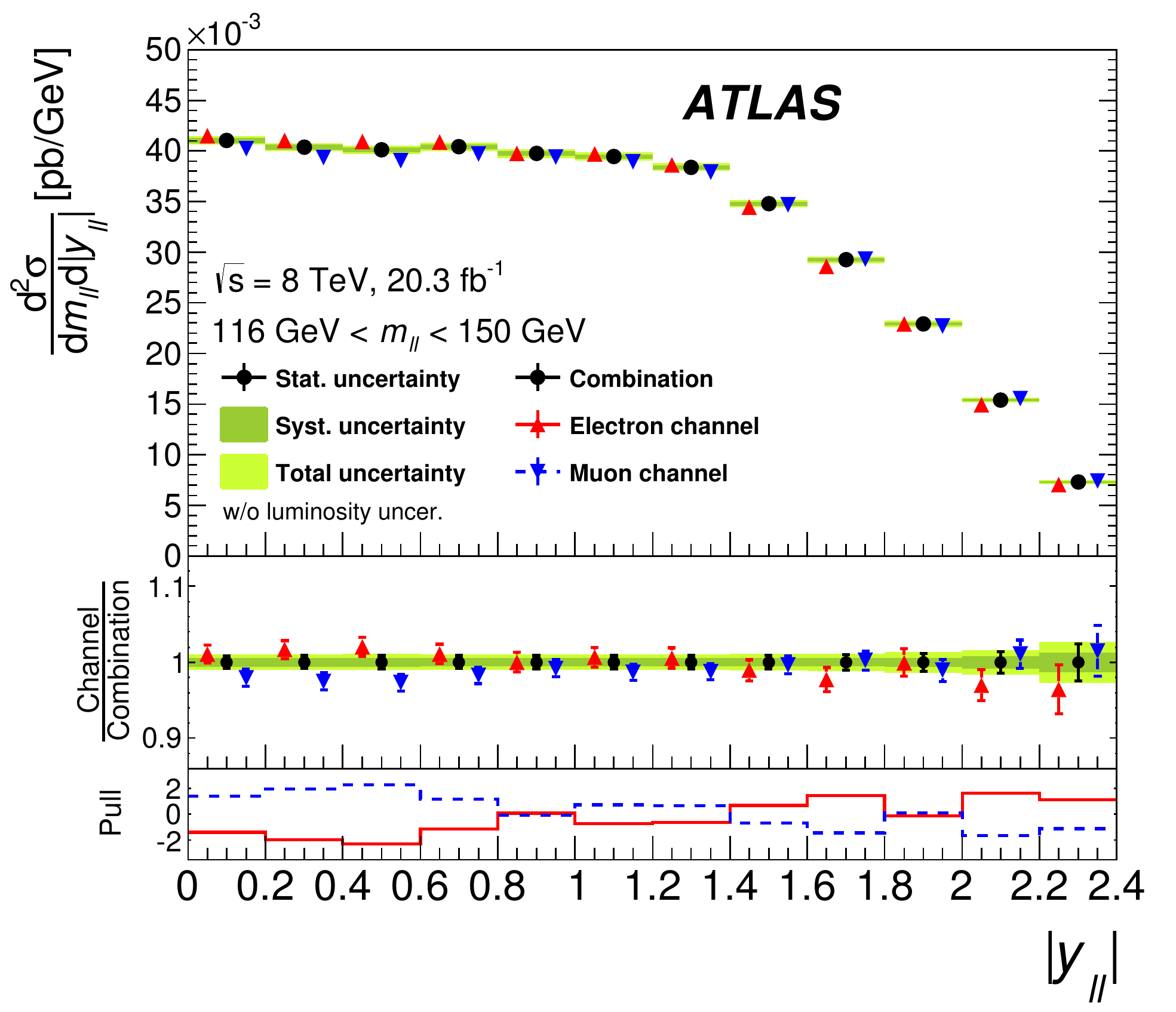}
\includegraphics[width=0.49\textwidth]{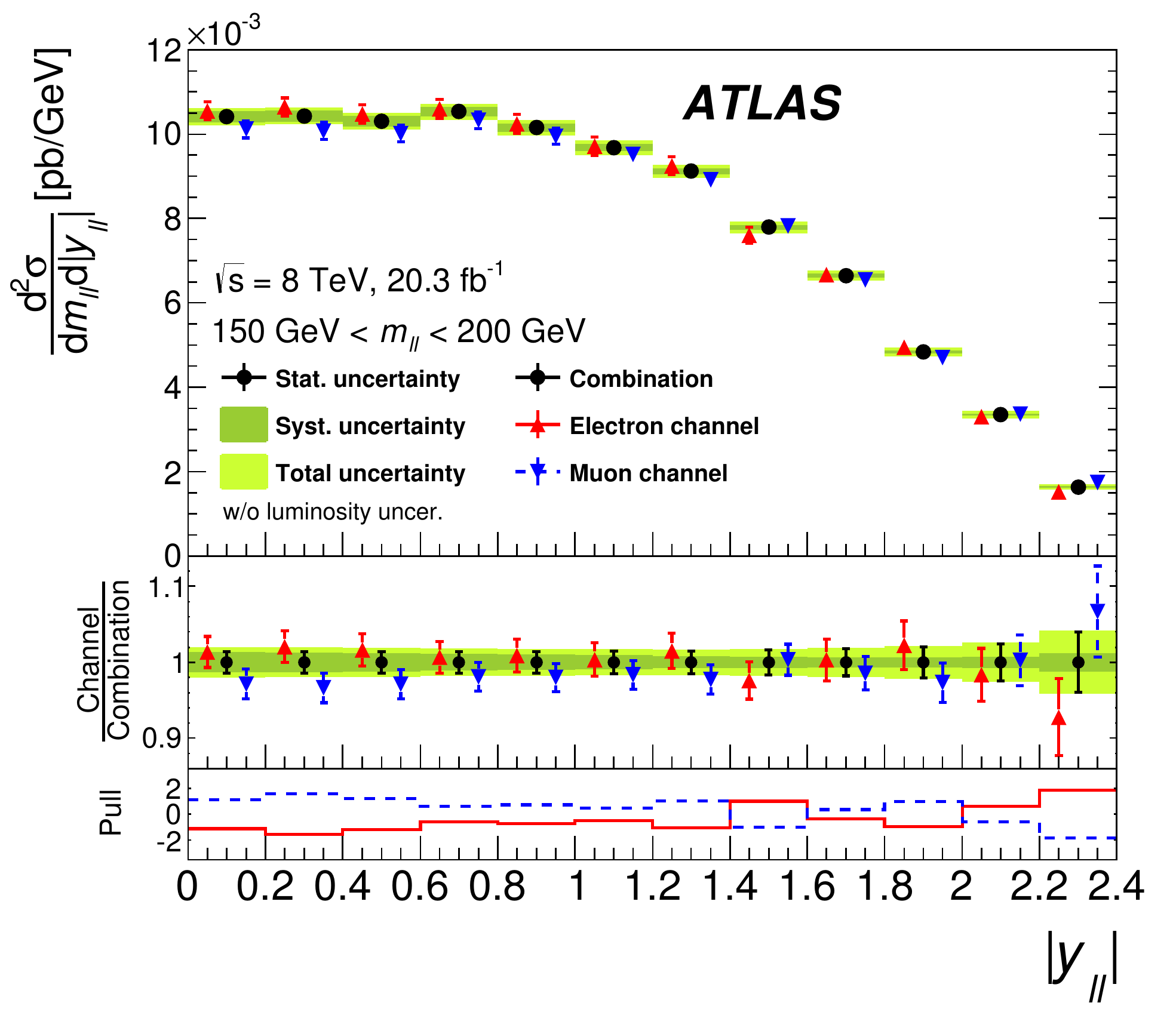}
\includegraphics[width=0.49\textwidth]{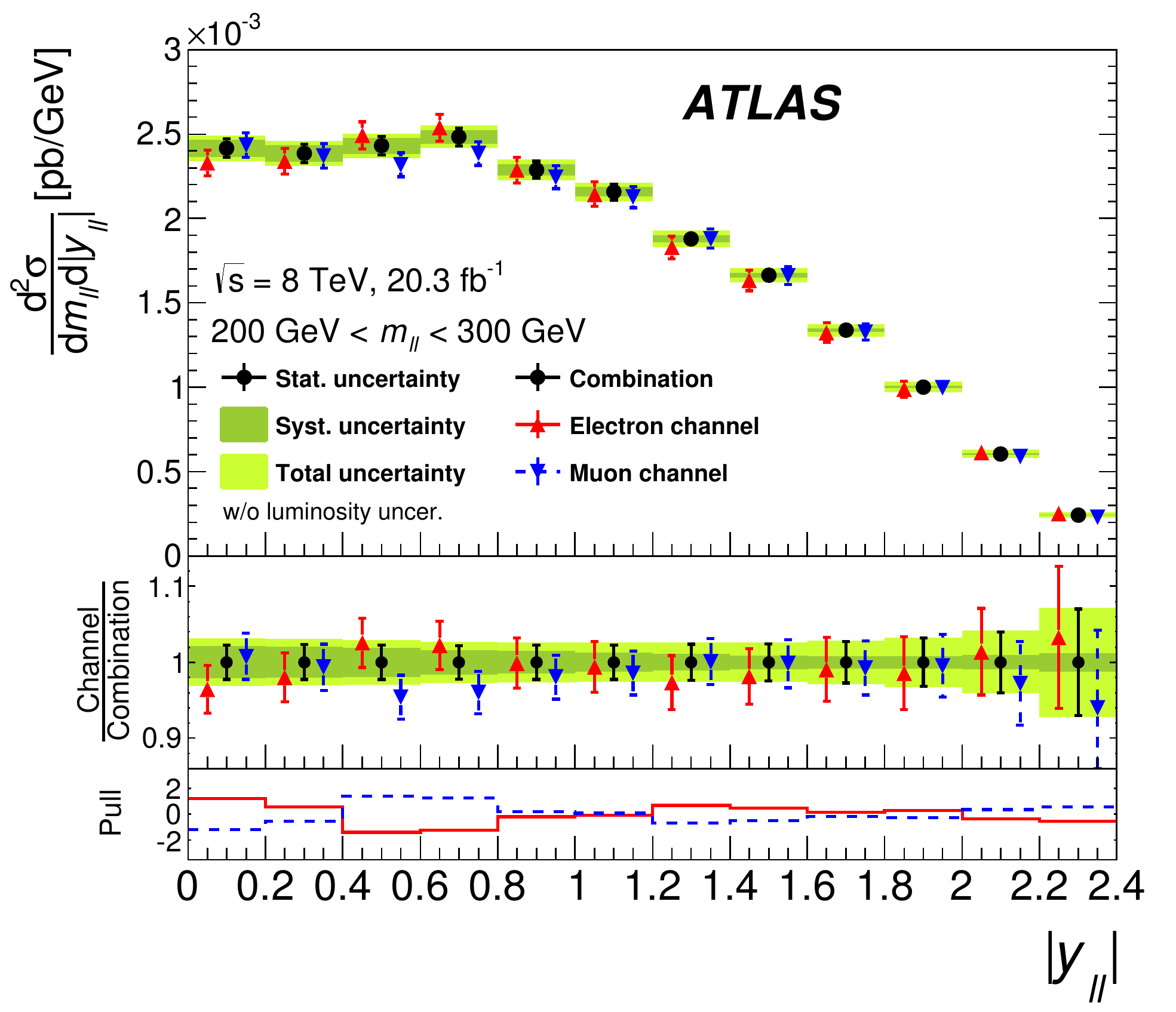}
\includegraphics[width=0.49\textwidth]{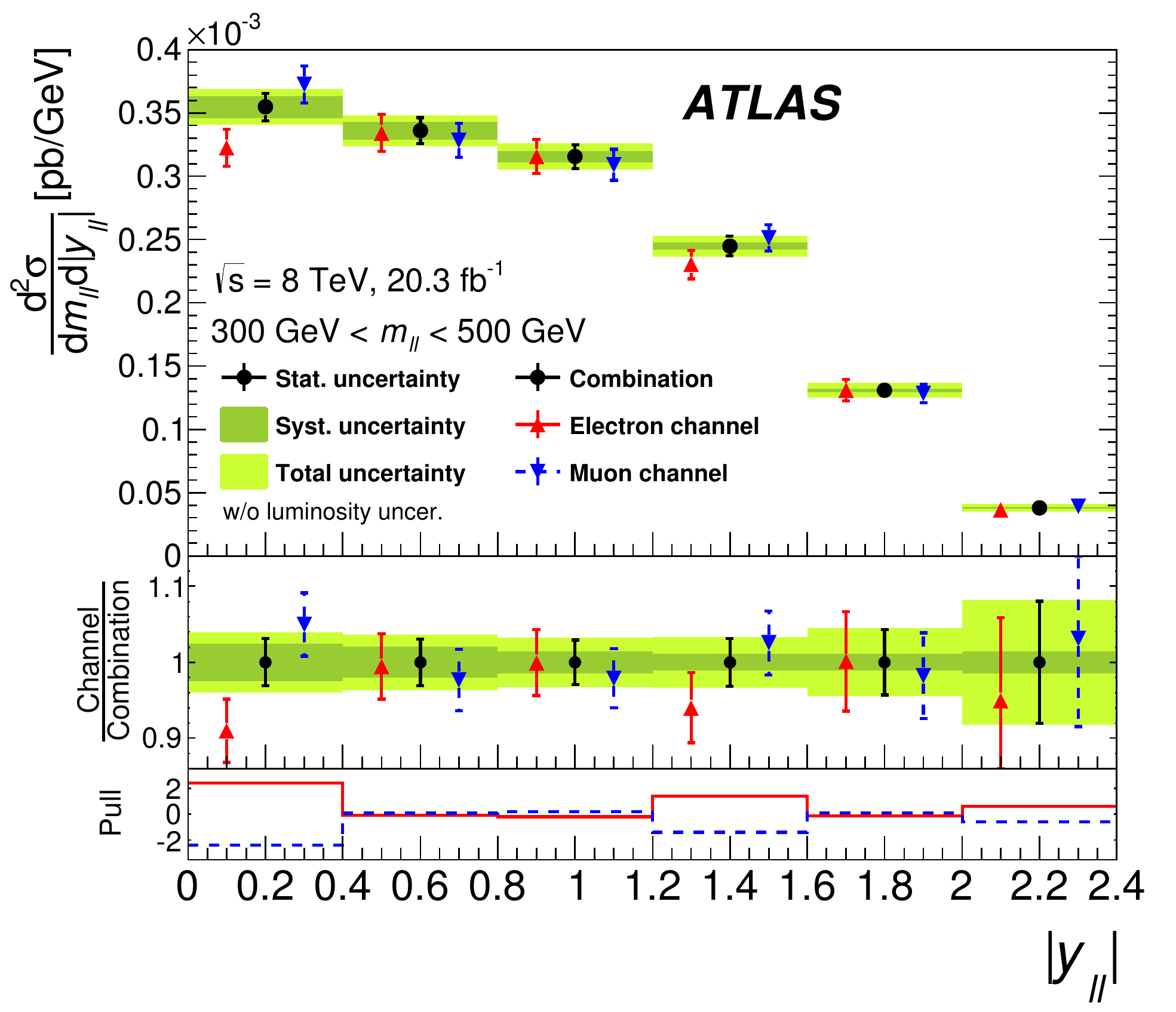}
\includegraphics[width=0.49\textwidth]{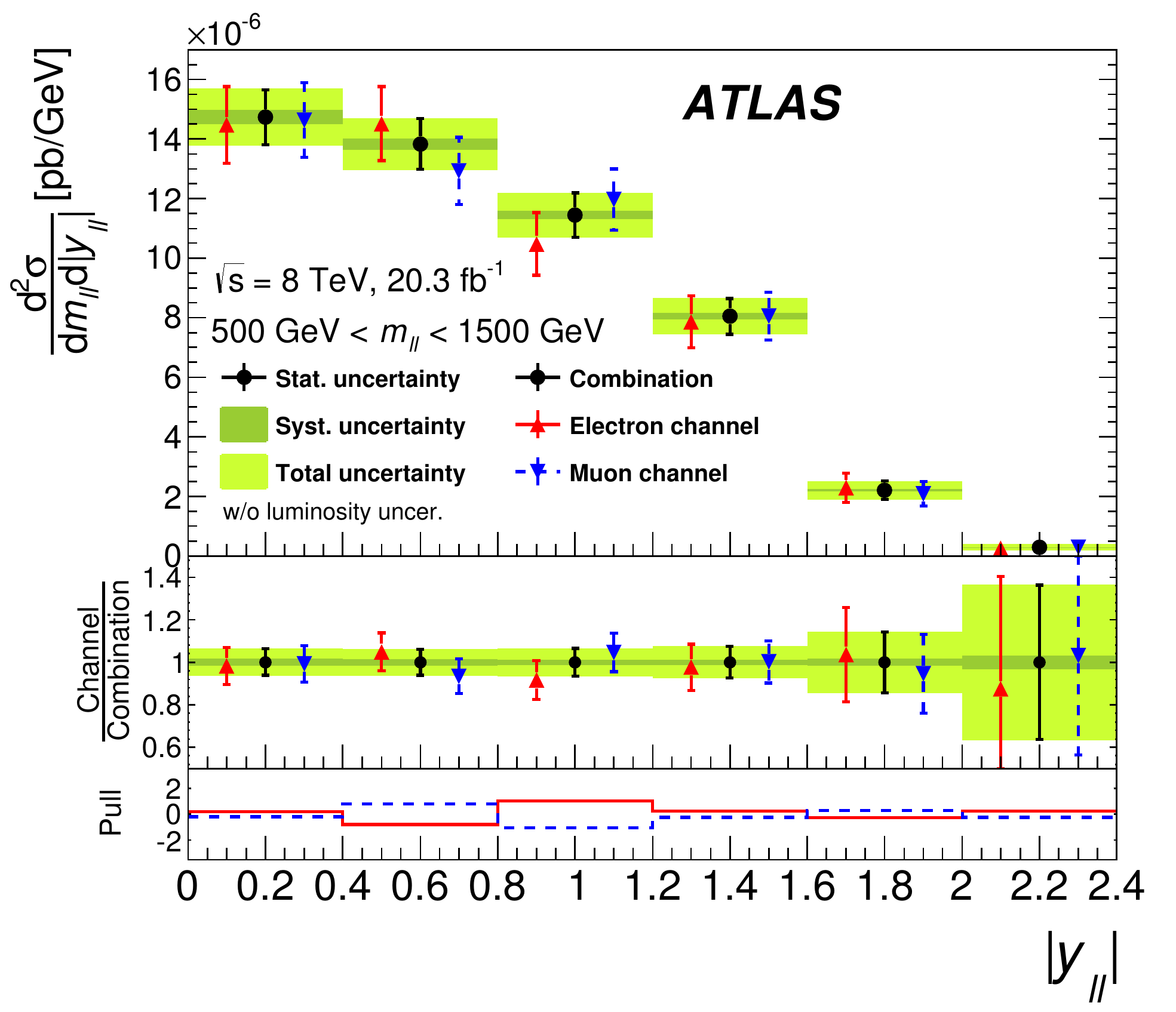}
\end{center}
\caption{Comparison of the electron (red points), muon (blue points)
  and combined (black points) fiducial Born-level cross sections,
  differential in invariant mass $m_{\ell\ell}$ and absolute dilepton
  rapidity $|y_{\ell\ell}|$.  The error bars
  represent the statistical uncertainty. The inner shaded band
  represents the systematic uncertainty on the combined cross
  sections, and the outer shaded band represents the total measurement
  uncertainty (excluding the luminosity uncertainty).  The central
  panel shows the ratio of each measurement channel to the combined
  data, and the lower panel shows the pull of the electron (red) and
  muon (blue) channel measurements with respect to the combined data.}
\label{fig:combinedRapidity}
\end{figure}

\begin{figure}[p]
\begin{center}
\includegraphics[width=0.49\textwidth]{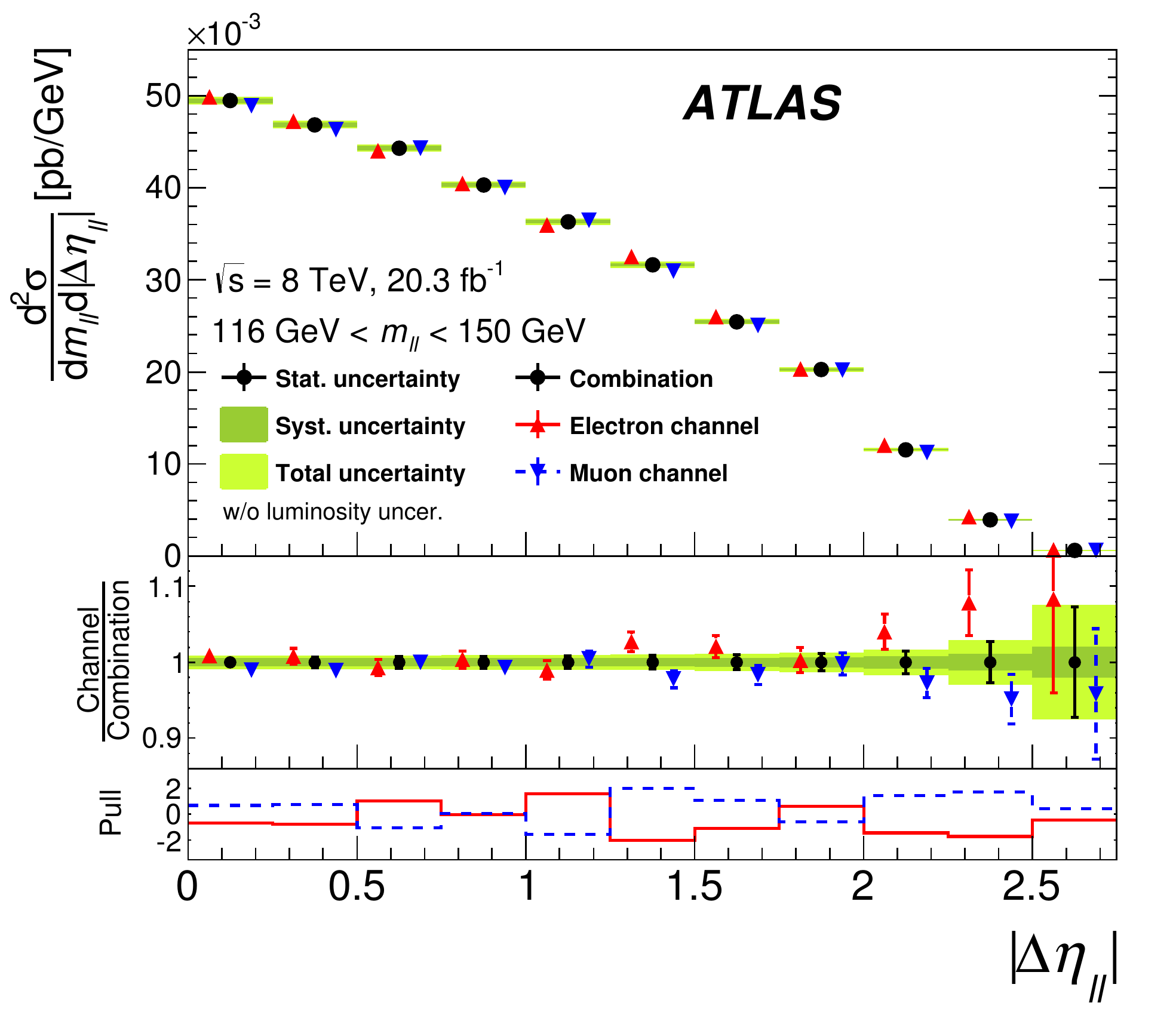}
\includegraphics[width=0.49\textwidth]{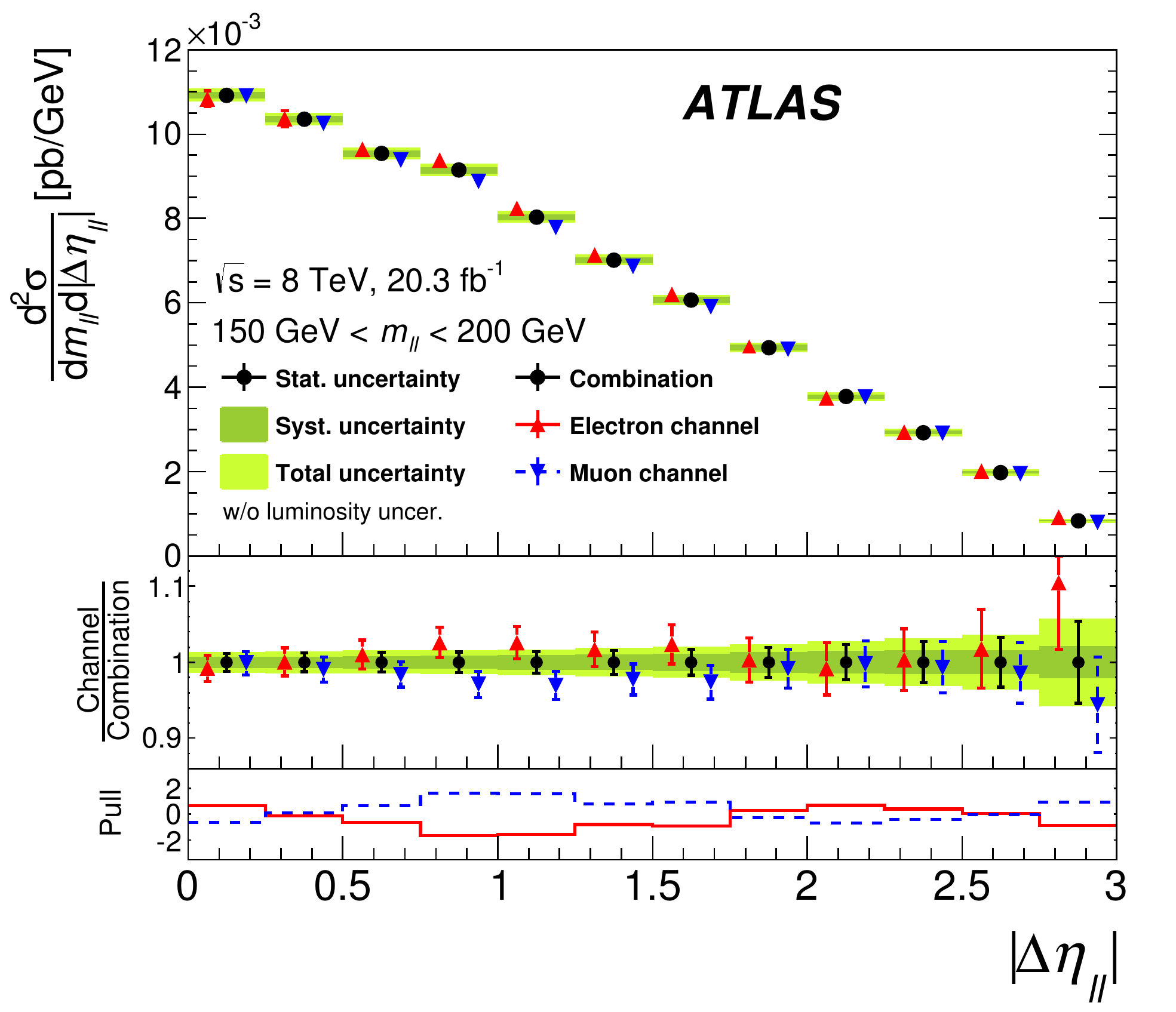}
\includegraphics[width=0.49\textwidth]{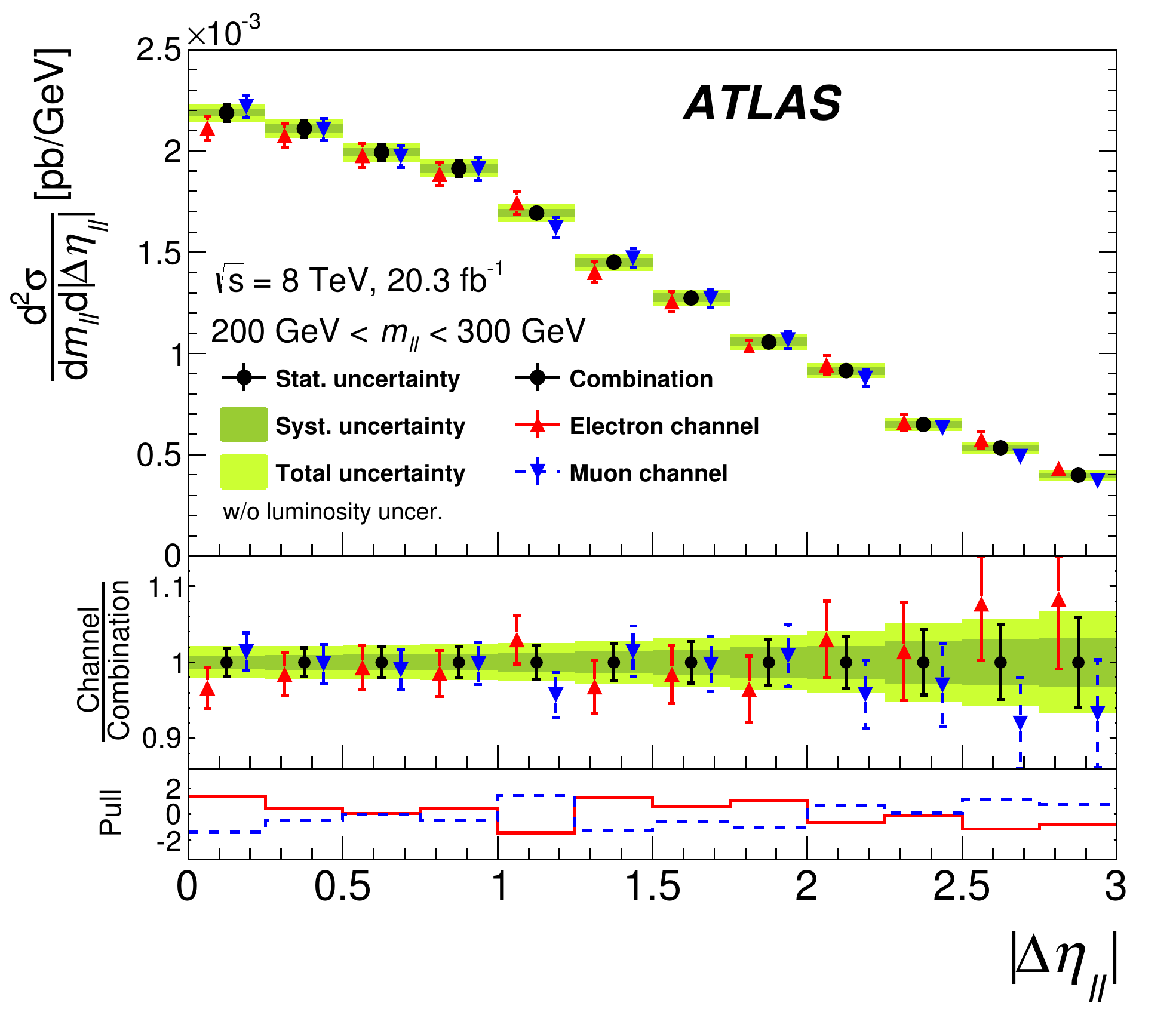}
\includegraphics[width=0.49\textwidth]{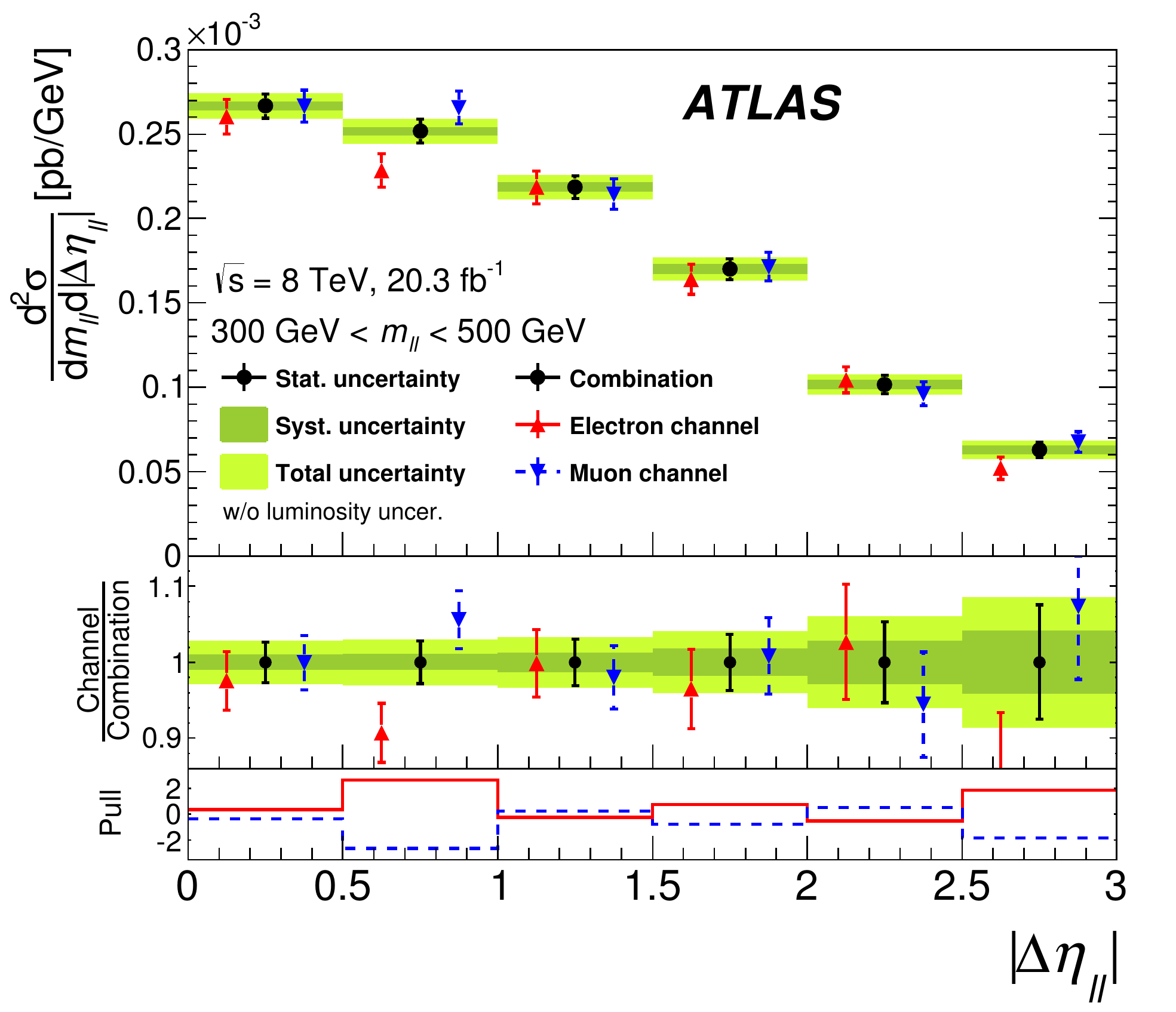}
\includegraphics[width=0.49\textwidth]{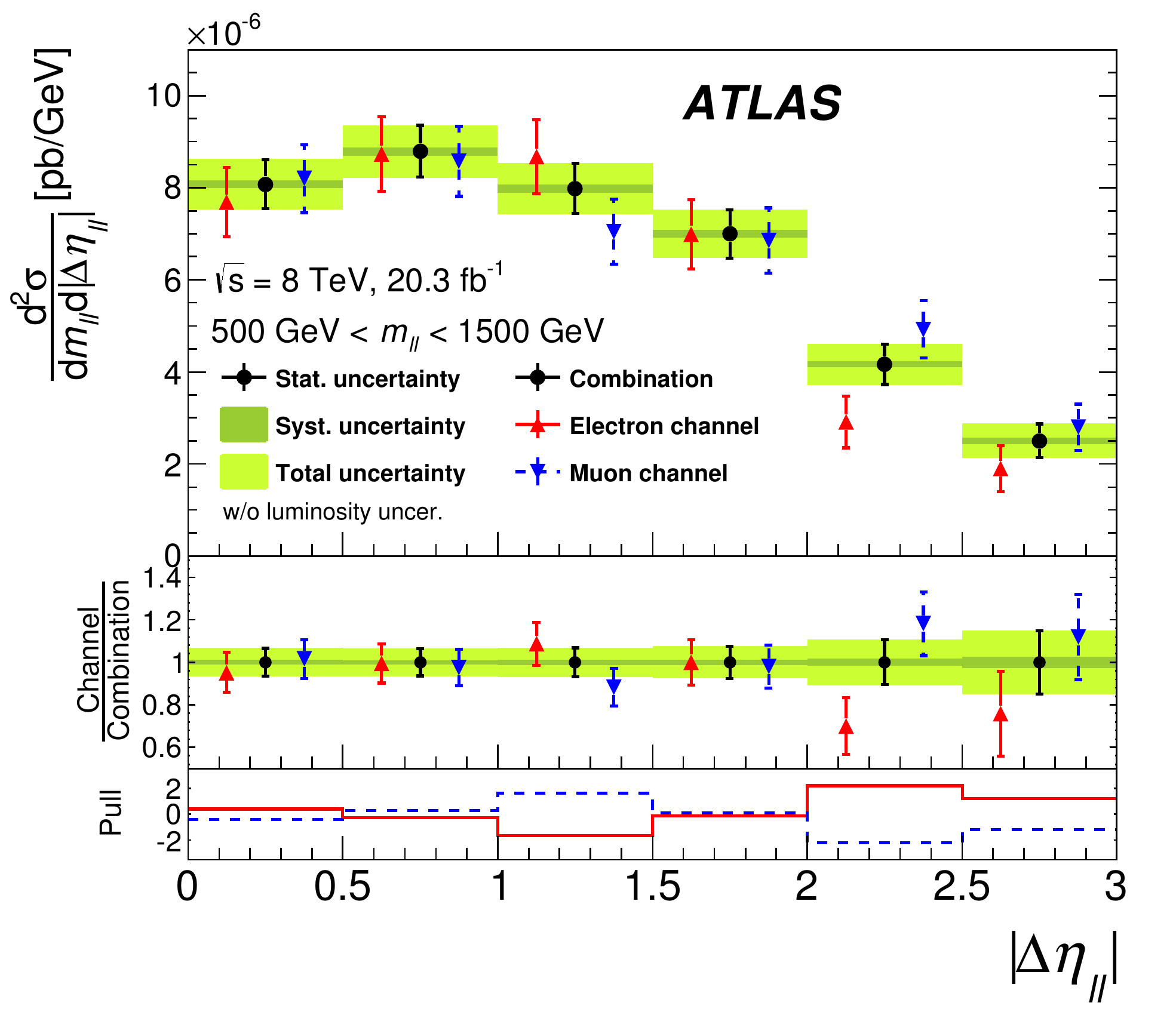}
\end{center}
\caption{Comparison of the electron (red points), muon (blue points)
  and combined (black points) fiducial Born-level cross sections,
  differential in invariant mass $m_{\ell\ell}$ and absolute dilepton
  pseudorapidity separation $|\Delta \eta_{\ell\ell}|$.  The
  error bars represent the statistical uncertainty. The inner shaded
  band represents the systematic uncertainty on the combined cross
  sections, and the outer shaded band represents the total measurement
  uncertainty (excluding the luminosity uncertainty).  The central
  panel shows the ratio of each measurement channel to the combined
  data, and the lower panel shows the pull of the electron (red) and
  muon (blue) channel measurements with respect to the combined data.}
\label{fig:combinedDeta}
\end{figure}

The results of the combination are shown in
figures~\ref{fig:combinedMass}--\ref{fig:combinedDeta}. In each figure
the upper panels show the measured Born-level cross sections for the
electron channel, muon channel and the combination. The ratio of the
individual channels to the combined measurement is also shown as well
as the pulls from the two channels, defined as the single-channel
measurement subtracted from the combined result in units of the
uncertainty. No coherent trends between the measurements are observed
and the pulls are found to be typically below two standard deviations,
and everywhere below three standard deviations.

The single-differential cross section shown in figure~\ref{fig:combinedMass}
falls rapidly over five orders of magnitude as $m_{\ell\ell}$
increases by about a factor of ten. In figure~\ref{fig:combinedRapidity}
the cross sections differential in $m_{\ell\ell}$ and $|y_{\ell\ell}|$
show a marked narrowing of the rapidity plateau width as $m_{\ell\ell}$
increases. The measurements obtained as a function of
$|\Delta\eta_{\ell\ell}|$ and $m_{\ell\ell}$ are shown in
figure~\ref{fig:combinedDeta}. For all $m_{\ell\ell}$, the cross
sections are largest where the absolute magnitude of the lepton
pseudorapidity separation is close to zero, and are observed to fall
as the separation increases.

\FloatBarrier
%-------------------------------------------------------------------------------
\section{Comparison to theoretical predictions}
\label{sec:theory_comp}
%-------------------------------------------------------------------------------

The combined fiducial cross sections at Born-level are compared to
NNLO pQCD calculations using various PDFs. The calculations use the
FEWZ~3.1 framework (see section~\ref{sec:MC}), and include NLO EW
corrections in the $G_\mu$~\cite{Dittmaier:2009cr} electroweak scheme.
The dynamic scale $m_{\ell \ell}$ is used in the calculation, and the
renormalisation and factorisation scales are set to $\mu_{\mathrm{R}}
= \mu_{\mathrm{F}} = m_{\ell \ell}$.
The calculation includes the contribution of the non-resonant PI
process, $\gamma \gamma \rightarrow \ell \ell$. This contribution is
estimated at leading order (LO) using the photon PDF from the
NNPDF2.3qed PDF set~\cite{Ball:2013hta}.

\begin{figure}[t]
\begin{center}
\includegraphics[width=0.495\textwidth]{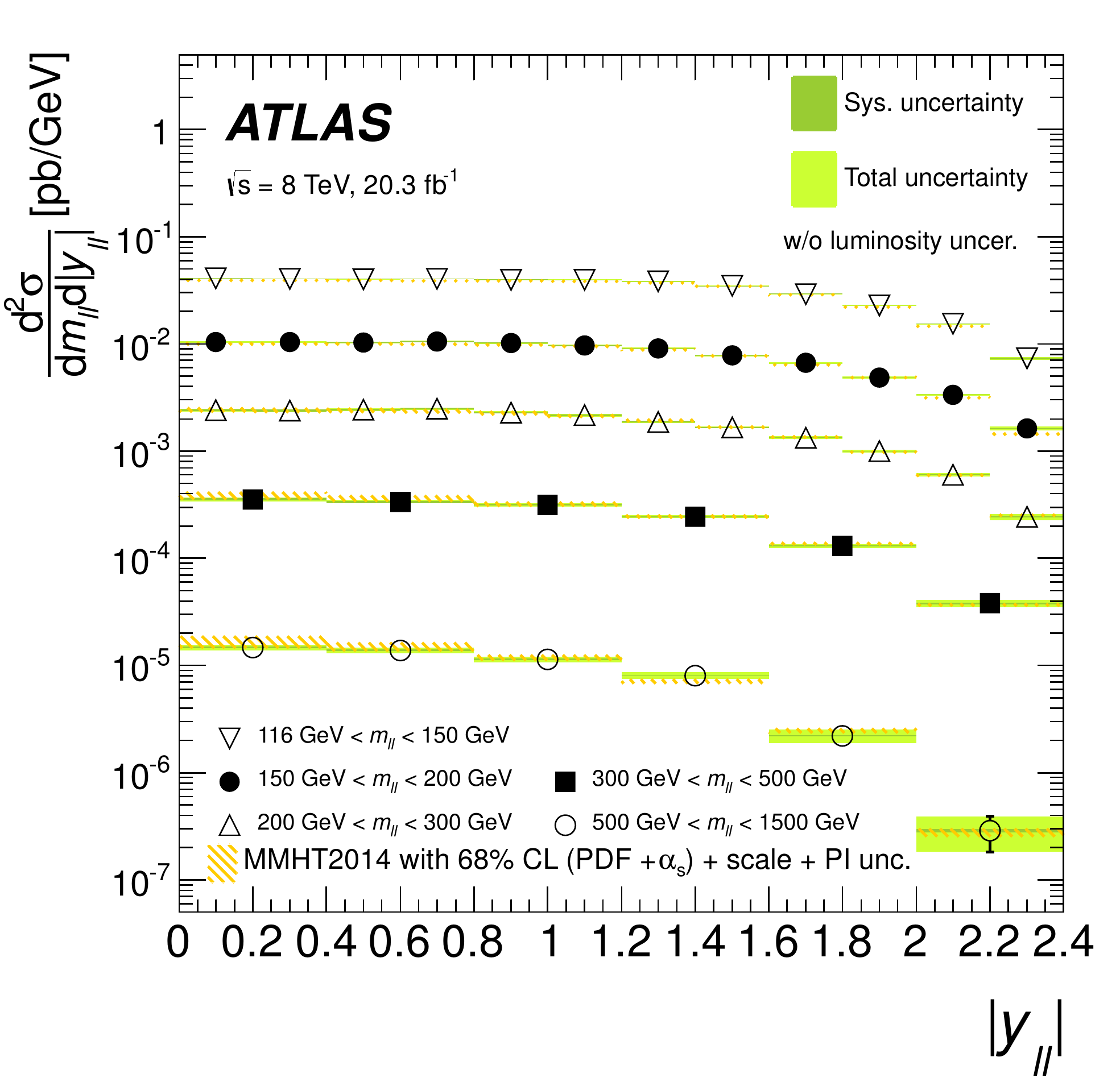}
\includegraphics[width=0.495\textwidth]{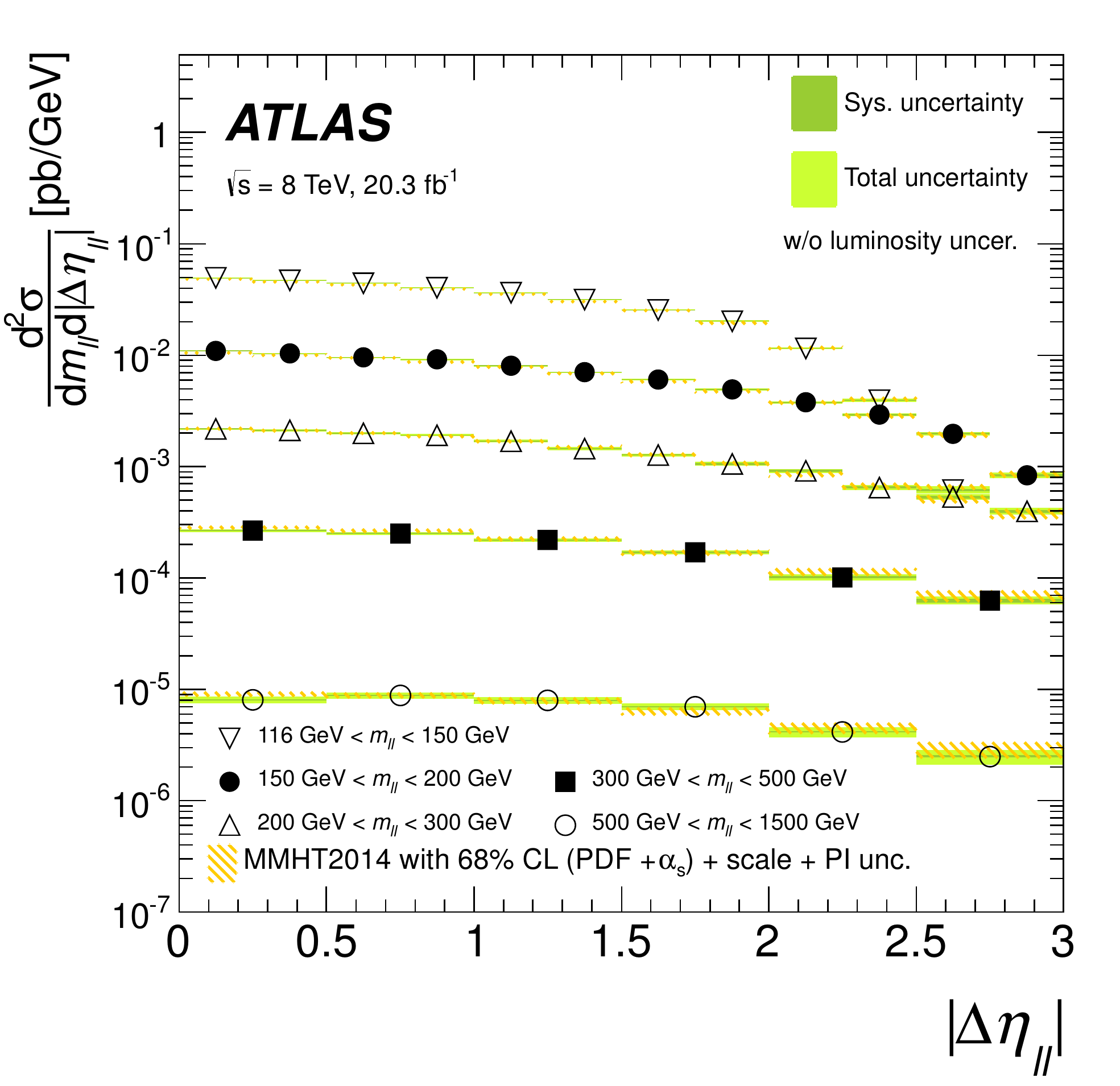}
\end{center}
\caption{The combined double-differential cross sections as a function
  of invariant mass \mll\ and absolute dilepton rapidity
  $|y_{\ell\ell}|$ (left panel) and as a function of \mll\ and
  absolute dilepton pseudorapidity separation $|\Delta
  \eta_{\ell\ell}|$ (right panel) at the Born-level within the
  fiducial region with statistical, systematic and total
  uncertainties, excluding the $1.9\%$ uncertainty on the luminosity.
  Data are compared to combined NNLO pQCD and NLO EW calculations
  using the MMHT2014 PDF, where the uncertainty band displays the
  combined 68\% confidence level (CL) PDF and $\alpha_{\rm S}$
  variation, the renormalisation and factorisation scale uncertainties
  and the PI uncertainty.  }
\label{fig:2DTheoryMMHTData}
\end{figure}

\begin{figure}[t]
\begin{center}
\includegraphics[width=0.495\textwidth]{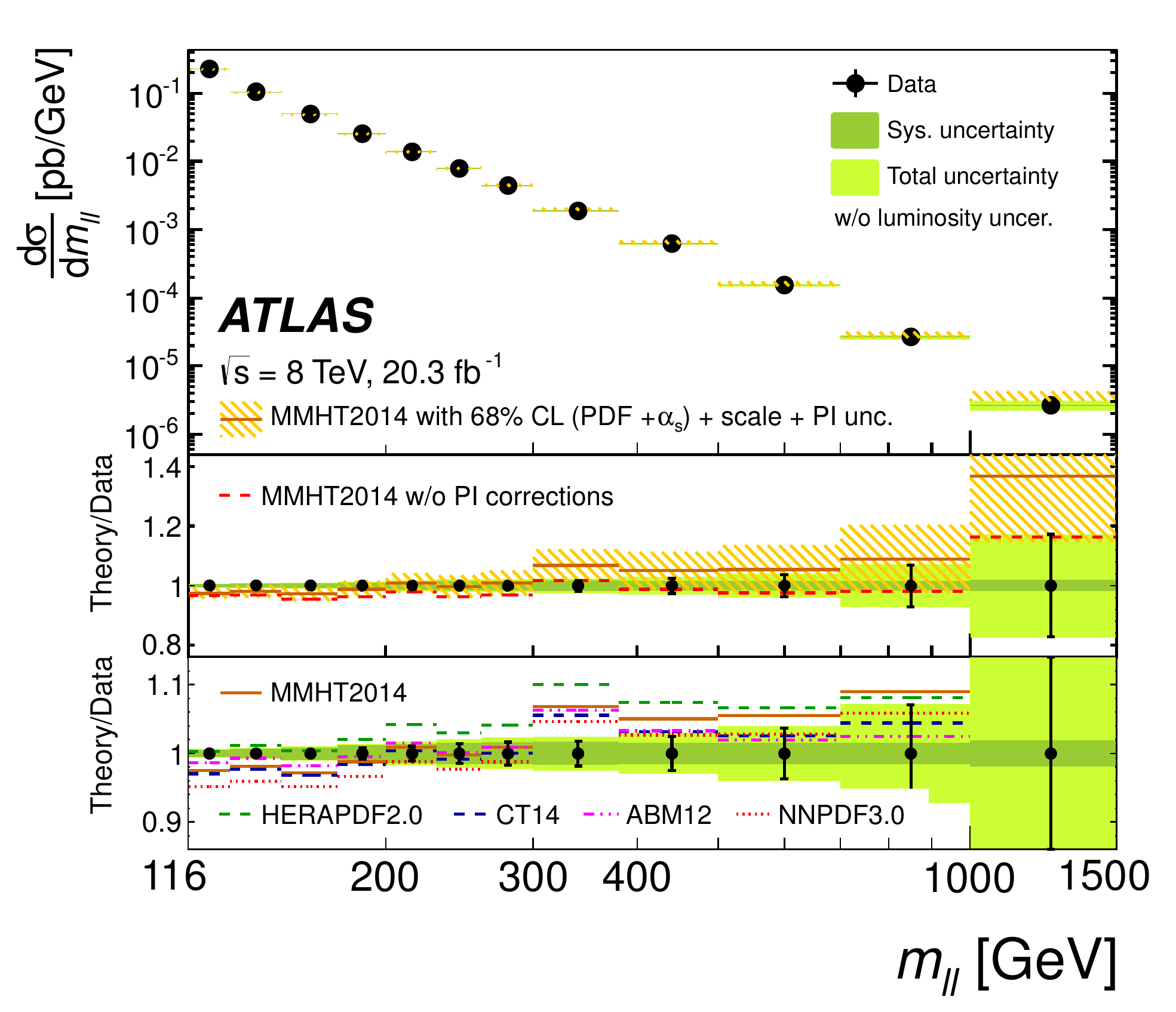}
\includegraphics[width=0.495\textwidth]{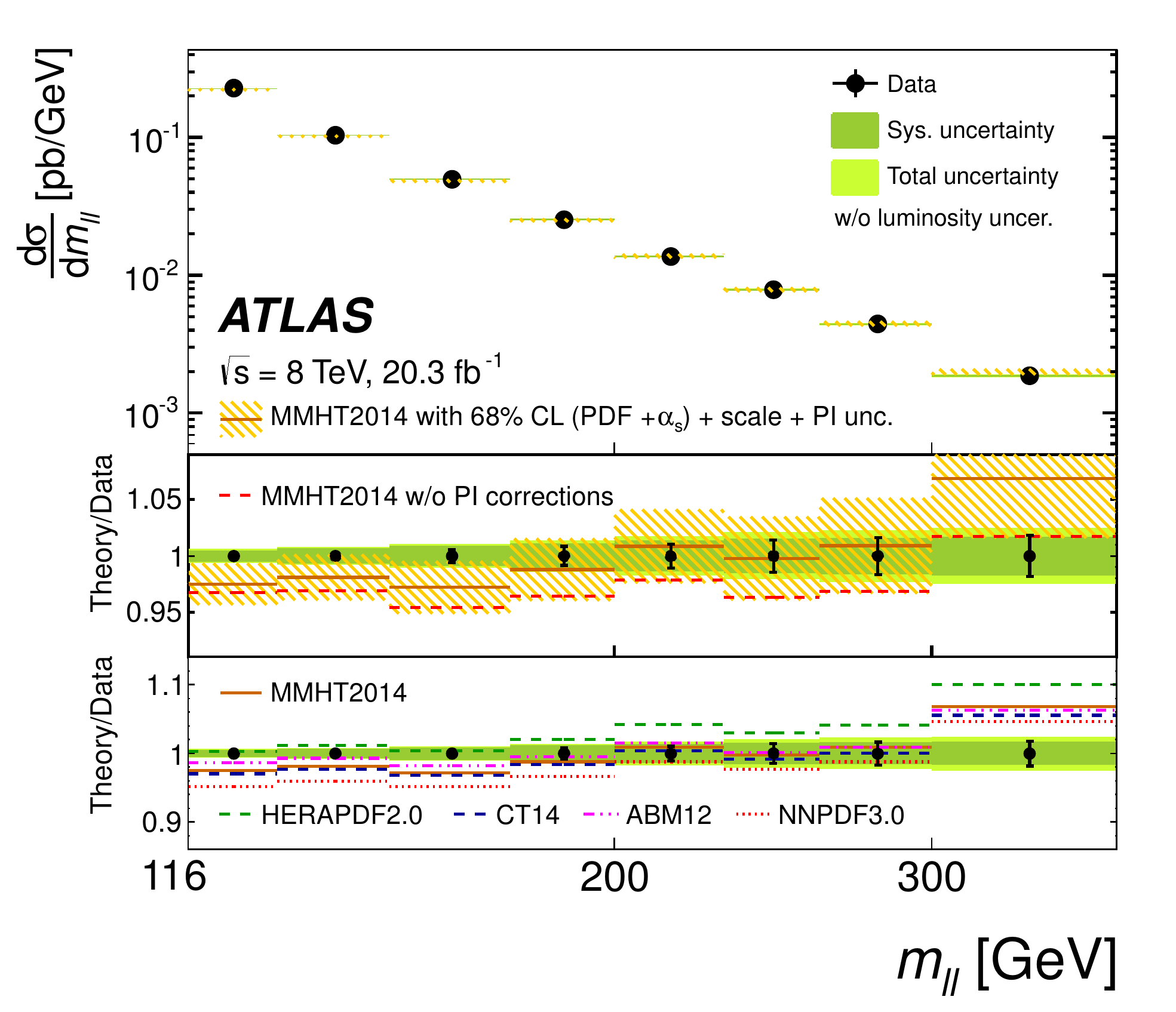}
\end{center}
\caption{The combined single-differential cross section 
as a function of invariant mass \mll\ at the
  Born-level within the fiducial region with statistical,
  systematic and total uncertainties, excluding the $1.9\%$
  uncertainty on the luminosity. Data are compared to combined NNLO
  pQCD and NLO EW calculations using the MMHT2014 PDF, where the
  uncertainty band displays the combined 68\% confidence level (CL)
  PDF and $\alpha_{\rm S}$ variation, the renormalisation and factorisation
  scale uncertainties and the PI uncertainty.  The two ratio panels
  show the ratio of the calculation, both with and without the PI contribution
  with respect to data (middle panel), as well as the ratio for calculations
  using different PDFs (bottom panel).
On the right, the results are shown for a restricted range of $m_{\ell\ell}$.
}
\label{fig:TheoryDataMass}
\end{figure}

Theoretical predictions using the MMHT14 NNLO PDF
set~\cite{Harland-Lang:2014zoa} are compared to the combined
double-differential fiducial cross sections at Born-level as a
function of $m_{\ell\ell}$ and either $|y_{\ell\ell}|$ or $|\Delta
\eta_{\ell\ell}|$ in figure~\ref{fig:2DTheoryMMHTData}.  The
single-differential cross section as a function of $m_{\ell\ell}$ is
shown in figure \ref{fig:TheoryDataMass}.  The uncertainty bands
assigned to the calculations show the combined 68\% confidence level
(CL) PDF and $\alpha_{\rm S}$ variation, the renormalisation and
factorisation scale uncertainties and the PI uncertainty.  The
$\alpha_{\rm S}$ uncertainty is determined by varying $\alpha_{\rm S}$
by 0.001 with respect to its default value of 0.118. The scale
uncertainties are defined by the envelope of variations in which the
scales are changed by factors of two subject to the constraint $0.5
\le \mu_{\mathrm{R}} / \mu_{\mathrm{F}} \le 2$. The relative
uncertainty on the PI contribution represents the region covering 68\%
of the NNPDF2.3qed MC replicas,\footnote{The NNPDF Collaboration
  provides a large number of MC replicas in order to represent the
  PDF, where the central PDF is given by the mean of all replicas and
  the uncertainty is defined as the region covering 68\% of all MC
  replicas.} corresponding to rather large values of 62\%--92\%
depending on the bin.

\begin{figure}
\begin{center}
\includegraphics[width=1.0\textwidth]{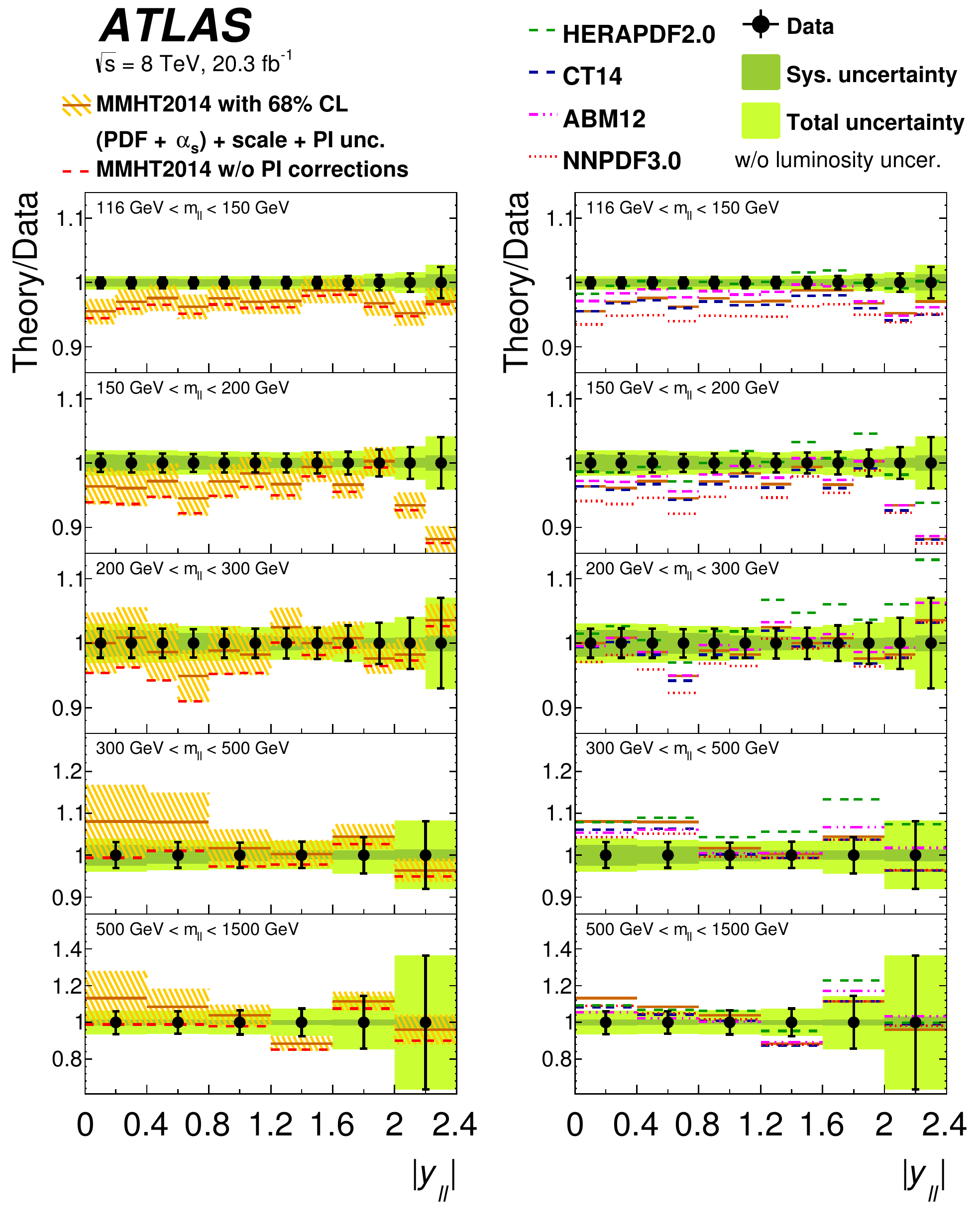}
\end{center}
\caption{The ratio of theoretical NNLO pQCD and NLO EW calculations to
  the combined double-differential cross section as a function of
  invariant mass \mll\ and absolute dilepton rapidity $|y_{\ell\ell}|$
  at the Born-level within the fiducial region with statistical,
  systematic and total uncertainties, excluding the $1.9\%$
  uncertainty on the luminosity.  The calculations are shown for the
  MMHT14 PDF with and without the PI contribution on the left side and for
  MMHT14, HERAPDF2.0, CT10, ABM12 and NNPDF3.0 on the right side. The
  uncertainty band on the left side displays the combined 68\%
  confidence level (CL) PDF and $\alpha_{\rm S}$ variation, the
  renormalisation and factorisation scale uncertainties and the PI
  uncertainty.}
\label{fig:2DTheoryDataRapiditySummary}
\end{figure}

\begin{figure}
\begin{center}
\includegraphics[width=1.0\textwidth]{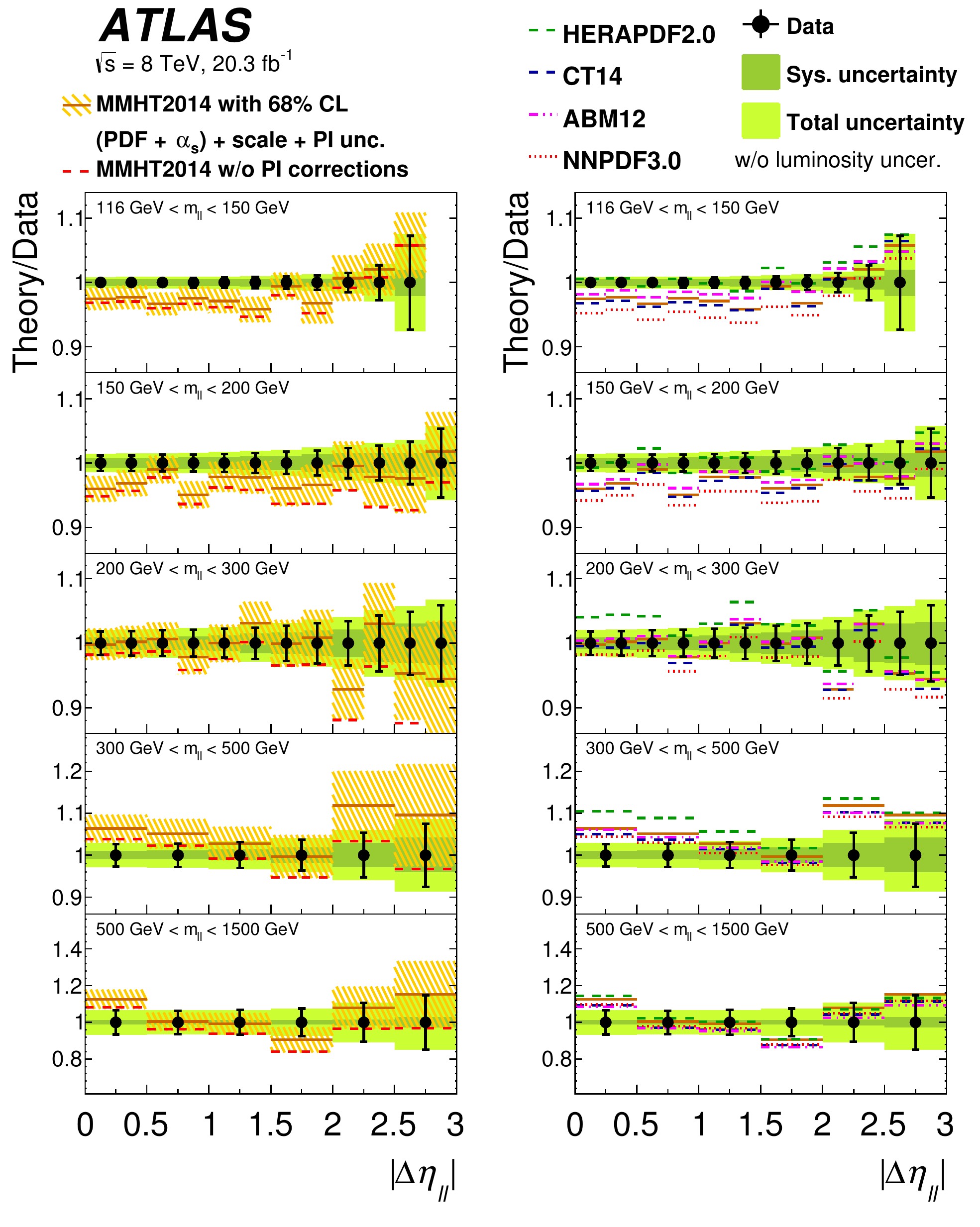}
\end{center}
\caption{The ratio of theoretical NNLO pQCD and NLO EW calculations to
  the combined double-differential cross section as a function of
  invariant mass \mll\ and absolute dilepton pseudorapidity separation
  $|\Delta \eta_{\ell\ell}|$ at Born-level within the fiducial region
  with statistical, systematic and total uncertainties, excluding the
  $1.9\%$ uncertainty on the luminosity.  The calculations are shown
  for the MMHT14 PDF with and without the PI contribution on the left side
  and for MMHT14, HERAPDF2.0, CT10, ABM12 and NNPDF3.0 on the right
  side. The uncertainty band on the left side displays the combined
  68\% confidence level (CL) PDF and $\alpha_{\rm S}$ variation, the
  renormalisation and factorisation scale uncertainties and the PI
  uncertainty.}
\label{fig:2DTheoryDataDeta}
\end{figure}

Figures ~\ref{fig:2DTheoryDataRapiditySummary} and
\ref{fig:2DTheoryDataDeta} show the ratio of theoretical predictions
for the double-differential cross sections to the combined
measurements.  The corresponding ratios for the single-differential
cross section are shown in the middle and lower panels of
figure~\ref{fig:TheoryDataMass}.  The PI contribution reaches up to
$15\%$ in certain regions of phase space, as can be seen in the middle
panels of figure~\ref{fig:TheoryDataMass} and the left panels of
figures~\ref{fig:2DTheoryDataRapiditySummary} and
\ref{fig:2DTheoryDataDeta} where a comparison is shown between the
ratios for the default calculation and a calculation neglecting the PI
contribution.
In the regions where the PI contribution is large the PI uncertainty
dominates the total uncertainty band, otherwise the PDF uncertainty is
dominant.  Thus it can be inferred from, e.g., the middle panel of
figure~\ref{fig:TheoryDataMass}, that the PDF (PI) uncertainty
dominates the uncertainty band at small (large) $m_{\ell\ell}$.  The
uncertainties are similar at about $m_{\ell\ell}\simeq~$200~\GeV.

The uncertainties on the theoretical calculation are in general larger
than those on the measurement for most of the phase space, indicating
that the data presented here have the potential to constrain the
theoretical predictions.  The calculation including the PI
contribution is in general in agreement with the data. However, it
seems to undershoot the data at small $m_{\ell\ell}$ as can be seen in
all three figures.

The change in the theoretical prediction when replacing the MMHT PDF
by other NNLO PDFs such as HERAPDF2.0~\cite{Abramowicz:2015mha},
CT14~\cite{Dulat:2015mca}, ABM12~\cite{Alekhin:2013nda} or
NNPDF3.0~\cite{Ball:2014uwa} is shown in the lower panels of
figure~\ref{fig:TheoryDataMass} and the right panels of figures~\ref
{fig:2DTheoryDataRapiditySummary} and \ref{fig:2DTheoryDataDeta}. For
ease of visibility, no separate uncertainty bands are shown for each
individual PDF. However, they have been calculated and found to be
smaller (ABM12), larger (CT14, NNPDF3.0) or even much larger
(HERAPDF2.0) than the ones from MMHT14 when scaled to the 68\% CL.
The calculations using the various PDFs generally agree with the data,
with the one using NNPDF3.0 showing the least agreement at low
$m_{\ell\ell}$.  In particular at low $m_{\ell\ell}$, the differences
between the predictions in all three figures are larger than the total
uncertainty of the measurement, indicating the sensitivity of the data
to the PDFs, and the potential to constrain them.

\begin{table}[tb]
\begin{center}
%\resizebox{0.5\textwidth}{!}{
\begin{tabular}{r|rrr}
\hline\hline
	   & $m_{\ell\ell}$ & $|y_{\ell\ell}|$ & $|\Delta \eta_{\ell\ell}|$ \\
\hline
 MMHT2014  & 18.2/12  & 59.3/48  & 62.8/47	 \\
 CT14      & 16.0/12  & 51.0/48  & 61.3/47	 \\
 NNPDF3.0  & 20.0/12  & 57.6/48  & 62.1/47	 \\
 HERAPDF2.0& 15.1/12  & 55.5/48  & 60.8/47       \\
 ABM12     & 14.1/12  & 57.9/48  & 53.5/47       \\
\hline
\end{tabular} 
%}%end of resizebox
\caption{The $\chi^2/dof$ values for the compatibility of data and theory after the minimisation procedure. 
%The PDF uncertainty is decomposed into the eigenvectors.
}
\label{table:chi2}
\end{center}
\end{table}

In order to quantitatively assess the compatibility between data and
theory, a $\chi^2$ minimisation procedure as implemented in the
xFitter package \cite{Alekhin:2014irh} is used. All correlated and
uncorrelated experimental uncertainties (see
tables~\ref{table:combXsec_Mass}, ~\ref{table:combXsec_dMdy}
and~\ref{table:combXsec_dMdeta}), the luminosity uncertainty and the
theoretical uncertainties are included in the $\chi^2$ function. The
correlated theoretical uncertainties include the uncertainties on the
respective PDF, the PI contribution, $\alpha_{\rm S}$, and the scale.
The PDF uncertainties for all the PDF sets except for the photon PDF
are further decomposed into the full set of eigenvectors, where in the
case of NNPDF3.0 the replica representation is transformed into an
eigenvector representation.
Uncorrelated theoretical uncertainties arising from the statistical
precision of the calculations are also taken into account but are at
the level of 0.1\% for the calculation of the DY and 0.2\% for the PI
cross sections.  All correlated uncertainties are included as nuisance
parameters.  After minimisation, the central values of the nuisance
parameters may be shifted from unity and their uncertainties may be
reduced. A sizeable shift and reduction in uncertainty indicates that
the data can constrain the respective nuisance parameter.

The $\chi^2$ minimisation is performed separately for all five PDF
sets and for all three distributions shown in
figures~\ref{fig:TheoryDataMass},
~\ref{fig:2DTheoryDataRapiditySummary} and \ref{fig:2DTheoryDataDeta}.
The $\chi^2$ values after minimisation are given in table
\ref{table:chi2}.  They indicate general compatibility between the
data and the theory, mainly due to the fact that the theoretical
uncertainties for all PDF sets except for ABM12 are larger than the
ones for MMHT14 shown in the figures.  The overall best agreement with
the data is found for the calculations using ABM12, especially when
taking into account their smaller PDF uncertainties.

The impact of this data on existing PDF sets can be inferred using
Bayesian reweighting \cite{Giele:1998gw}, which has been further
developed and validated by the NNPDF collaboration in the context of
their MC replicas~\cite{Ball:2010gb,Ball:2011gg}.  Each replica is
used together with this data set in the $\chi^2$ minimisation
described above, and a weight is assigned based on the resulting
$\chi^2$.  Replicas not describing the data well get a smaller weight
assigned, and a new PDF is derived by including the weights in the
calculation of the central value and the PDF uncertainty, i.e., without 
performing a new global PDF fit.

\begin{figure}
\begin{center}
\includegraphics[width=0.495\textwidth]{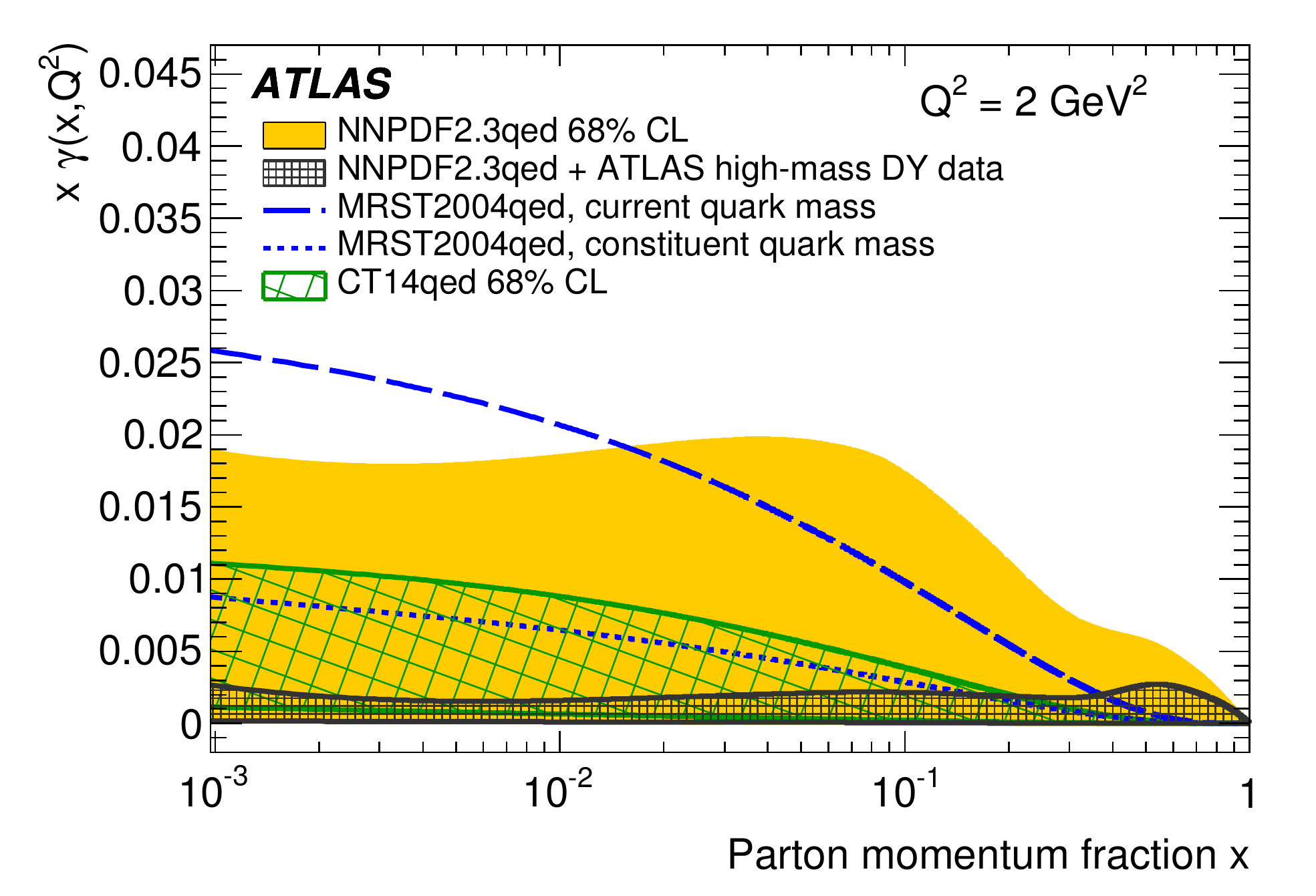}
\includegraphics[width=0.495\textwidth]{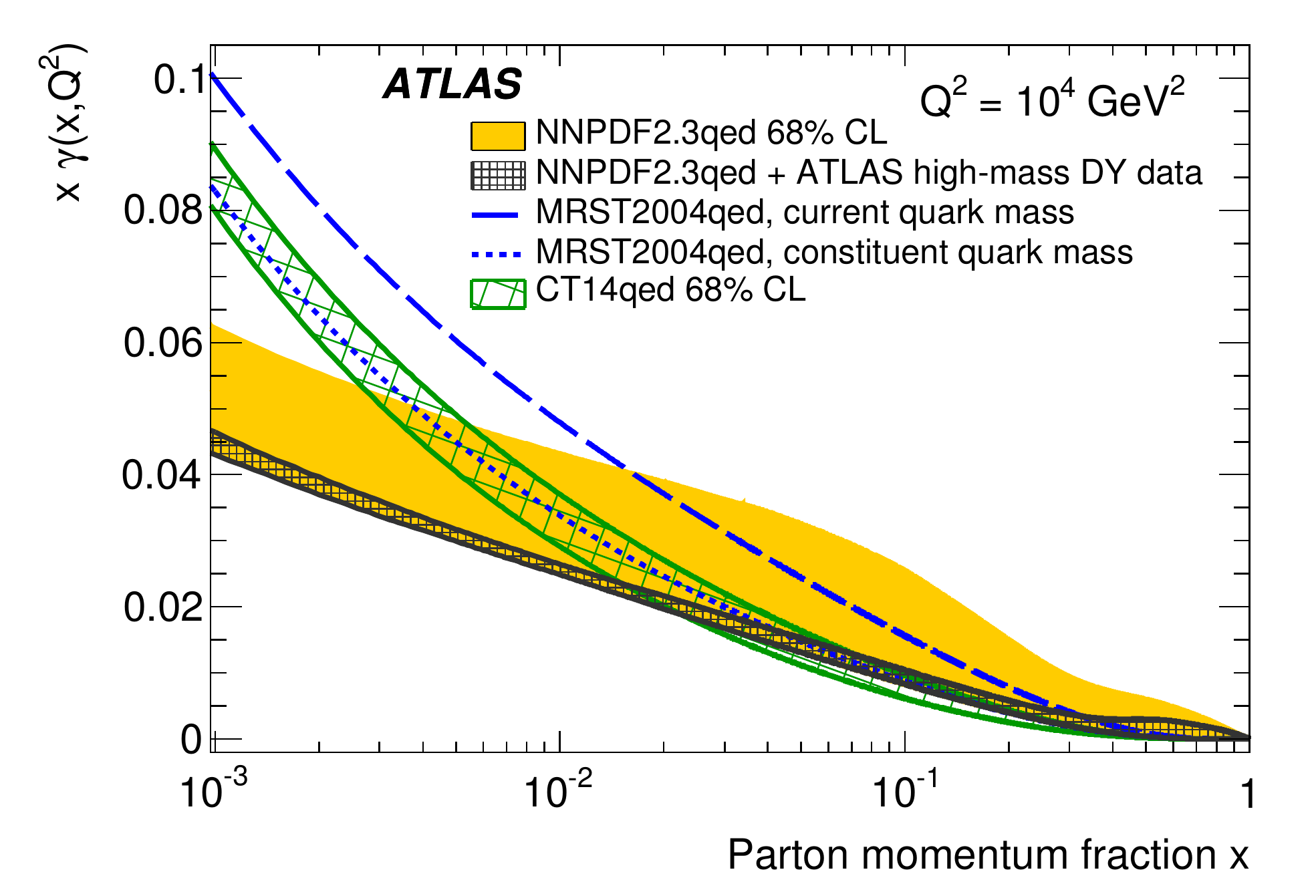}
\end{center}
\caption{The 68\% confidence level interval of the NNPDF2.3qed NNLO
  photon PDF as a function of momentum fraction $x$ at the input scale
  $Q_0^2 = 2$~\GeV$^2$ (left panel) and $Q^2 = 10^4$~\GeV$^2$ (right
  panel) before (yellow solid area)~\cite{Ball:2013hta} and after
  (grey shaded area) inclusion of the double-differential cross
  section measurement as a function of invariant mass $m_{\ell\ell}$
  and absolute dilepton rapidity $|y_{\ell\ell}|$.  Also shown
  is the MRST2004qed photon PDF in a current quark (blue dashed line)
  and a constituent quark (blue dotted line) mass
  scheme~\cite{MRST2004QED}, and the 68\% CL band (green shaded area)
  for the CT14qed photon PDF~\cite{Schmidt:2015zda}.  }
\label{fig:photonPDF}
\end{figure}

Bayesian reweighting is used for the 100 MC replicas representing the
NNPDF2.3qed photon PDF obtained in NNLO fits. The results of the
reweighting are shown in figure~\ref{fig:photonPDF}. The solid yellow area
represents the 68\% confidence level interval of the NNPDF2.3qed
photon PDF at $Q^2 = 2$~\GeV$^2$ and at $Q^2 =
10^4$~\GeV$^2$, which are also displayed in ref.~\cite{Ball:2013hta}.
In addition, the MRST2004qed photon PDF~\cite{MRST2004QED} using two
different quark mass schemes (see section \ref{sec:sys}) and the
recent CT14qed photon PDF~\cite{Schmidt:2015zda} are displayed, which
exhibit a different PDF evolution compared to NNPDF2.3qed as can be
seen when comparing the two plots at the input scale and at $Q^2 =
10^4$~\GeV$^2$.  The shaded area indicates the new PDF after inclusion
of this data by means of Bayesian reweighting, where the $\chi^2$
minimisations are performed for the double-differential cross section
as a function of $m_{\ell\ell}$ and $|y_{\ell\ell}|$ using the
prediction based on the MMHT14 PDF.
The reduction of uncertainties is rather large and confirms the strong
sensitivity of this data to the photon PDF.  Using the
double-differential cross section as a function of $m_{\ell\ell}$ and
$|\Delta \eta_{\ell\ell}|$ instead, a slightly smaller impact is
found. This can be explained by the fact that the contributions from
the PI process are largest in the regions of small $|y_{\ell\ell}|$
and large $|\Delta \eta_{\ell\ell}|$, where the uncertainties of the
measurement are smallest for $|y_{\ell\ell}|$ but largest for $|\Delta
\eta_{\ell\ell}|$.

Inspection of the optimised experimental nuisance parameters of those minimisations 
with the best $\chi^2$ values    
shows that the largest pulls and uncertainty reductions are found for the luminosity.
Larger values for the data luminosity by about 1.1 and 1.2 
standard deviations are favoured for the minimisations in
$|y_{\ell\ell}|$ and $|\Delta \eta_{\ell\ell}|$ respectively, leading
to smaller values for the experimental cross section.  For the MMHT14
PDF, the largest pulls and reduction of uncertainty by about 25\% is
found for an eigenvector (``eigenvector 21'') particularly sensitive
to the sea and strange sea quark distribution, where previous ATLAS
data on on-shell $W$ and $Z$ production~\cite{STDM-2011-06} is already
the most constraining data set in one eigenvector
direction~\cite{Harland-Lang:2014zoa}.

\FloatBarrier

\FloatBarrier
%-------------------------------------------------------------------------------
\section{Conclusion}
\label{sec:conclusion}
%-------------------------------------------------------------------------------

The double-differential fiducial cross sections ${\rm d}^2\sigma/{\rm
  d}m_{\ell\ell}{\rm d}|y_{\ell\ell}|$ and ${\rm d}^2\sigma/{\rm
  d}m_{\ell\ell}{\rm d}|\Delta\eta_{\ell\ell}|$ for the Drell--Yan and
photon induced production of dileptons in the invariant mass range
$116<m_{\ell\ell}<1500$~\GeV\ are measured,
as well as the single-differential fiducial cross section
${\rm d}\sigma/{\rm d}m_{\ell\ell}$.
The measurements are performed in the electron and muon channels using
$20.3$~fb$^{-1}$ of integrated luminosity collected by the ATLAS detector at the LHC
in $pp$ collisions at $\sqrt{s}=8$~TeV.
The two measurements are combined taking into
account the systematic uncertainty correlations. 
The combined cross sections achieve an experimental
precision of better than $1\%$ at low $m_{\ell\ell}$, 
excluding the overall uncertainty in the luminosity measurement of $1.9\%$.

The fiducial cross sections are compared to fixed order theoretical
predictions at NNLO accuracy using a selection of recent PDF sets. The
calculations are performed using renormalisation and factorisation
scales equal to $m_{\ell\ell}$, and are corrected for additional
higher-order electroweak radiative effects. Theoretical uncertainties
arising from the PDFs, the choice of scale and a variation of
$\alpha_{\rm S}$ are calculated and are found to be larger than the
measurement uncertainties, indicating the data has the potential to
constrain PDFs. This is confirmed by a $\chi^2$ minimisation procedure
comparing data to theoretical predictions, which in combination with a
Bayesian reweighting method shows a dramatic reduction of the
uncertainties on the photon PDF.

%%\input{tab}

%-------------------------------------------------------------------------------
\section*{Acknowledgements}
%-------------------------------------------------------------------------------

% Acknowledgements for papers with collision data
% Version 15-Apr-2016

% Standard acknowledgements start here
%----------------------------------------------
We thank CERN for the very successful operation of the LHC, as well as the
support staff from our institutions without whom ATLAS could not be
operated efficiently.

We acknowledge the support of ANPCyT, Argentina; YerPhI, Armenia; ARC, Australia; BMWFW and FWF, Austria; ANAS, Azerbaijan; SSTC, Belarus; CNPq and FAPESP, Brazil; NSERC, NRC and CFI, Canada; CERN; CONICYT, Chile; CAS, MOST and NSFC, China; COLCIENCIAS, Colombia; MSMT CR, MPO CR and VSC CR, Czech Republic; DNRF and DNSRC, Denmark; IN2P3-CNRS, CEA-DSM/IRFU, France; GNSF, Georgia; BMBF, HGF, and MPG, Germany; GSRT, Greece; RGC, Hong Kong SAR, China; ISF, I-CORE and Benoziyo Center, Israel; INFN, Italy; MEXT and JSPS, Japan; CNRST, Morocco; FOM and NWO, Netherlands; RCN, Norway; MNiSW and NCN, Poland; FCT, Portugal; MNE/IFA, Romania; MES of Russia and NRC KI, Russian Federation; JINR; MESTD, Serbia; MSSR, Slovakia; ARRS and MIZ\v{S}, Slovenia; DST/NRF, South Africa; MINECO, Spain; SRC and Wallenberg Foundation, Sweden; SERI, SNSF and Cantons of Bern and Geneva, Switzerland; MOST, Taiwan; TAEK, Turkey; STFC, United Kingdom; DOE and NSF, United States of America. In addition, individual groups and members have received support from BCKDF, the Canada Council, CANARIE, CRC, Compute Canada, FQRNT, and the Ontario Innovation Trust, Canada; EPLANET, ERC, FP7, Horizon 2020 and Marie Sklodowska-Curie Actions, European Union; Investissements d'Avenir Labex and Idex, ANR, R{\'e}gion Auvergne and Fondation Partager le Savoir, France; DFG and AvH Foundation, Germany; Herakleitos, Thales and Aristeia programmes co-financed by EU-ESF and the Greek NSRF; BSF, GIF and Minerva, Israel; BRF, Norway; Generalitat de Catalunya, Generalitat Valenciana, Spain; the Royal Society and Leverhulme Trust, United Kingdom.

The crucial computing support from all WLCG partners is acknowledged
gratefully, in particular from CERN and the ATLAS Tier-1 facilities at
TRIUMF (Canada), NDGF (Denmark, Norway, Sweden), CC-IN2P3 (France),
KIT/GridKA (Germany), INFN-CNAF (Italy), NL-T1 (Netherlands), PIC (Spain),
ASGC (Taiwan), RAL (UK) and BNL (USA) and in the Tier-2 facilities
worldwide.
%----------------------------------------------

%The \texttt{atlaslatex} package contains the acknowledgements that were valid 
%at the time of the release you are using.
%These can be found in the \texttt{acknowledgements} subdirectory.
%When your ATLAS paper or PUB/CONF note is ready to be published,
%download the latest set of acknowledgements from:\\
%\url{https://twiki.cern.ch/twiki/bin/view/AtlasProtected/PubComAcknowledgements}

%The supporting notes for the analysis should also contain a list of contributors.
%This information should usually be included in \texttt{mydocument-metadata.tex}.
%The list should be printed either here or before the table of contents.

%-------------------------------------------------------------------------------
\clearpage
\appendix
\part*{Appendix}
\addcontentsline{toc}{part}{Appendix}
\begin{table}[!htb]
\begin{center}
\resizebox{\textwidth}{!}{
\begin{tabular}{r|rrrr|rrrrrrrrrrrrrr|r}
\hline\hline
$m_{ee}$ & $\frac{\text{d}\sigma}{\text{d}m_{ee}}$  & 
$\delta^{\rm stat}$           &  $\delta^{\rm sys}$          & $\delta^{\rm tot}$ &
$\delta^{\rm trig}_{\rm cor}$     &  $\delta^{\rm trig}_{\rm unc}$   &
$\delta^{\rm reco}_{\rm cor}$     &  $\delta^{\rm id}_{\rm cor}$   &
$\delta^{\rm iso}_{\rm cor}$      &  $\delta^{\rm iso}_{\rm unc}$  &
$\delta^{\rm Eres}_{\rm cor}$    &
$\delta^{\rm Escale}_{\rm cor}$  &
$\delta^{\rm mult.}_{\rm cor}$      &  $\delta^{\rm mult.}_{\rm unc}$    &
$\delta^{\rm top}_{\rm cor}$      &  $\delta^{\rm diboson}_{\rm cor}$    &  $\delta^{\rm bgMC}_{\rm unc}$    &
$\delta^{\rm MC}_{\rm unc}$ & $k_{\rm dressed}$ \\
{[\GeV]}  & [pb/\GeV] & [\%] & [\%] & [\%] & [\%] & [\%]& [\%] & [\%]& [\%] & [\%]& [\%] & [\%]& [\%]& [\%] & [\%] & [\%] & [\%] & [\%] & \\
\hline
116--130 & $2.31 \times 10^{-1}$ &  0.5 &  0.8 &  1.0 & -0.1 &  0.0 &  0.0 & -0.3 &  0.0 &  0.0 &  0.1 &  0.5 & -0.5 &  0.1 & -0.3 & -0.1 &  0.0 &  0.1 & 1.047 \\ 
130--150 & $1.05 \times 10^{-1}$ &  0.7 &  1.0 &  1.2 & -0.1 &  0.0 & -0.1 & -0.4 &  0.0 &  0.1 &  0.1 &  0.4 & -0.7 &  0.2 & -0.5 & -0.2 &  0.1 &  0.1 & 1.046 \\ 
150--175 & $5.06 \times 10^{-2}$ &  0.8 &  1.3 &  1.6 &  0.0 &  0.1 & -0.1 & -0.5 &  0.0 &  0.1 &  0.1 &  0.4 & -0.8 &  0.3 & -0.7 & -0.2 &  0.1 &  0.1 & 1.047 \\ 
175--200 & $2.60 \times 10^{-2}$ &  1.2 &  1.6 &  2.0 & -0.1 &  0.1 & -0.1 & -0.6 &  0.0 &  0.1 &  0.0 &  0.5 & -0.9 &  0.3 & -0.9 & -0.3 &  0.2 &  0.1 & 1.052 \\ 
200--230 & $1.39 \times 10^{-2}$ &  1.5 &  2.0 &  2.5 & -0.1 &  0.1 & -0.1 & -0.7 &  0.0 &  0.2 &  0.1 &  0.7 & -1.2 &  0.4 & -1.1 & -0.4 &  0.2 &  0.2 & 1.053 \\ 
230--260 & $7.95 \times 10^{-3}$ &  2.0 &  2.2 &  3.0 & -0.1 &  0.1 & -0.2 & -0.7 & -0.1 &  0.2 &  0.1 &  1.0 & -1.1 &  0.4 & -1.3 & -0.4 &  0.3 &  0.2 & 1.056 \\ 
260--300 & $4.43 \times 10^{-3}$ &  2.4 &  2.3 &  3.3 & -0.1 &  0.1 & -0.2 & -0.7 & -0.1 &  0.2 &  0.1 &  0.9 & -1.3 &  0.5 & -1.3 & -0.6 &  0.4 &  0.2 & 1.058 \\ 
300--380 & $1.84 \times 10^{-3}$ &  2.6 &  2.5 &  3.6 & -0.1 &  0.2 & -0.2 & -0.8 & -0.1 &  0.3 &  0.1 &  1.3 & -1.1 &  0.4 & -1.4 & -0.6 &  0.4 &  0.2 & 1.063 \\ 
380--500 & $5.99 \times 10^{-4}$ &  3.6 &  2.7 &  4.5 & -0.1 &  0.2 & -0.2 & -0.8 & -0.2 &  0.5 &  0.1 &  1.6 & -1.4 &  0.5 & -1.1 & -0.6 &  0.5 &  0.2 & 1.067 \\ 
500--700 & $1.52 \times 10^{-4}$ &  5.3 &  2.6 &  6.0 & -0.1 &  0.2 & -0.2 & -0.8 & -0.2 &  0.7 &  0.1 &  2.0 & -0.7 &  0.5 & -0.7 & -0.6 &  0.5 &  0.3 & 1.075 \\ 
700--1000 & $2.64 \times 10^{-5}$ & 10.2 &  3.3 & 10.7 & -0.2 &  0.4 & -0.2 & -0.8 & -0.3 &  1.4 &  0.1 &  2.3 & -0.6 &  0.8 & -0.4 & -0.6 &  0.7 &  0.4 & 1.085 \\ 
1000--1500 & $3.23 \times 10^{-6}$ & 22.5 &  5.8 & 23.2 & -0.7 &  0.9 & -0.2 & -0.8 & -0.3 &  3.5 &  0.0 &  2.8 & -1.9 &  1.6 & -0.3 & -0.6 &  2.1 &  0.2 & 1.100 \\ 
\hline
\end{tabular} }%end of resizebox
\caption{The electron channel Born-level single-differential cross section $\frac{\text{d}\sigma}{\text{d}m_{ee}}$. 
The measurements are listed together with the statistical ($\delta^{\rm stat}$), systematic ($\delta^{\rm sys}$) and total ($\delta^{\rm tot}$) uncertainties.
In addition the contributions from the individual correlated ({\rm cor}) and uncorrelated ({\rm unc}) systematic error sources are also provided consisting of the
trigger efficiency ($\delta^{\rm trig}$), electron reconstruction efficiency ($\delta^{\rm reco}$), electron identification efficiency ($\delta^{\rm id}$),
the isolation efficiency ($\delta^{\rm iso}$), 
the electron energy resolution ($\delta^{\rm Eres}$), the electron energy scale ($\delta^{\rm Escale}$), 
the multijet and $W$+jets background ($\delta^{\rm mult.}$), 
the top and diboson background normalisation ($\delta^{\rm top},\delta^{\rm diboson}$), 
the top and diboson background MC statistical uncertainty ($\delta^{\rm bgMC}$), 
and the signal MC statistical uncertainty ($\delta^{\rm MC}$).
The ratio of the dressed-level to Born-level predictions ($k_{\rm dressed}$) is also provided.
The luminosity uncertainty of 1.9\% is not shown and not included in the overall systematic and total uncertainties.}
\label{table:elecXsec_Mass}
\end{center}
\end{table}

\begin{table}[!htb]
\begin{center}
\resizebox{\textwidth}{!}{
\begin{tabular}{rc|rrrr|rrrrrrrrrrrrrr|r}
\hline\hline
$m_{ee}$  &  $|y_{ee}|$ & $\frac{\text{d}^2\sigma}{\text{d}m_{ee}\text{d}|y_{ee}|}$  & 
$\delta^{\rm stat}$           &  $\delta^{\rm sys}$          & $\delta^{\rm tot}$ &
$\delta^{\rm trig}_{\rm cor}$     &  $\delta^{\rm trig}_{\rm unc}$   &
$\delta^{\rm reco}_{\rm cor}$     &  $\delta^{\rm id}_{\rm cor}$   &
$\delta^{\rm iso}_{\rm cor}$      &  $\delta^{\rm iso}_{\rm unc}$  &
$\delta^{\rm Eres}_{\rm cor}$    &
$\delta^{\rm Escale}_{\rm cor}$  &
$\delta^{\rm mult.}_{\rm cor}$      &  $\delta^{\rm mult.}_{\rm unc}$    &
$\delta^{\rm top}_{\rm cor}$      &  $\delta^{\rm diboson}_{\rm cor}$    &  $\delta^{\rm bgMC}_{\rm unc}$    &
$\delta^{\rm MC}_{\rm unc}$ & $k_{\rm dressed}$ \\
{[GeV]} &  & [pb/GeV] & [\%] & [\%] & [\%] & [\%] & [\%]& [\%] & [\%]& [\%]& [\%] & [\%]& [\%] & [\%]& [\%] & [\%] & [\%] & [\%] & [\%] & \\
\hline
116--150 & 0.0--0.2 & $4.15 \times 10^{-2}$ &  1.1 &  0.8 &  1.4 & -0.1 &  0.0 & -0.1 & -0.3 &  0.0 &  0.0 &  0.1 &  0.2 & -0.3 &  0.1 & -0.5 & -0.1 &  0.1 &  0.2 & 1.048 \\ 
116--150 & 0.2--0.4 & $4.11 \times 10^{-2}$ &  1.2 &  0.8 &  1.4 & -0.1 &  0.0 & -0.1 & -0.3 &  0.0 &  0.0 &  0.1 &  0.3 & -0.3 &  0.1 & -0.5 & -0.1 &  0.1 &  0.2 & 1.048 \\ 
116--150 & 0.4--0.6 & $4.09 \times 10^{-2}$ &  1.2 &  0.9 &  1.5 & -0.1 &  0.0 & -0.1 & -0.3 &  0.0 &  0.0 &  0.3 &  0.3 & -0.4 &  0.1 & -0.5 & -0.1 &  0.1 &  0.3 & 1.047 \\ 
116--150 & 0.6--0.8 & $4.09 \times 10^{-2}$ &  1.2 &  0.9 &  1.5 & -0.1 &  0.0 & -0.1 & -0.3 &  0.0 &  0.0 &  0.2 &  0.4 & -0.3 &  0.1 & -0.4 & -0.1 &  0.1 &  0.3 & 1.048 \\ 
116--150 & 0.8--1.0 & $3.97 \times 10^{-2}$ &  1.3 &  0.9 &  1.6 & -0.1 &  0.0 & -0.1 & -0.3 &  0.0 &  0.0 &  0.2 &  0.5 & -0.3 &  0.2 & -0.4 & -0.1 &  0.1 &  0.3 & 1.047 \\ 
116--150 & 1.0--1.2 & $3.97 \times 10^{-2}$ &  1.3 &  1.0 &  1.6 & -0.1 &  0.0 & -0.1 & -0.3 &  0.0 &  0.0 &  0.1 &  0.6 & -0.5 &  0.2 & -0.3 & -0.1 &  0.1 &  0.3 & 1.047 \\ 
116--150 & 1.2--1.4 & $3.86 \times 10^{-2}$ &  1.3 &  1.2 &  1.8 & -0.1 &  0.0 & -0.1 & -0.3 &  0.0 &  0.0 &  0.3 &  0.7 & -0.6 &  0.2 & -0.3 & -0.1 &  0.1 &  0.3 & 1.046 \\ 
116--150 & 1.4--1.6 & $3.44 \times 10^{-2}$ &  1.4 &  1.3 &  1.9 & -0.1 &  0.0 & -0.1 & -0.4 &  0.0 &  0.0 &  0.2 &  0.8 & -0.7 &  0.2 & -0.2 & -0.1 &  0.1 &  0.3 & 1.046 \\ 
116--150 & 1.6--1.8 & $2.86 \times 10^{-2}$ &  1.6 &  1.5 &  2.2 & -0.1 &  0.0 & -0.1 & -0.5 &  0.0 &  0.0 &  0.2 &  1.0 & -0.9 &  0.3 & -0.2 & -0.1 &  0.1 &  0.4 & 1.044 \\ 
116--150 & 1.8--2.0 & $2.29 \times 10^{-2}$ &  1.8 &  1.6 &  2.4 & -0.1 &  0.0 & -0.1 & -0.6 &  0.0 &  0.1 &  0.3 &  1.1 & -0.9 &  0.3 & -0.1 & -0.1 &  0.1 &  0.4 & 1.043 \\ 
116--150 & 2.0--2.2 & $1.49 \times 10^{-2}$ &  2.1 &  2.0 &  2.9 & -0.1 &  0.0 & -0.1 & -0.6 &  0.0 &  0.1 &  0.4 &  0.8 & -1.5 &  0.4 & -0.1 & -0.1 &  0.1 &  0.5 & 1.044 \\ 
116--150 & 2.2--2.4 & $7.05 \times 10^{-3}$ &  3.3 &  3.1 &  4.5 &  0.0 &  0.0 & -0.2 & -0.7 &  0.0 &  0.1 &  0.4 &  1.2 & -2.5 &  0.6 & -0.1 & -0.1 &  0.2 &  0.8 & 1.045 \\ 
\hline
150--200 & 0.0--0.2 & $1.06 \times 10^{-2}$ &  2.0 &  1.5 &  2.5 & -0.1 &  0.1 & -0.1 & -0.5 &  0.0 &  0.1 &  0.1 &  0.2 & -0.6 &  0.2 & -1.2 & -0.3 &  0.3 &  0.2 & 1.052 \\ 
150--200 & 0.2--0.4 & $1.06 \times 10^{-2}$ &  2.0 &  1.5 &  2.6 & -0.1 &  0.1 & -0.1 & -0.5 &  0.0 &  0.1 &  0.1 &  0.3 & -0.6 &  0.3 & -1.1 & -0.3 &  0.3 &  0.2 & 1.050 \\ 
150--200 & 0.4--0.6 & $1.05 \times 10^{-2}$ &  2.1 &  1.5 &  2.6 & -0.1 &  0.1 & -0.1 & -0.5 &  0.0 &  0.1 &  0.1 &  0.2 & -0.6 &  0.3 & -1.1 & -0.4 &  0.3 &  0.3 & 1.052 \\ 
150--200 & 0.6--0.8 & $1.06 \times 10^{-2}$ &  2.1 &  1.5 &  2.6 & -0.1 &  0.1 & -0.1 & -0.5 &  0.0 &  0.1 &  0.2 &  0.5 & -0.7 &  0.3 & -1.0 & -0.3 &  0.3 &  0.3 & 1.053 \\ 
150--200 & 0.8--1.0 & $1.02 \times 10^{-2}$ &  2.1 &  1.5 &  2.6 & -0.1 &  0.1 & -0.1 & -0.5 &  0.0 &  0.1 &  0.1 &  0.4 & -0.8 &  0.4 & -0.8 & -0.3 &  0.3 &  0.3 & 1.050 \\ 
150--200 & 1.0--1.2 & $9.71 \times 10^{-3}$ &  2.2 &  1.7 &  2.8 & -0.1 &  0.1 & -0.1 & -0.5 &  0.0 &  0.1 &  0.2 &  0.6 & -1.1 &  0.4 & -0.7 & -0.3 &  0.3 &  0.3 & 1.050 \\ 
150--200 & 1.2--1.4 & $9.25 \times 10^{-3}$ &  2.3 &  1.5 &  2.7 & -0.1 &  0.1 & -0.1 & -0.5 &  0.0 &  0.1 &  0.1 &  0.7 & -0.8 &  0.3 & -0.6 & -0.3 &  0.3 &  0.3 & 1.048 \\ 
150--200 & 1.4--1.6 & $7.60 \times 10^{-3}$ &  2.5 &  1.8 &  3.1 &  0.0 &  0.1 & -0.1 & -0.6 &  0.0 &  0.1 &  0.1 &  1.0 & -1.0 &  0.4 & -0.5 & -0.3 &  0.3 &  0.3 & 1.046 \\ 
150--200 & 1.6--1.8 & $6.66 \times 10^{-3}$ &  2.8 &  1.9 &  3.3 &  0.0 &  0.1 & -0.2 & -0.7 &  0.0 &  0.1 &  0.2 &  1.2 & -1.1 &  0.4 & -0.3 & -0.2 &  0.3 &  0.4 & 1.043 \\ 
150--200 & 1.8--2.0 & $4.94 \times 10^{-3}$ &  3.1 &  1.7 &  3.6 &  0.0 &  0.1 & -0.2 & -0.8 &  0.0 &  0.1 &  0.3 &  0.8 & -1.0 &  0.4 & -0.2 & -0.2 &  0.3 &  0.5 & 1.043 \\ 
150--200 & 2.0--2.2 & $3.30 \times 10^{-3}$ &  3.5 &  1.9 &  4.0 &  0.0 &  0.1 & -0.4 & -0.8 &  0.0 &  0.1 &  0.2 &  0.5 & -1.4 &  0.4 & -0.1 & -0.1 &  0.3 &  0.5 & 1.038 \\ 
150--200 & 2.2--2.4 & $1.52 \times 10^{-3}$ &  5.5 &  3.2 &  6.3 & -0.1 &  0.1 & -0.6 & -0.9 &  0.0 &  0.1 &  0.2 &  1.3 & -2.4 &  0.8 & -0.1 & -0.1 &  0.3 &  0.8 & 1.038 \\ 
\hline
200--300 & 0.0--0.2 & $2.33 \times 10^{-3}$ &  3.2 &  2.5 &  4.1 & -0.1 &  0.1 & -0.2 & -0.7 & -0.1 &  0.2 &  0.1 &  0.5 & -0.9 &  0.5 & -1.9 & -0.5 &  0.6 &  0.3 & 1.063 \\ 
200--300 & 0.2--0.4 & $2.34 \times 10^{-3}$ &  3.2 &  2.4 &  4.0 & -0.1 &  0.1 & -0.2 & -0.7 & -0.1 &  0.2 &  0.2 &  0.4 & -0.9 &  0.5 & -1.8 & -0.5 &  0.6 &  0.3 & 1.063 \\ 
200--300 & 0.4--0.6 & $2.49 \times 10^{-3}$ &  3.2 &  2.4 &  4.0 & -0.1 &  0.1 & -0.2 & -0.7 & -0.1 &  0.2 &  0.1 &  0.5 & -1.3 &  0.6 & -1.6 & -0.6 &  0.6 &  0.3 & 1.063 \\ 
200--300 & 0.6--0.8 & $2.54 \times 10^{-3}$ &  3.1 &  2.3 &  3.9 & -0.1 &  0.1 & -0.1 & -0.7 & -0.1 &  0.2 &  0.1 &  0.8 & -1.2 &  0.6 & -1.4 & -0.5 &  0.5 &  0.3 & 1.060 \\ 
200--300 & 0.8--1.0 & $2.29 \times 10^{-3}$ &  3.3 &  2.3 &  4.0 & -0.1 &  0.1 & -0.1 & -0.7 & -0.1 &  0.2 &  0.2 &  1.0 & -1.1 &  0.6 & -1.3 & -0.5 &  0.5 &  0.3 & 1.056 \\ 
200--300 & 1.0--1.2 & $2.14 \times 10^{-3}$ &  3.4 &  2.4 &  4.1 & -0.1 &  0.1 & -0.2 & -0.7 & -0.1 &  0.2 &  0.2 &  1.3 & -1.3 &  0.5 & -1.0 & -0.5 &  0.5 &  0.4 & 1.053 \\ 
200--300 & 1.2--1.4 & $1.83 \times 10^{-3}$ &  3.6 &  2.4 &  4.4 & -0.1 &  0.1 & -0.2 & -0.7 & -0.1 &  0.2 &  0.1 &  1.4 & -1.4 &  0.5 & -0.8 & -0.4 &  0.5 &  0.4 & 1.049 \\ 
200--300 & 1.4--1.6 & $1.63 \times 10^{-3}$ &  3.7 &  2.1 &  4.3 & -0.1 &  0.1 & -0.2 & -0.8 & -0.1 &  0.2 &  0.1 &  1.2 & -1.2 &  0.4 & -0.6 & -0.3 &  0.5 &  0.4 & 1.044 \\ 
200--300 & 1.6--1.8 & $1.32 \times 10^{-3}$ &  4.2 &  2.3 &  4.8 & -0.1 &  0.1 & -0.3 & -0.8 & -0.1 &  0.2 &  0.3 &  1.5 & -1.2 &  0.4 & -0.4 & -0.3 &  0.5 &  0.5 & 1.041 \\ 
200--300 & 1.8--2.0 & $9.87 \times 10^{-4}$ &  4.8 &  2.4 &  5.4 & -0.1 &  0.1 & -0.4 & -0.9 & -0.1 &  0.2 &  0.3 &  1.5 & -1.2 &  0.5 & -0.2 & -0.2 &  0.5 &  0.6 & 1.044 \\ 
200--300 & 2.0--2.2 & $6.13 \times 10^{-4}$ &  5.6 &  2.3 &  6.1 & -0.1 &  0.1 & -0.6 & -1.0 & -0.1 &  0.2 &  0.3 &  0.7 & -1.6 &  0.5 & -0.1 & -0.1 &  0.4 &  0.6 & 1.044 \\ 
200--300 & 2.2--2.4 & $2.51 \times 10^{-4}$ &  9.1 &  3.2 &  9.6 & -0.1 &  0.2 & -0.9 & -1.1 & -0.1 &  0.2 &  0.7 &  1.4 & -1.8 &  1.1 & -0.1 & -0.1 &  0.5 &  1.1 & 1.042 \\ 
\hline
300--500 & 0.0--0.4 & $3.23 \times 10^{-4}$ &  4.6 &  3.3 &  5.7 & -0.1 &  0.2 & -0.2 & -0.8 & -0.1 &  0.4 &  0.1 &  0.9 & -1.8 &  0.6 & -2.2 & -0.8 &  0.8 &  0.3 & 1.080 \\ 
300--500 & 0.4--0.8 & $3.34 \times 10^{-4}$ &  4.3 &  2.8 &  5.1 & -0.1 &  0.2 & -0.2 & -0.8 & -0.1 &  0.4 &  0.1 &  1.4 & -1.1 &  0.6 & -1.6 & -0.7 &  0.7 &  0.3 & 1.072 \\ 
300--500 & 0.8--1.2 & $3.16 \times 10^{-4}$ &  4.3 &  2.8 &  5.2 & -0.1 &  0.2 & -0.2 & -0.8 & -0.1 &  0.4 &  0.2 &  2.0 & -0.9 &  0.5 & -1.1 & -0.6 &  0.7 &  0.3 & 1.058 \\ 
300--500 & 1.2--1.6 & $2.30 \times 10^{-4}$ &  4.9 &  2.9 &  5.7 & -0.1 &  0.2 & -0.2 & -0.8 & -0.1 &  0.4 &  0.1 &  2.0 & -1.6 &  0.5 & -0.6 & -0.4 &  0.6 &  0.4 & 1.053 \\ 
300--500 & 1.6--2.0 & $1.31 \times 10^{-4}$ &  6.5 &  3.2 &  7.3 & -0.1 &  0.2 & -0.4 & -0.9 & -0.2 &  0.4 &  0.2 &  2.8 & -0.3 &  0.4 & -0.2 & -0.2 &  0.5 &  0.6 & 1.047 \\ 
300--500 & 2.0--2.4 & $3.62 \times 10^{-5}$ & 11.5 &  3.5 & 12.0 & -0.1 &  0.2 & -0.6 & -1.0 & -0.2 &  0.4 &  0.4 &  2.5 & -1.3 &  1.0 &  0.0 & -0.1 &  0.8 &  0.9 & 1.046 \\ 
\hline
500--1500 & 0.0--0.4 & $1.45 \times 10^{-5}$ &  8.9 &  2.8 &  9.4 & -0.2 &  0.3 & -0.2 & -0.8 & -0.2 &  1.0 &  0.1 &  1.5 & -0.7 &  0.8 & -1.0 & -0.7 &  1.0 &  0.3 & 1.096 \\ 
500--1500 & 0.4--0.8 & $1.45 \times 10^{-5}$ &  8.5 &  2.9 &  9.0 & -0.2 &  0.3 & -0.2 & -0.8 & -0.2 &  1.0 &  0.1 &  2.1 & -0.3 &  0.6 & -0.6 & -0.6 &  0.7 &  0.5 & 1.083 \\ 
500--1500 & 0.8--1.2 & $1.05 \times 10^{-5}$ & 10.0 &  3.5 & 10.6 & -0.1 &  0.3 & -0.2 & -0.8 & -0.2 &  0.9 &  0.1 &  2.7 & -1.1 &  0.8 & -0.5 & -0.5 &  0.9 &  0.5 & 1.067 \\ 
500--1500 & 1.2--1.6 & $7.86 \times 10^{-6}$ & 11.1 &  3.6 & 11.7 & -0.1 &  0.2 & -0.2 & -0.8 & -0.2 &  0.9 &  0.1 &  3.2 & -0.3 &  0.7 & -0.1 & -0.2 &  0.4 &  0.4 & 1.055 \\ 
500--1500 & 1.6--2.0 & $2.29 \times 10^{-6}$ & 21.4 &  4.3 & 21.8 & -0.1 &  0.2 & -0.4 & -0.9 & -0.3 &  0.8 &  0.3 &  3.9 & -0.4 &  0.7 & -0.1 & -0.2 &  0.7 &  0.9 & 1.056 \\ 
500--1500 & 2.0--2.4 & $2.51 \times 10^{-7}$ & 60.4 &  7.8 & 60.9 & -0.1 &  0.2 & -0.6 & -1.0 & -0.3 &  0.8 &  1.1 &  5.7 & -2.7 &  2.7 & -0.1 & -0.1 &  2.3 &  2.4 & 1.067 \\ 
\hline
\hline
\end{tabular} }%end of resizebox
\caption{The electron channel Born-level double-differential cross section $\frac{\text{d}^2\sigma}{\text{d}m_{ee}\text{d}|y_{ee}|}$. 
The measurements are listed together with the statistical ($\delta^{\rm stat}$), systematic ($\delta^{\rm sys}$) and total ($\delta^{\rm tot}$) uncertainties.
In addition the contributions from the individual correlated ({\rm cor}) and uncorrelated ({\rm unc}) systematic error sources are also provided consisting of the
trigger efficiency ($\delta^{\rm trig}$), electron reconstruction efficiency ($\delta^{\rm reco}$), electron identification efficiency ($\delta^{\rm id}$),
the isolation efficiency ($\delta^{\rm iso}$), 
the electron energy resolution ($\delta^{\rm Eres}$), the electron energy scale ($\delta^{\rm Escale}$), 
the multijet and $W$+jets background ($\delta^{\rm mult.}$), 
the top and diboson background normalisation ($\delta^{\rm top},\delta^{\rm diboson}$), 
the top and diboson background MC statistical uncertainty ($\delta^{\rm bgMC}$), 
and the signal MC statistical uncertainty ($\delta^{\rm MC}$).
The ratio of the dressed-level to Born-level predictions ($k_{\rm dressed}$) is also provided.
The luminosity uncertainty of 1.9\% is not shown and not included in the overall systematic and total uncertainties.}
\label{table:elecXsec_dMdy}
\end{center}
\end{table}

\begin{table}[!htb]
\begin{center}
\resizebox{\textwidth}{!}{
\begin{tabular}{rc|rrrr|rrrrrrrrrrrrrr|r}
\hline\hline
$m_{ee}$  &  $|\Delta \eta_{ee}|$ & $\frac{\text{d}^2\sigma}{\text{d}m_{ee}\text{d}|\Delta\eta_{ee}|}$  & 
$\delta^{\rm stat}$           &  $\delta^{\rm sys}$          & $\delta^{\rm tot}$ &
$\delta^{\rm trig}_{\rm cor}$     &  $\delta^{\rm trig}_{\rm unc}$   &
$\delta^{\rm reco}_{\rm cor}$     &  $\delta^{\rm id}_{\rm cor}$   &
$\delta^{\rm iso}_{\rm cor}$      &  $\delta^{\rm iso}_{\rm unc}$  &
$\delta^{\rm Eres}_{\rm cor}$    &
$\delta^{\rm Escale}_{\rm cor}$  &
$\delta^{\rm mult.}_{\rm cor}$      &  $\delta^{\rm mult.}_{\rm unc}$    &
$\delta^{\rm top}_{\rm cor}$      &  $\delta^{\rm diboson}_{\rm cor}$    &  $\delta^{\rm bgMC}_{\rm unc}$    &
$\delta^{\rm MC}_{\rm unc}$ & $k_{\rm dressed}$ \\
{[GeV]} &  & [pb/GeV] & [\%] & [\%] & [\%] & [\%] & [\%]& [\%] & [\%]& [\%] & [\%]& [\%]& [\%] & [\%]& [\%] & [\%] & [\%] & [\%] & [\%] & \\
\hline
116--150 & 0.00--0.25 & $4.99 \times 10^{-2}$ &  1.0 &  1.0 &  1.4 &  0.0 &  0.1 & -0.1 & -0.4 &  0.0 &  0.1 &  0.2 &  0.6 & -0.5 &  0.1 & -0.3 & -0.1 &  0.1 &  0.2 & 1.043 \\ 
116--150 & 0.25--0.50 & $4.72 \times 10^{-2}$ &  1.0 &  1.1 &  1.5 & -0.1 &  0.0 & -0.1 & -0.4 &  0.0 &  0.1 &  0.2 &  0.7 & -0.5 &  0.1 & -0.3 & -0.1 &  0.1 &  0.2 & 1.044 \\ 
116--150 & 0.50--0.75 & $4.40 \times 10^{-2}$ &  1.1 &  1.0 &  1.5 & -0.1 &  0.0 & -0.1 & -0.4 &  0.0 &  0.1 &  0.1 &  0.7 & -0.5 &  0.1 & -0.3 & -0.1 &  0.1 &  0.2 & 1.044 \\ 
116--150 & 0.75--1.00 & $4.05 \times 10^{-2}$ &  1.1 &  0.9 &  1.4 & -0.1 &  0.0 & -0.1 & -0.3 &  0.0 &  0.0 &  0.1 &  0.5 & -0.4 &  0.1 & -0.3 & -0.1 &  0.1 &  0.3 & 1.045 \\ 
116--150 & 1.00--1.25 & $3.59 \times 10^{-2}$ &  1.2 &  0.9 &  1.5 & -0.1 &  0.0 & -0.1 & -0.3 &  0.0 &  0.0 &  0.2 &  0.5 & -0.5 &  0.2 & -0.4 & -0.1 &  0.1 &  0.3 & 1.048 \\ 
116--150 & 1.25--1.50 & $3.25 \times 10^{-2}$ &  1.3 &  0.9 &  1.5 & -0.1 &  0.0 &  0.0 & -0.3 &  0.0 &  0.0 &  0.1 &  0.4 & -0.5 &  0.2 & -0.4 & -0.1 &  0.1 &  0.3 & 1.050 \\ 
116--150 & 1.50--1.75 & $2.60 \times 10^{-2}$ &  1.4 &  1.0 &  1.8 & -0.1 &  0.0 &  0.0 & -0.2 &  0.0 &  0.0 &  0.2 &  0.4 & -0.6 &  0.2 & -0.4 & -0.2 &  0.1 &  0.3 & 1.050 \\ 
116--150 & 1.75--2.00 & $2.03 \times 10^{-2}$ &  1.7 &  1.1 &  2.0 & -0.1 &  0.0 &  0.0 & -0.3 &  0.0 &  0.0 &  0.1 &  0.5 & -0.7 &  0.3 & -0.4 & -0.2 &  0.2 &  0.4 & 1.055 \\ 
116--150 & 2.00--2.25 & $1.20 \times 10^{-2}$ &  2.2 &  1.4 &  2.6 & -0.1 &  0.0 &  0.0 & -0.3 &  0.0 &  0.0 &  0.2 &  0.5 & -0.9 &  0.5 & -0.5 & -0.3 &  0.3 &  0.4 & 1.055 \\ 
116--150 & 2.25--2.50 & $4.25 \times 10^{-3}$ &  4.0 &  2.2 &  4.5 & -0.1 &  0.0 &  0.0 & -0.3 &  0.0 &  0.0 &  0.2 &  0.6 & -1.4 &  0.9 & -0.8 & -0.3 &  0.5 &  0.6 & 1.047 \\ 
116--150 & 2.50--2.75 & $6.70 \times 10^{-4}$ & 11.4 &  5.3 & 12.5 & -0.1 &  0.0 & -0.1 & -0.4 &  0.0 &  0.0 &  0.4 &  0.5 & -3.6 &  2.4 & -1.4 & -0.8 &  1.8 &  1.7 & 1.044 \\ 
\hline
150--200 & 0.00--0.25 & $1.08 \times 10^{-2}$ &  1.7 &  1.3 &  2.2 & -0.1 &  0.1 & -0.1 & -0.7 &  0.0 &  0.1 &  0.1 &  0.7 & -0.7 &  0.2 & -0.6 & -0.1 &  0.2 &  0.2 & 1.042 \\ 
150--200 & 0.25--0.50 & $1.04 \times 10^{-2}$ &  1.8 &  1.3 &  2.2 & -0.1 &  0.1 & -0.1 & -0.6 &  0.0 &  0.1 &  0.1 &  0.7 & -0.6 &  0.2 & -0.6 & -0.2 &  0.2 &  0.2 & 1.042 \\ 
150--200 & 0.50--0.75 & $9.63 \times 10^{-3}$ &  1.9 &  1.3 &  2.3 &  0.0 &  0.1 & -0.1 & -0.6 &  0.0 &  0.1 &  0.1 &  0.6 & -0.7 &  0.2 & -0.7 & -0.2 &  0.2 &  0.2 & 1.043 \\ 
150--200 & 0.75--1.00 & $9.38 \times 10^{-3}$ &  2.0 &  1.3 &  2.4 &  0.0 &  0.1 & -0.1 & -0.6 &  0.0 &  0.1 &  0.1 &  0.6 & -0.6 &  0.2 & -0.7 & -0.2 &  0.2 &  0.3 & 1.044 \\ 
150--200 & 1.00--1.25 & $8.24 \times 10^{-3}$ &  2.0 &  1.4 &  2.5 &  0.0 &  0.1 & -0.1 & -0.5 &  0.0 &  0.1 &  0.1 &  0.5 & -0.7 &  0.2 & -0.8 & -0.2 &  0.3 &  0.3 & 1.046 \\ 
150--200 & 1.25--1.50 & $7.14 \times 10^{-3}$ &  2.2 &  1.4 &  2.7 &  0.0 &  0.1 & -0.1 & -0.5 &  0.0 &  0.1 &  0.1 &  0.4 & -0.8 &  0.3 & -0.9 & -0.3 &  0.3 &  0.3 & 1.049 \\ 
150--200 & 1.50--1.75 & $6.21 \times 10^{-3}$ &  2.5 &  1.5 &  2.9 & -0.1 &  0.0 & -0.1 & -0.4 &  0.0 &  0.1 &  0.1 &  0.3 & -0.9 &  0.3 & -1.0 & -0.3 &  0.3 &  0.3 & 1.054 \\ 
150--200 & 1.75--2.00 & $4.95 \times 10^{-3}$ &  2.9 &  1.9 &  3.4 & -0.1 &  0.0 & -0.1 & -0.3 &  0.0 &  0.0 &  0.1 &  0.3 & -1.2 &  0.5 & -1.1 & -0.5 &  0.5 &  0.4 & 1.058 \\ 
150--200 & 2.00--2.25 & $3.74 \times 10^{-3}$ &  3.5 &  2.1 &  4.1 & -0.1 &  0.0 &  0.0 & -0.3 &  0.0 &  0.0 &  0.2 &  0.3 & -1.3 &  0.7 & -1.3 & -0.6 &  0.6 &  0.4 & 1.064 \\ 
150--200 & 2.25--2.50 & $2.94 \times 10^{-3}$ &  4.0 &  2.5 &  4.8 & -0.1 &  0.0 &  0.0 & -0.3 &  0.0 &  0.0 &  0.1 &  0.3 & -1.7 &  1.0 & -1.2 & -0.6 &  0.6 &  0.5 & 1.071 \\ 
150--200 & 2.50--2.75 & $2.01 \times 10^{-3}$ &  5.1 &  2.9 &  5.9 & -0.1 &  0.0 &  0.0 & -0.3 &  0.0 &  0.0 &  0.1 &  0.4 & -1.8 &  1.6 & -1.2 & -0.7 &  0.8 &  0.7 & 1.073 \\ 
150--200 & 2.75--3.00 & $9.24 \times 10^{-4}$ &  8.0 &  4.9 &  9.4 & -0.1 &  0.0 &  0.0 & -0.4 &  0.0 &  0.0 &  0.6 &  0.7 & -3.4 &  2.7 & -1.4 & -0.8 &  1.2 &  1.0 & 1.070 \\ 
\hline
200--300 & 0.00--0.25 & $2.11 \times 10^{-3}$ &  2.8 &  1.8 &  3.3 & -0.1 &  0.2 & -0.2 & -0.8 & -0.1 &  0.2 &  0.1 &  1.1 & -0.7 &  0.2 & -0.7 & -0.2 &  0.3 &  0.3 & 1.043 \\ 
200--300 & 0.25--0.50 & $2.08 \times 10^{-3}$ &  2.9 &  1.9 &  3.4 & -0.1 &  0.2 & -0.2 & -0.8 & -0.1 &  0.2 &  0.1 &  1.2 & -0.7 &  0.2 & -0.8 & -0.2 &  0.4 &  0.3 & 1.044 \\ 
200--300 & 0.50--0.75 & $1.98 \times 10^{-3}$ &  3.0 &  1.9 &  3.6 & -0.1 &  0.1 & -0.2 & -0.8 & -0.1 &  0.2 &  0.1 &  1.1 & -0.8 &  0.2 & -0.9 & -0.2 &  0.4 &  0.3 & 1.044 \\ 
200--300 & 0.75--1.00 & $1.89 \times 10^{-3}$ &  3.1 &  1.9 &  3.6 & -0.1 &  0.1 & -0.2 & -0.8 & -0.1 &  0.2 &  0.1 &  1.0 & -0.9 &  0.2 & -0.9 & -0.3 &  0.4 &  0.3 & 1.047 \\ 
200--300 & 1.00--1.25 & $1.74 \times 10^{-3}$ &  3.1 &  1.8 &  3.6 & -0.1 &  0.1 & -0.2 & -0.8 & -0.1 &  0.2 &  0.1 &  0.8 & -0.8 &  0.2 & -1.0 & -0.3 &  0.4 &  0.3 & 1.048 \\ 
200--300 & 1.25--1.50 & $1.40 \times 10^{-3}$ &  3.6 &  2.2 &  4.2 & -0.1 &  0.1 & -0.2 & -0.7 & -0.1 &  0.2 &  0.1 &  0.7 & -1.2 &  0.3 & -1.3 & -0.4 &  0.6 &  0.3 & 1.049 \\ 
200--300 & 1.50--1.75 & $1.25 \times 10^{-3}$ &  3.9 &  2.2 &  4.5 & -0.1 &  0.1 & -0.1 & -0.7 &  0.0 &  0.1 &  0.1 &  0.6 & -1.1 &  0.4 & -1.4 & -0.4 &  0.6 &  0.4 & 1.057 \\ 
200--300 & 1.75--2.00 & $1.02 \times 10^{-3}$ &  4.6 &  2.5 &  5.2 &  0.0 &  0.1 & -0.1 & -0.6 &  0.0 &  0.1 &  0.1 &  0.5 & -1.2 &  0.5 & -1.8 & -0.6 &  0.8 &  0.4 & 1.060 \\ 
200--300 & 2.00--2.25 & $9.44 \times 10^{-4}$ &  4.9 &  2.8 &  5.6 &  0.0 &  0.1 & -0.1 & -0.5 &  0.0 &  0.1 &  0.1 &  0.4 & -1.5 &  0.7 & -1.8 & -0.7 &  0.9 &  0.5 & 1.068 \\ 
200--300 & 2.25--2.50 & $6.59 \times 10^{-4}$ &  6.3 &  3.8 &  7.4 &  0.0 &  0.1 & -0.1 & -0.5 &  0.0 &  0.1 &  0.1 &  0.4 & -2.2 &  1.2 & -2.3 & -1.1 &  1.3 &  0.6 & 1.078 \\ 
200--300 & 2.50--2.75 & $5.75 \times 10^{-4}$ &  7.0 &  3.6 &  7.8 & -0.1 &  0.0 & -0.1 & -0.4 &  0.0 &  0.1 &  0.2 &  0.4 & -1.9 &  1.4 & -2.1 & -1.2 &  1.4 &  0.7 & 1.087 \\ 
200--300 & 2.75--3.00 & $4.31 \times 10^{-4}$ &  8.5 &  5.2 & 10.0 & -0.1 &  0.0 & -0.1 & -0.3 &  0.0 &  0.0 &  0.4 &  0.5 & -3.5 &  2.4 & -2.2 & -1.5 &  1.7 &  1.0 & 1.110 \\ 
\hline
300--500 & 0.00--0.50 & $2.60 \times 10^{-4}$ &  3.9 &  2.4 &  4.6 & -0.1 &  0.2 & -0.2 & -0.8 & -0.2 &  0.5 &  0.1 &  1.8 & -0.7 &  0.2 & -0.6 & -0.2 &  0.5 &  0.3 & 1.048 \\ 
300--500 & 0.50--1.00 & $2.28 \times 10^{-4}$ &  4.3 &  2.3 &  4.9 & -0.1 &  0.2 & -0.2 & -0.8 & -0.2 &  0.4 &  0.1 &  1.6 & -0.9 &  0.2 & -0.8 & -0.3 &  0.6 &  0.3 & 1.048 \\ 
300--500 & 1.00--1.50 & $2.18 \times 10^{-4}$ &  4.4 &  2.3 &  5.0 & -0.1 &  0.2 & -0.2 & -0.8 & -0.1 &  0.4 &  0.1 &  1.4 & -0.9 &  0.3 & -1.0 & -0.4 &  0.7 &  0.3 & 1.057 \\ 
300--500 & 1.50--2.00 & $1.64 \times 10^{-4}$ &  5.4 &  2.7 &  6.1 & -0.1 &  0.2 & -0.2 & -0.8 & -0.1 &  0.3 &  0.1 &  1.2 & -1.4 &  0.4 & -1.5 & -0.5 &  0.9 &  0.4 & 1.064 \\ 
300--500 & 2.00--2.50 & $1.04 \times 10^{-4}$ &  7.4 &  3.5 &  8.2 & -0.1 &  0.1 & -0.2 & -0.8 & -0.1 &  0.2 &  0.2 &  0.8 & -1.7 &  0.8 & -2.3 & -0.8 &  1.4 &  0.6 & 1.082 \\ 
300--500 & 2.50--3.00 & $5.21 \times 10^{-5}$ & 12.7 &  6.5 & 14.3 & -0.1 &  0.1 & -0.2 & -0.7 &  0.0 &  0.1 &  0.2 &  1.0 & -3.2 &  2.3 & -4.1 & -2.2 &  2.9 &  0.8 & 1.107 \\ 
\hline
500--1500 & 0.00--0.50 & $7.69 \times 10^{-6}$ &  9.8 &  3.1 & 10.3 & -0.2 &  0.3 & -0.2 & -0.8 & -0.3 &  1.3 &  0.1 &  2.4 & -0.6 &  0.6 & -0.2 & -0.3 &  0.7 &  0.3 & 1.054 \\ 
500--1500 & 0.50--1.00 & $8.74 \times 10^{-6}$ &  9.3 &  2.9 &  9.7 & -0.2 &  0.3 & -0.2 & -0.8 & -0.3 &  1.2 &  0.1 &  2.3 & -0.3 &  0.4 & -0.3 & -0.3 &  0.8 &  0.3 & 1.058 \\ 
500--1500 & 1.00--1.50 & $8.68 \times 10^{-6}$ &  9.3 &  2.7 &  9.7 & -0.1 &  0.3 & -0.2 & -0.8 & -0.2 &  1.0 &  0.0 &  2.2 & -0.1 &  0.4 & -0.4 & -0.3 &  0.6 &  0.4 & 1.063 \\ 
500--1500 & 1.50--2.00 & $6.99 \times 10^{-6}$ & 10.8 &  2.7 & 11.1 & -0.1 &  0.2 & -0.2 & -0.8 & -0.2 &  0.7 &  0.1 &  1.9 & -1.1 &  0.6 & -0.5 & -0.5 &  0.7 &  0.4 & 1.078 \\ 
500--1500 & 2.00--2.50 & $2.92 \times 10^{-6}$ & 19.2 &  4.1 & 19.6 & -0.1 &  0.2 & -0.2 & -0.8 & -0.2 &  0.5 &  0.2 &  1.7 & -0.5 &  1.7 & -1.6 & -1.2 &  2.6 &  0.5 & 1.095 \\ 
500--1500 & 2.50--3.00 & $1.90 \times 10^{-6}$ & 26.3 &  6.0 & 27.0 & -0.1 &  0.2 & -0.3 & -0.8 & -0.1 &  0.4 &  0.1 &  1.6 & -0.9 &  3.0 & -2.6 & -1.9 &  3.9 &  1.2 & 1.120 \\ 
\hline
\hline
\end{tabular} }%end of resizebox
\caption{The electron channel Born-level double-differential cross section $\frac{\text{d}^2\sigma}{\text{d}m_{ee}\text{d}|\Delta \eta_{ee}|}$. 
The measurements are listed together with the statistical ($\delta^{\rm stat}$), systematic ($\delta^{\rm sys}$) and total ($\delta^{\rm tot}$) uncertainties.
In addition the contributions from the individual correlated ({\rm cor}) and uncorrelated ({\rm unc}) systematic error sources are also provided consisting of the
trigger efficiency ($\delta^{\rm trig}$), electron reconstruction efficiency ($\delta^{\rm reco}$), electron identification efficiency ($\delta^{\rm id}$),
the isolation efficiency ($\delta^{\rm iso}$), 
the electron energy resolution ($\delta^{\rm Eres}$), the electron energy scale ($\delta^{\rm Escale}$), 
the multijet and $W$+jets background ($\delta^{\rm mult.}$), 
the top and diboson background normalisation ($\delta^{\rm top},\delta^{\rm diboson}$), 
the top and diboson background MC statistical uncertainty ($\delta^{\rm bgMC}$), 
and the signal MC statistical uncertainty ($\delta^{\rm MC}$).
The ratio of the dressed-level to Born-level predictions ($k_{\rm dressed}$) is also provided.
The luminosity uncertainty of 1.9\% is not shown and not included in the overall systematic and total uncertainties.}
\label{table:elecXsec_dMdeta}
\end{center}
\end{table}

\begin{table}[!htb]
\begin{center}
\resizebox{\textwidth}{!}{
\begin{tabular}{r|rrrr|rrrrrrrrrrrr|r}
\hline\hline
$m_{\mu\mu}$  &  $\frac{\text{d}\sigma}{\text{d}m_{\mu\mu}}$  & 
$\delta^{\rm stat}$           &  $\delta^{\rm sys}$          & $\delta^{\rm tot}$ &
$\delta^{\rm trig}_{\rm cor}$     & 
$\delta^{\rm reco}_{\rm cor}$     & 
$\delta^{\rm MSres}_{\rm cor}$    &  $\delta^{\rm IDres}_{\rm cor}$  &
$\delta^{\rm pT}_{\rm cor}$       &  $\delta^{\rm iso}_{\rm cor}$   &
$\delta^{\rm top}_{\rm cor}$      &  $\delta^{\rm diboson}_{\rm cor}$&
$\delta^{\rm bgMC}_{\rm unc}$      &  $\delta^{\rm mult.}_{\rm cor}$  &
$\delta^{\rm mult.}_{\rm unc}$ &  $\delta^{\rm MC}_{\rm unc}$ & $k_{\rm dressed}$ \\
{[GeV]} & [pb/GeV] & [\%] & [\%] & [\%] & [\%] & [\%]& [\%] & [\%]& [\%] & [\%]& [\%] & [\%]& [\%] & [\%] & [\%] & [\%] &  \\
\hline
$  116-  130$ & $2.25 \times 10^{-1}$ &  0.5 &  0.6 &  0.8 & -0.1 & -0.4 & -0.1 & -0.1 & -0.4 & -0.1 & -0.3 & -0.1 &  0.0 &  0.0 &  0.1 &  0.1 & 1.055 \\ 
$  130-  150$ & $1.04 \times 10^{-1}$ &  0.6 &  0.7 &  0.9 & -0.1 & -0.4 & -0.1 &  0.0 & -0.3 & -0.1 & -0.5 & -0.1 &  0.1 &  0.0 &  0.1 &  0.1 & 1.047 \\ 
$  150-  175$ & $4.94 \times 10^{-2}$ &  0.8 &  0.9 &  1.2 & -0.1 & -0.4 &  0.0 &  0.0 & -0.2 & -0.1 & -0.7 & -0.2 &  0.1 & -0.1 &  0.1 &  0.1 & 1.043 \\ 
$  175-  200$ & $2.51 \times 10^{-2}$ &  1.1 &  1.2 &  1.6 & -0.1 & -0.5 &  0.0 &  0.0 & -0.2 & -0.1 & -1.0 & -0.3 &  0.1 & -0.1 &  0.2 &  0.1 & 1.040 \\ 
$  200-  230$ & $1.37 \times 10^{-2}$ &  1.4 &  1.5 &  2.0 & -0.1 & -0.5 &  0.0 & -0.1 & -0.2 & -0.1 & -1.2 & -0.4 &  0.2 & -0.1 &  0.3 &  0.2 & 1.037 \\ 
$  230-  260$ & $7.87 \times 10^{-3}$ &  1.8 &  1.6 &  2.5 & -0.1 & -0.5 &  0.0 &  0.1 & -0.3 & -0.1 & -1.3 & -0.4 &  0.3 & -0.1 &  0.5 &  0.2 & 1.036 \\ 
$  260-  300$ & $4.45 \times 10^{-3}$ &  2.1 &  1.7 &  2.7 & -0.1 & -0.6 &  0.0 & -0.1 & -0.2 & -0.2 & -1.4 & -0.5 &  0.3 & -0.1 &  0.5 &  0.2 & 1.037 \\ 
$  300-  380$ & $1.90 \times 10^{-3}$ &  2.3 &  1.9 &  3.0 & -0.1 & -0.6 &  0.1 &  0.0 & -0.3 & -0.2 & -1.4 & -0.6 &  0.4 & -0.2 &  0.7 &  0.2 & 1.035 \\ 
$  380-  500$ & $6.40 \times 10^{-4}$ &  3.2 &  1.8 &  3.7 & -0.1 & -0.7 & -0.1 & -0.1 & -0.2 & -0.3 & -1.2 & -0.5 &  0.5 & -0.1 &  0.8 &  0.2 & 1.037 \\ 
$  500-  700$ & $1.54 \times 10^{-4}$ &  5.0 &  2.0 &  5.4 & -0.1 & -0.8 & -0.1 &  0.0 & -0.2 & -0.4 & -0.9 & -0.5 &  0.6 & -1.3 &  0.0 &  0.2 & 1.036 \\ 
$  700- 1000$ & $2.66 \times 10^{-5}$ &  9.6 &  2.1 &  9.8 & -0.1 & -0.8 & -0.5 & -0.1 & -0.4 & -0.5 & -0.5 & -0.5 &  0.8 & -1.3 &  0.0 &  0.4 & 1.040 \\ 
$ 1000- 1500$ & $2.17 \times 10^{-6}$ & 26.0 &  2.7 & 26.2 & -0.1 & -1.1 & -0.1 & -1.0 & -0.3 & -0.6 & -0.4 & -0.6 &  1.5 & -1.4 &  0.0 &  0.4 & 1.043 \\ 
\hline
\hline
\end{tabular} }%end of resizebox
\caption{The muon channel Born-level single-differential cross section $\frac{\text{d}\sigma}{\text{d}m_{\mu\mu}}$. 
The measurements are listed together with the statistical ($\delta^{\rm stat}$), systematic ($\delta^{\rm sys}$) and total ($\delta^{\rm tot}$) uncertainties.
In addition the contributions from the individual correlated ({\rm cor}) and uncorrelated ({\rm unc}) systematic error sources are also provided consisting of the
trigger efficiency ($\delta^{\rm trig}$), muon reconstruction efficiency ($\delta^{\rm reco}$), the MS resolution ($\delta^{\rm MSres}$),
the ID resolution ($\delta^{\rm IDres}$), the muon transverse momentum scale ($\delta^{\rm pT}$), 
the isolation efficiency ($\delta^{\rm iso}$), 
the top and diboson background normalisation ($\delta^{\rm top},\delta^{\rm diboson}$), 
the top and diboson background MC statistical uncertainty ($\delta^{\rm bgMC}$),
the multijet background ($\delta^{\rm mult}$) and the signal MC statistical uncertainty ($\delta^{\rm MC}$).
The ratio of the dressed-level to Born-level predictions ($k_{\rm dressed}$) is also provided.
The luminosity uncertainty of 1.9\% is not shown and not included in the overall systematic and total uncertainties.}
\label{table:muonXsec_Mass}
\end{center}
\end{table}

\begin{table}[!htb]
\begin{center}
\resizebox{\textwidth}{!}{
\begin{tabular}{rc|rrrr|rrrrrrrrrrrr|r}
\hline\hline
$m_{\mu\mu}$  &  $|y_{\mu\mu}|$ &
$\frac{\text{d}^2\sigma}{\text{d}m_{\mu\mu}\text{d}|y_{\mu\mu}|}$  & 
$\delta^{\rm stat}$           &  $\delta^{\rm sys}$          & $\delta^{\rm tot}$ &
$\delta^{\rm trig}_{\rm cor}$     &  
$\delta^{\rm reco}_{\rm cor}$     &  
$\delta^{\rm MSres}_{\rm cor}$    &  $\delta^{\rm IDres}_{\rm cor}$  &
$\delta^{\rm pT}_{\rm cor}$       &  $\delta^{\rm iso}_{\rm unc}$   &
$\delta^{\rm top}_{\rm cor}$      &  $\delta^{\rm diboson}_{\rm cor}$&
$\delta^{\rm bgMC}_{\rm unc}$      &  $\delta^{\rm mult.}_{\rm cor}$  &
$\delta^{\rm mult.}_{\rm unc}$ &  $\delta^{\rm MC}_{\rm unc}$ & $k_{\rm dressed}$ \\
{[GeV]} &  & [pb/GeV] & [\%] & [\%] & [\%] & [\%] & [\%]& [\%] & [\%]& [\%] & [\%]& [\%] & [\%]& [\%] & [\%] & [\%] & [\%] &   \\
\hline
\hline
$  116-  150$ & $ 0.0- 0.2$ & $4.02 \times 10^{-2}$ &  1.1 &  0.8 &  1.4 & -0.1 & -0.4 & -0.1 & -0.1 & -0.2 & -0.1 & -0.5 & -0.1 &  0.1 & -0.1 &  0.1 &  0.2 & 1.054 \\ 
$  116-  150$ & $ 0.2- 0.4$ & $3.94 \times 10^{-2}$ &  1.1 &  0.8 &  1.4 & -0.1 & -0.4 & -0.1 &  0.0 & -0.3 & -0.1 & -0.5 & -0.1 &  0.1 & -0.1 &  0.1 &  0.2 & 1.052 \\ 
$  116-  150$ & $ 0.4- 0.6$ & $3.90 \times 10^{-2}$ &  1.1 &  0.7 &  1.4 & -0.1 & -0.4 &  0.0 &  0.0 & -0.3 & -0.1 & -0.5 & -0.1 &  0.1 & -0.1 &  0.1 &  0.2 & 1.054 \\ 
$  116-  150$ & $ 0.6- 0.8$ & $3.97 \times 10^{-2}$ &  1.1 &  0.7 &  1.3 & -0.1 & -0.4 & -0.2 &  0.0 & -0.2 & -0.1 & -0.5 & -0.1 &  0.1 &  0.0 &  0.1 &  0.2 & 1.054 \\ 
$  116-  150$ & $ 0.8- 1.0$ & $3.94 \times 10^{-2}$ &  1.1 &  0.7 &  1.3 & -0.1 & -0.4 & -0.1 & -0.1 & -0.3 & -0.1 & -0.4 & -0.1 &  0.1 &  0.0 &  0.1 &  0.2 & 1.054 \\ 
$  116-  150$ & $ 1.0- 1.2$ & $3.89 \times 10^{-2}$ &  1.1 &  0.7 &  1.3 & -0.1 & -0.3 & -0.1 &  0.0 & -0.3 & -0.1 & -0.3 & -0.1 &  0.1 &  0.0 &  0.1 &  0.2 & 1.054 \\ 
$  116-  150$ & $ 1.2- 1.4$ & $3.79 \times 10^{-2}$ &  1.1 &  0.7 &  1.3 & -0.1 & -0.4 & -0.1 &  0.0 & -0.4 & -0.1 & -0.2 & -0.1 &  0.1 &  0.0 &  0.0 &  0.2 & 1.053 \\ 
$  116-  150$ & $ 1.4- 1.6$ & $3.47 \times 10^{-2}$ &  1.1 &  0.7 &  1.3 & -0.1 & -0.4 &  0.1 & -0.1 & -0.4 & -0.1 & -0.2 & -0.1 &  0.1 &  0.0 &  0.0 &  0.3 & 1.050 \\ 
$  116-  150$ & $ 1.6- 1.8$ & $2.93 \times 10^{-2}$ &  1.3 &  0.7 &  1.4 & -0.1 & -0.4 &  0.0 & -0.1 & -0.4 & -0.1 & -0.1 & -0.1 &  0.1 &  0.0 &  0.0 &  0.3 & 1.050 \\ 
$  116-  150$ & $ 1.8- 2.0$ & $2.27 \times 10^{-2}$ &  1.5 &  0.8 &  1.7 & -0.1 & -0.5 &  0.1 & -0.1 & -0.5 & -0.1 & -0.1 & -0.1 &  0.1 &  0.0 &  0.0 &  0.4 & 1.049 \\ 
$  116-  150$ & $ 2.0- 2.2$ & $1.56 \times 10^{-2}$ &  1.9 &  1.0 &  2.1 & -0.1 & -0.6 & -0.4 &  0.1 & -0.5 & -0.1 & -0.1 & -0.1 &  0.1 &  0.0 &  0.0 &  0.5 & 1.047 \\ 
$  116-  150$ & $ 2.2- 2.4$ & $7.42 \times 10^{-3}$ &  3.3 &  1.7 &  3.7 & -0.1 & -0.7 & -0.7 &  0.2 & -0.9 & -0.1 &  0.0 & -0.1 &  0.2 &  0.0 &  0.0 &  1.0 & 1.045 \\ 
\hline
$  150-  200$ & $ 0.0- 0.2$ & $1.01 \times 10^{-2}$ &  2.0 &  1.5 &  2.5 & -0.1 & -0.4 &  0.1 &  0.0 & -0.2 & -0.1 & -1.3 & -0.3 &  0.3 & -0.1 &  0.2 &  0.2 & 1.045 \\ 
$  150-  200$ & $ 0.2- 0.4$ & $1.01 \times 10^{-2}$ &  2.0 &  1.5 &  2.5 & -0.1 & -0.4 &  0.0 & -0.1 & -0.2 & -0.1 & -1.2 & -0.3 &  0.3 & -0.1 &  0.2 &  0.2 & 1.043 \\ 
$  150-  200$ & $ 0.4- 0.6$ & $1.00 \times 10^{-2}$ &  2.0 &  1.4 &  2.4 & -0.1 & -0.4 & -0.1 & -0.1 & -0.2 & -0.1 & -1.2 & -0.3 &  0.3 & -0.1 &  0.2 &  0.2 & 1.043 \\ 
$  150-  200$ & $ 0.6- 0.8$ & $1.03 \times 10^{-2}$ &  1.9 &  1.2 &  2.3 & -0.1 & -0.4 & -0.1 &  0.1 & -0.2 & -0.1 & -1.0 & -0.3 &  0.3 & -0.1 &  0.1 &  0.2 & 1.044 \\ 
$  150-  200$ & $ 0.8- 1.0$ & $9.95 \times 10^{-3}$ &  1.9 &  1.1 &  2.2 & -0.1 & -0.4 &  0.1 &  0.0 & -0.1 & -0.1 & -0.9 & -0.3 &  0.3 & -0.1 &  0.1 &  0.2 & 1.044 \\ 
$  150-  200$ & $ 1.0- 1.2$ & $9.52 \times 10^{-3}$ &  1.9 &  1.0 &  2.2 & -0.1 & -0.4 &  0.0 &  0.0 & -0.2 & -0.1 & -0.7 & -0.2 &  0.2 & -0.1 &  0.1 &  0.2 & 1.041 \\ 
$  150-  200$ & $ 1.2- 1.4$ & $8.91 \times 10^{-3}$ &  1.9 &  0.9 &  2.1 & -0.1 & -0.4 & -0.1 &  0.0 & -0.2 & -0.1 & -0.6 & -0.3 &  0.2 & -0.1 &  0.1 &  0.2 & 1.043 \\ 
$  150-  200$ & $ 1.4- 1.6$ & $7.82 \times 10^{-3}$ &  2.1 &  0.8 &  2.2 & -0.1 & -0.5 &  0.1 & -0.1 & -0.3 & -0.1 & -0.4 & -0.2 &  0.2 & -0.1 &  0.0 &  0.3 & 1.040 \\ 
$  150-  200$ & $ 1.6- 1.8$ & $6.54 \times 10^{-3}$ &  2.2 &  0.8 &  2.4 & -0.1 & -0.5 &  0.1 &  0.1 & -0.3 & -0.1 & -0.3 & -0.1 &  0.2 &  0.0 &  0.0 &  0.3 & 1.037 \\ 
$  150-  200$ & $ 1.8- 2.0$ & $4.71 \times 10^{-3}$ &  2.7 &  0.8 &  2.8 & -0.1 & -0.6 & -0.1 &  0.0 & -0.3 & -0.1 & -0.2 & -0.1 &  0.2 &  0.0 &  0.0 &  0.4 & 1.038 \\ 
$  150-  200$ & $ 2.0- 2.2$ & $3.36 \times 10^{-3}$ &  3.3 &  1.0 &  3.5 & -0.1 & -0.6 & -0.1 & -0.1 & -0.3 & -0.1 & -0.1 & -0.1 &  0.2 &  0.0 &  0.0 &  0.5 & 1.041 \\ 
$  150-  200$ & $ 2.2- 2.4$ & $1.74 \times 10^{-3}$ &  5.6 &  1.5 &  5.8 & -0.1 & -0.7 & -0.2 &  0.3 & -0.6 & -0.1 & -0.1 & -0.1 &  0.3 &  0.0 &  0.0 &  1.1 & 1.038 \\ 
\hline
$  200-  300$ & $ 0.0- 0.2$ & $2.43 \times 10^{-3}$ &  3.0 &  2.4 &  3.9 & -0.1 & -0.5 & -0.2 & -0.1 & -0.2 & -0.1 & -2.1 & -0.5 &  0.5 & -0.2 &  0.5 &  0.2 & 1.038 \\ 
$  200-  300$ & $ 0.2- 0.4$ & $2.37 \times 10^{-3}$ &  3.1 &  2.3 &  3.8 & -0.1 & -0.5 & -0.1 & -0.1 & -0.2 & -0.1 & -2.0 & -0.5 &  0.6 & -0.2 &  0.4 &  0.2 & 1.040 \\ 
$  200-  300$ & $ 0.4- 0.6$ & $2.32 \times 10^{-3}$ &  3.0 &  2.2 &  3.7 & -0.1 & -0.5 &  0.1 &  0.0 & -0.2 & -0.1 & -1.9 & -0.6 &  0.6 & -0.2 &  0.4 &  0.3 & 1.037 \\ 
$  200-  300$ & $ 0.6- 0.8$ & $2.38 \times 10^{-3}$ &  2.9 &  1.9 &  3.5 & -0.1 & -0.5 &  0.1 &  0.0 & -0.2 & -0.1 & -1.6 & -0.5 &  0.5 & -0.1 &  0.3 &  0.3 & 1.037 \\ 
$  200-  300$ & $ 0.8- 1.0$ & $2.24 \times 10^{-3}$ &  3.0 &  1.6 &  3.4 & -0.1 & -0.5 &  0.0 & -0.1 & -0.3 & -0.1 & -1.3 & -0.4 &  0.5 & -0.1 &  0.2 &  0.3 & 1.035 \\ 
$  200-  300$ & $ 1.0- 1.2$ & $2.13 \times 10^{-3}$ &  3.0 &  1.3 &  3.2 & -0.1 & -0.5 &  0.0 & -0.1 & -0.2 & -0.1 & -1.0 & -0.4 &  0.4 & -0.1 &  0.1 &  0.3 & 1.037 \\ 
$  200-  300$ & $ 1.2- 1.4$ & $1.88 \times 10^{-3}$ &  3.0 &  1.2 &  3.3 & -0.1 & -0.5 &  0.2 & -0.1 & -0.2 & -0.1 & -0.8 & -0.3 &  0.4 & -0.1 &  0.1 &  0.3 & 1.033 \\ 
$  200-  300$ & $ 1.4- 1.6$ & $1.66 \times 10^{-3}$ &  3.2 &  1.0 &  3.3 & -0.1 & -0.5 & -0.1 &  0.0 & -0.2 & -0.1 & -0.5 & -0.2 &  0.4 & -0.1 &  0.0 &  0.3 & 1.034 \\ 
$  200-  300$ & $ 1.6- 1.8$ & $1.33 \times 10^{-3}$ &  3.5 &  0.9 &  3.7 & -0.1 & -0.6 &  0.0 & -0.1 & -0.2 & -0.1 & -0.3 & -0.2 &  0.4 &  0.0 &  0.0 &  0.4 & 1.035 \\ 
$  200-  300$ & $ 1.8- 2.0$ & $9.97 \times 10^{-4}$ &  4.1 &  1.0 &  4.3 & -0.1 & -0.7 &  0.1 & -0.1 & -0.3 & -0.1 & -0.2 & -0.2 &  0.4 &  0.0 &  0.0 &  0.5 & 1.036 \\ 
$  200-  300$ & $ 2.0- 2.2$ & $5.87 \times 10^{-4}$ &  5.7 &  1.2 &  5.8 & -0.1 & -0.7 & -0.1 & -0.2 & -0.5 & -0.1 & -0.1 & -0.1 &  0.4 &  0.0 &  0.0 &  0.7 & 1.037 \\ 
$  200-  300$ & $ 2.2- 2.4$ & $2.29 \times 10^{-4}$ & 10.9 &  2.1 & 11.1 & -0.1 & -0.9 & -0.3 & -0.3 & -0.5 & -0.1 & -0.1 & -0.1 &  0.8 &  0.0 &  0.0 &  1.6 & 1.040 \\ 
\hline
$  300-  500$ & $ 0.0- 0.4$ & $3.72 \times 10^{-4}$ &  4.0 &  2.9 &  4.9 & -0.1 & -0.6 &  0.2 &  0.1 & -0.2 & -0.2 & -2.2 & -0.8 &  0.7 & -0.5 &  1.2 &  0.2 & 1.036 \\ 
$  300-  500$ & $ 0.4- 0.8$ & $3.28 \times 10^{-4}$ &  4.1 &  2.5 &  4.8 & -0.1 & -0.6 & -0.2 & -0.1 & -0.3 & -0.2 & -1.9 & -0.7 &  0.8 & -0.4 &  0.7 &  0.2 & 1.036 \\ 
$  300-  500$ & $ 0.8- 1.2$ & $3.09 \times 10^{-4}$ &  4.0 &  1.6 &  4.2 & -0.1 & -0.6 &  0.1 & -0.1 & -0.3 & -0.2 & -1.1 & -0.5 &  0.6 & -0.1 &  0.2 &  0.2 & 1.034 \\ 
$  300-  500$ & $ 1.2- 1.6$ & $2.51 \times 10^{-4}$ &  4.1 &  1.1 &  4.2 & -0.1 & -0.6 &  0.0 &  0.1 & -0.3 & -0.2 & -0.5 & -0.3 &  0.5 & -0.1 &  0.0 &  0.3 & 1.035 \\ 
$  300-  500$ & $ 1.6- 2.0$ & $1.29 \times 10^{-4}$ &  5.7 &  1.2 &  5.8 & -0.1 & -0.8 & -0.2 & -0.1 & -0.3 & -0.2 & -0.2 & -0.3 &  0.6 &  0.0 &  0.0 &  0.4 & 1.040 \\ 
$  300-  500$ & $ 2.0- 2.4$ & $3.93 \times 10^{-5}$ & 11.2 &  1.9 & 11.4 & -0.1 & -1.0 & -0.3 & -0.1 & -0.5 & -0.2 & -0.1 & -0.1 &  0.7 &  0.0 &  0.0 &  1.3 & 1.037 \\ 
\hline
$  500- 1500$ & $ 0.0- 0.4$ & $1.46 \times 10^{-5}$ &  8.6 &  2.3 &  8.9 & -0.1 & -0.7 & -0.1 &  0.1 & -0.2 & -0.4 & -1.4 & -0.8 &  1.1 & -0.7 &  0.0 &  0.3 & 1.036 \\ 
$  500- 1500$ & $ 0.4- 0.8$ & $1.29 \times 10^{-5}$ &  8.7 &  1.9 &  8.9 & -0.1 & -0.7 & -0.2 & -0.2 & -0.2 & -0.4 & -1.0 & -0.6 &  1.0 & -0.4 &  0.0 &  0.3 & 1.036 \\ 
$  500- 1500$ & $ 0.8- 1.2$ & $1.20 \times 10^{-5}$ &  8.5 &  1.4 &  8.6 & -0.1 & -0.8 & -0.3 & -0.2 & -0.3 & -0.4 & -0.5 & -0.3 &  0.7 & -0.1 &  0.0 &  0.3 & 1.038 \\ 
$  500- 1500$ & $ 1.2- 1.6$ & $8.06 \times 10^{-6}$ & 10.0 &  1.3 & 10.0 & -0.1 & -0.8 &  0.1 & -0.1 & -0.3 & -0.4 & -0.1 & -0.2 &  0.6 &  0.0 &  0.0 &  0.4 & 1.038 \\ 
$  500- 1500$ & $ 1.6- 2.0$ & $2.09 \times 10^{-6}$ & 19.5 &  2.1 & 19.6 & -0.1 & -1.1 & -0.9 & -0.5 & -0.2 & -0.4 & -0.1 & -0.1 &  0.5 &  0.0 &  0.0 &  1.3 & 1.042 \\ 
$  500- 1500$ & $ 2.0- 2.4$ & $2.96 \times 10^{-7}$ & 45.3 & 12.1 & 46.9 & -0.1 & -2.0 & -1.9 & -2.7 & -0.5 & -0.4 &  0.0 & -0.1 &  0.6 &  0.0 &  0.0 & 11.4 & 1.053 \\ 
\hline
\hline
\end{tabular} }%end of resizebox
\caption{The muon channel Born-level double-differential cross section $\frac{\text{d}^2\sigma}{\text{d}m_{\mu\mu}\text{d}|y_{\mu\mu}|}$. 
The measurements are listed together with the statistical ($\delta^{\rm stat}$), systematic ($\delta^{\rm sys}$) and total ($\delta^{\rm tot}$) uncertainties.
In addition the contributions from the individual correlated ({\rm cor}) and uncorrelated ({\rm unc}) systematic error sources are also provided consisting of the
trigger efficiency ($\delta^{\rm trig}$), muon reconstruction efficiency ($\delta^{\rm reco}$), the MS resolution ($\delta^{\rm MSres}$),
the ID resolution ($\delta^{\rm IDres}$), the muon transverse momentum scale ($\delta^{\rm pT}$), 
the isolation efficiency ($\delta^{\rm iso}$), 
the top and diboson background normalisation ($\delta^{\rm top},\delta^{\rm diboson}$), 
the top and diboson background MC statistical uncertainty ($\delta^{\rm bgMC}$),
the multijet background ($\delta^{\rm mult}$) and the signal MC statistical uncertainty ($\delta^{\rm MC}$).
The ratio of the dressed-level to Born-level predictions ($k_{\rm dressed}$) is also provided.
The luminosity uncertainty of 1.9\% is not shown and not included in the overall systematic and total uncertainties.}
\label{table:muonXsec_dMdy}
\end{center}
\end{table}

\begin{table}[!htb]
\begin{center}
\resizebox{\textwidth}{!}{
\begin{tabular}{rc|rrrr|rrrrrrrrrrrr|r}
\hline\hline
$m_{\mu\mu}$  &  $|\Delta \eta_{\mu\mu}|$ &
$\frac{\text{d}^2\sigma}{\text{d}m_{\mu\mu}\text{d}|\Delta\eta_{\mu\mu}|}$  & 
$\delta^{\rm stat}$           &  $\delta^{\rm sys}$          & $\delta^{\rm tot}$ &
$\delta^{\rm trig}_{\rm cor}$     &  
$\delta^{\rm reco}_{\rm cor}$     &  
$\delta^{\rm MSres}_{\rm cor}$    &  $\delta^{\rm IDres}_{\rm cor}$  &
$\delta^{\rm pT}_{\rm cor}$       &  $\delta^{\rm iso}_{\rm unc}$   &
$\delta^{\rm top}_{\rm cor}$      &  $\delta^{\rm diboson}_{\rm cor}$&
$\delta^{\rm bgMC}_{\rm unc}$      &  $\delta^{\rm mult.}_{\rm cor}$ &
$\delta^{\rm multi.}_{\rm unc}$ &  $\delta^{\rm MC}_{\rm unc}$ & $k_{\rm dressed}$\\
{[GeV]} &  & [pb/GeV] & [\%] & [\%] & [\%] & [\%] & [\%]& [\%] & [\%]& [\%] & [\%]& [\%] & [\%]& [\%] & [\%] & [\%] & [\%] & \\
\hline
\hline
$  116-  150$ & $  0.00-  0.25$ & $4.89 \times 10^{-2}$ &  0.9 &  0.6 &  1.1 & -0.1 & -0.4 & -0.1 & -0.1 & -0.3 & -0.1 & -0.3 & -0.1 &  0.1 &  0.0&  0.0 &  0.2 & 1.046 \\ 
$  116-  150$ & $  0.25-  0.50$ & $4.63 \times 10^{-2}$ &  0.9 &  0.6 &  1.1 & -0.1 & -0.4 & -0.1 & -0.1 & -0.3 & -0.1 & -0.3 & -0.1 &  0.1 &  0.0&  0.0 &  0.2 & 1.047 \\ 
$  116-  150$ & $  0.50-  0.75$ & $4.43 \times 10^{-2}$ &  0.9 &  0.7 &  1.1 & -0.1 & -0.4 & -0.1 &  0.0 & -0.3 & -0.1 & -0.3 & -0.1 &  0.1 &  0.0&  0.0 &  0.2 & 1.047 \\ 
$  116-  150$ & $  0.75-  1.00$ & $4.00 \times 10^{-2}$ &  1.0 &  0.7 &  1.2 & -0.1 & -0.4 & -0.1 & -0.1 & -0.4 & -0.1 & -0.3 & -0.1 &  0.1 &  0.0&  0.0 &  0.2 & 1.048 \\ 
$  116-  150$ & $  1.00-  1.25$ & $3.65 \times 10^{-2}$ &  1.0 &  0.7 &  1.2 & -0.1 & -0.4 & -0.1 &  0.0 & -0.3 & -0.1 & -0.3 & -0.1 &  0.1 &  0.0&  0.0 &  0.2 & 1.052 \\ 
$  116-  150$ & $  1.25-  1.50$ & $3.09 \times 10^{-2}$ &  1.1 &  0.7 &  1.3 & -0.1 & -0.4 &  0.0 & -0.1 & -0.3 & -0.1 & -0.4 & -0.1 &  0.1 &  0.0&  0.1 &  0.3 & 1.053 \\ 
$  116-  150$ & $  1.50-  1.75$ & $2.50 \times 10^{-2}$ &  1.3 &  0.8 &  1.5 & -0.1 & -0.4 & -0.1 &  0.0 & -0.4 & -0.1 & -0.4 & -0.1 &  0.1 & -0.1&  0.1 &  0.3 & 1.057 \\ 
$  116-  150$ & $  1.75-  2.00$ & $2.02 \times 10^{-2}$ &  1.4 &  0.8 &  1.6 & -0.1 & -0.4 & -0.1 & -0.1 & -0.4 & -0.1 & -0.4 & -0.1 &  0.1 & -0.1&  0.1 &  0.3 & 1.067 \\ 
$  116-  150$ & $  2.00-  2.25$ & $1.12 \times 10^{-2}$ &  1.9 &  1.0 &  2.2 & -0.1 & -0.4 & -0.1 & -0.2 & -0.4 & -0.1 & -0.6 & -0.2 &  0.2 & -0.1&  0.2 &  0.4 & 1.079 \\ 
$  116-  150$ & $  2.25-  2.50$ & $3.75 \times 10^{-3}$ &  3.5 &  1.3 &  3.7 & -0.1 & -0.4 & -0.1 & -0.1 & -0.2 & -0.1 & -0.9 & -0.3 &  0.5 & -0.3&  0.2 &  0.6 & 1.088 \\ 
$  116-  150$ & $  2.50-  2.75$ & $5.93 \times 10^{-4}$ &  9.0 &  2.8 &  9.4 & -0.1 & -0.4 & -0.4 & -0.3 & -0.2 & -0.1 & -1.6 & -0.7 &  1.5 & -0.3&  0.4 &  1.3 & 1.099 \\ 
\hline
$  150-  200$ & $  0.00-  0.25$ & $1.09 \times 10^{-2}$ &  1.6 &  0.8 &  1.8 & -0.1 & -0.5 &  0.1 & -0.1 & -0.2 & -0.1 & -0.5 & -0.1 &  0.2 &  0.0&  0.0 &  0.2 & 1.037 \\ 
$  150-  200$ & $  0.25-  0.50$ & $1.03 \times 10^{-2}$ &  1.6 &  0.8 &  1.8 & -0.1 & -0.5 &  0.0 &  0.0 & -0.2 & -0.1 & -0.6 & -0.1 &  0.2 &  0.0&  0.0 &  0.2 & 1.037 \\ 
$  150-  200$ & $  0.50-  0.75$ & $9.38 \times 10^{-3}$ &  1.7 &  0.9 &  1.9 & -0.1 & -0.4 &  0.0 &  0.0 & -0.2 & -0.1 & -0.6 & -0.2 &  0.2 &  0.0&  0.0 &  0.2 & 1.037 \\ 
$  150-  200$ & $  0.75-  1.00$ & $8.88 \times 10^{-3}$ &  1.8 &  0.9 &  2.0 & -0.1 & -0.4 & -0.1 &  0.0 & -0.2 & -0.1 & -0.7 & -0.2 &  0.2 & -0.1&  0.1 &  0.2 & 1.038 \\ 
$  150-  200$ & $  1.00-  1.25$ & $7.79 \times 10^{-3}$ &  1.9 &  1.1 &  2.2 & -0.1 & -0.4 &  0.0 &  0.0 & -0.2 & -0.1 & -0.8 & -0.2 &  0.2 & -0.1&  0.1 &  0.2 & 1.040 \\ 
$  150-  200$ & $  1.25-  1.50$ & $6.86 \times 10^{-3}$ &  2.1 &  1.2 &  2.4 & -0.1 & -0.5 &  0.0 & -0.1 & -0.2 & -0.1 & -0.9 & -0.3 &  0.3 & -0.1&  0.1 &  0.3 & 1.041 \\ 
$  150-  200$ & $  1.50-  1.75$ & $5.90 \times 10^{-3}$ &  2.3 &  1.3 &  2.6 & -0.1 & -0.5 &  0.0 &  0.0 & -0.2 & -0.1 & -1.0 & -0.3 &  0.3 & -0.1&  0.1 &  0.3 & 1.043 \\ 
$  150-  200$ & $  1.75-  2.00$ & $4.89 \times 10^{-3}$ &  2.6 &  1.5 &  3.0 & -0.1 & -0.5 &  0.0 &  0.1 & -0.2 & -0.1 & -1.2 & -0.4 &  0.4 & -0.1&  0.2 &  0.3 & 1.046 \\ 
$  150-  200$ & $  2.00-  2.25$ & $3.77 \times 10^{-3}$ &  3.0 &  1.7 &  3.4 & -0.1 & -0.5 &  0.0 & -0.1 & -0.3 & -0.1 & -1.3 & -0.4 &  0.5 & -0.2&  0.3 &  0.4 & 1.050 \\ 
$  150-  200$ & $  2.25-  2.50$ & $2.91 \times 10^{-3}$ &  3.4 &  1.8 &  3.8 & -0.1 & -0.5 &  0.1 &  0.1 & -0.2 & -0.1 & -1.3 & -0.6 &  0.6 & -0.3&  0.4 &  0.4 & 1.057 \\ 
$  150-  200$ & $  2.50-  2.75$ & $1.95 \times 10^{-3}$ &  4.0 &  1.8 &  4.4 & -0.1 & -0.4 &  0.1 &  0.0 & -0.2 & -0.1 & -1.3 & -0.6 &  0.7 & -0.3&  0.5 &  0.5 & 1.064 \\ 
$  150-  200$ & $  2.75-  3.00$ & $7.89 \times 10^{-4}$ &  6.7 &  2.8 &  7.2 & -0.1 & -0.5 & -0.1 &  0.0 & -0.3 & -0.1 & -1.7 & -1.1 &  1.3 & -0.4&  0.8 &  0.9 & 1.086 \\ 
\hline
$  200-  300$ & $  0.00-  0.25$ & $2.22 \times 10^{-3}$ &  2.5 &  1.0 &  2.7 & -0.1 & -0.5 & -0.1 & -0.1 & -0.2 & -0.1 & -0.6 & -0.1 &  0.3 &  0.0&  0.0 &  0.2 & 1.032 \\ 
$  200-  300$ & $  0.25-  0.50$ & $2.11 \times 10^{-3}$ &  2.6 &  1.0 &  2.7 & -0.1 & -0.5 &  0.0 &  0.0 & -0.2 & -0.1 & -0.7 & -0.1 &  0.3 &  0.0&  0.0 &  0.2 & 1.033 \\ 
$  200-  300$ & $  0.50-  0.75$ & $1.97 \times 10^{-3}$ &  2.7 &  1.1 &  2.9 & -0.1 & -0.5 &  0.0 & -0.1 & -0.2 & -0.1 & -0.8 & -0.2 &  0.3 & -0.1&  0.1 &  0.2 & 1.033 \\ 
$  200-  300$ & $  0.75-  1.00$ & $1.91 \times 10^{-3}$ &  2.8 &  1.2 &  3.0 & -0.1 & -0.5 &  0.1 & -0.1 & -0.2 & -0.1 & -0.9 & -0.2 &  0.4 & -0.1&  0.1 &  0.3 & 1.036 \\ 
$  200-  300$ & $  1.00-  1.25$ & $1.62 \times 10^{-3}$ &  3.1 &  1.4 &  3.4 & -0.1 & -0.5 &  0.0 &  0.0 & -0.2 & -0.1 & -1.1 & -0.3 &  0.5 &  0.0&  0.1 &  0.3 & 1.035 \\ 
$  200-  300$ & $  1.25-  1.50$ & $1.47 \times 10^{-3}$ &  3.3 &  1.6 &  3.7 & -0.1 & -0.5 & -0.1 &  0.1 & -0.2 & -0.1 & -1.3 & -0.3 &  0.5 & -0.1&  0.1 &  0.3 & 1.035 \\ 
$  200-  300$ & $  1.50-  1.75$ & $1.27 \times 10^{-3}$ &  3.6 &  1.9 &  4.1 & -0.1 & -0.6 &  0.0 & -0.1 & -0.1 & -0.1 & -1.6 & -0.5 &  0.6 & -0.2&  0.2 &  0.3 & 1.036 \\ 
$  200-  300$ & $  1.75-  2.00$ & $1.07 \times 10^{-3}$ &  4.1 &  2.3 &  4.7 & -0.1 & -0.6 & -0.2 & -0.1 & -0.2 & -0.1 & -1.9 & -0.7 &  0.8 & -0.1&  0.2 &  0.3 & 1.038 \\ 
$  200-  300$ & $  2.00-  2.25$ & $8.77 \times 10^{-4}$ &  4.6 &  2.5 &  5.3 & -0.1 & -0.6 & -0.1 & -0.1 & -0.2 & -0.1 & -2.1 & -0.7 &  0.8 & -0.2&  0.4 &  0.4 & 1.038 \\ 
$  200-  300$ & $  2.25-  2.50$ & $6.30 \times 10^{-4}$ &  5.6 &  3.3 &  6.5 & -0.2 & -0.6 & -0.1 &  0.1 & -0.3 & -0.1 & -2.7 & -0.9 &  1.1 & -0.5&  0.6 &  0.4 & 1.041 \\ 
$  200-  300$ & $  2.50-  2.75$ & $4.91 \times 10^{-4}$ &  6.5 &  3.8 &  7.6 & -0.2 & -0.6 & -0.1 & -0.1 & -0.3 & -0.1 & -2.9 & -1.4 &  1.5 & -0.6&  0.9 &  0.4 & 1.046 \\ 
$  200-  300$ & $  2.75-  3.00$ & $3.71 \times 10^{-4}$ &  7.6 &  4.1 &  8.6 & -0.1 & -0.6 & -0.1 &  0.1 & -0.3 & -0.1 & -3.0 & -1.4 &  1.7 & -0.6&  1.3 &  0.5 & 1.054 \\ 
\hline
$  300-  500$ & $  0.00-  0.50$ & $2.67 \times 10^{-4}$ &  3.6 &  1.1 &  3.7 & -0.1 & -0.6 & -0.1 &  0.1 & -0.2 & -0.2 & -0.6 & -0.2 &  0.4 & -0.2&  0.1 &  0.2 & 1.034 \\ 
$  300-  500$ & $  0.50-  1.00$ & $2.66 \times 10^{-4}$ &  3.6 &  1.1 &  3.8 & -0.1 & -0.6 & -0.1 &  0.0 & -0.2 & -0.2 & -0.6 & -0.2 &  0.4 &  0.0&  0.0 &  0.2 & 1.033 \\ 
$  300-  500$ & $  1.00-  1.50$ & $2.14 \times 10^{-4}$ &  4.2 &  1.4 &  4.5 & -0.1 & -0.6 &  0.0 &  0.0 & -0.2 & -0.2 & -1.0 & -0.3 &  0.6 & -0.1&  0.1 &  0.3 & 1.035 \\ 
$  300-  500$ & $  1.50-  2.00$ & $1.71 \times 10^{-4}$ &  5.0 &  2.1 &  5.4 & -0.1 & -0.6 &  0.3 &  0.1 & -0.3 & -0.2 & -1.6 & -0.5 &  0.8 & -0.3&  0.2 &  0.3 & 1.036 \\ 
$  300-  500$ & $  2.00-  2.50$ & $9.59 \times 10^{-5}$ &  7.3 &  3.6 &  8.2 & -0.2 & -0.7 & -0.1 &  0.0 & -0.3 & -0.2 & -2.8 & -1.1 &  1.5 & -0.6&  0.8 &  0.4 & 1.038 \\ 
$  300-  500$ & $  2.50-  3.00$ & $6.75 \times 10^{-5}$ &  9.0 &  4.6 & 10.1 & -0.2 & -0.7 &  0.1 & -0.1 & -0.5 & -0.2 & -3.5 & -1.3 &  2.0 & -0.7&  1.5 &  0.4 & 1.041 \\ 
\hline
$  500- 1500$ & $  0.00-  0.50$ & $8.20 \times 10^{-6}$ &  9.0 &  1.2 &  9.1 & -0.1 & -0.8 & -0.3 & -0.1 & -0.2 & -0.5 & -0.3 & -0.2 &  0.5 & -0.1&  0.0 &  0.4 & 1.035 \\ 
$  500- 1500$ & $  0.50-  1.00$ & $8.57 \times 10^{-6}$ &  8.8 &  1.2 &  8.9 & -0.1 & -0.8 &  0.0 & -0.2 & -0.2 & -0.5 & -0.3 & -0.2 &  0.5 &  0.0&  0.0 &  0.4 & 1.035 \\ 
$  500- 1500$ & $  1.00-  1.50$ & $7.05 \times 10^{-6}$ & 10.0 &  1.5 & 10.1 & -0.1 & -0.8 & -0.1 & -0.1 & -0.2 & -0.4 & -0.5 & -0.3 &  0.9 & -0.1&  0.0 &  0.3 & 1.036 \\ 
$  500- 1500$ & $  1.50-  2.00$ & $6.85 \times 10^{-6}$ & 10.4 &  1.6 & 10.5 & -0.1 & -0.7 & -0.3 &  0.0 & -0.3 & -0.4 & -0.7 & -0.3 &  1.0 & -0.1&  0.0 &  0.4 & 1.037 \\ 
$  500- 1500$ & $  2.00-  2.50$ & $4.92 \times 10^{-6}$ & 12.7 &  2.3 & 12.9 & -0.1 & -0.8 & -0.2 & -0.2 & -0.2 & -0.3 & -1.0 & -0.6 &  1.6 & -0.3&  0.0 &  0.5 & 1.042 \\ 
$  500- 1500$ & $  2.50-  3.00$ & $2.80 \times 10^{-6}$ & 17.9 &  3.7 & 18.3 & -0.1 & -0.8 & -0.3 &  0.2 & -0.4 & -0.3 & -1.9 & -1.1 &  2.5 & -0.8&  0.0 &  0.7 & 1.040 \\ 
\hline
\hline
\end{tabular} }%end of resizebox
\caption{the muon channel Born-level double-differential cross section $\frac{\text{d}^2\sigma}{\text{d}m_{\mu\mu}\text{d}|\Delta\eta_{\mu\mu}|}$.
The measurements are listed together with the statistical ($\delta^{\rm stat}$), systematic ($\delta^{\rm sys}$) and total ($\delta^{\rm tot}$) uncertainties.
In addition the contributions from the individual correlated ({\rm cor}) and uncorrelated ({\rm unc}) systematic error sources are also provided consisting of the
trigger efficiency ($\delta^{\rm trig}$), muon reconstruction efficiency ($\delta^{\rm reco}$), the MS resolution ($\delta^{\rm MSres}$),
the ID resolution ($\delta^{\rm IDres}$), the muon transverse momentum scale ($\delta^{\rm pT}$), 
the isolation efficiency ($\delta^{\rm iso}$), 
the top and diboson background normalisation ($\delta^{\rm top},\delta^{\rm diboson}$), 
the top and diboson background MC statistical uncertainty ($\delta^{\rm bgMC}$),
the multijet background ($\delta^{\rm mult}$) and the signal MC statistical uncertainty ($\delta^{\rm MC}$).
The ratio of the dressed-level to Born-level predictions ($k_{\rm dressed}$) is also provided.
The luminosity uncertainty of 1.9\% is not shown and not included in the overall systematic and total uncertainties.}
\label{table:muonXsec_dMdeta}
\end{center}
\end{table}

%-------------------------------------------------------------------------------

%-------------------------------------------------------------------------------
% If you use biblatex and either biber or bibtex to process the bibliography
% just say \printbibliography here
\clearpage
\printbibliography
% If you want to use the traditional BibTeX you need to use the syntax below.
%\bibliographystyle{bibtex/bst/atlasBibStyleWoTitle}
%\bibliography{HMDY_8TeV_paper,bibtex/bib/ATLAS}
%-------------------------------------------------------------------------------

\onecolumn
\newpage
% ATLAS Collaboration author list
% Data extracted on 13-Apr-2016 for paper reference STDM-2014-06
%\documentclass[11pt]{article}
%\usepackage{a4wide}\begin{document}
\begin{flushleft}
{\Large The ATLAS Collaboration}

\bigskip

G.~Aad$^\textrm{\scriptsize 87}$,
B.~Abbott$^\textrm{\scriptsize 114}$,
J.~Abdallah$^\textrm{\scriptsize 65}$,
O.~Abdinov$^\textrm{\scriptsize 12}$,
B.~Abeloos$^\textrm{\scriptsize 118}$,
R.~Aben$^\textrm{\scriptsize 108}$,
O.S.~AbouZeid$^\textrm{\scriptsize 138}$,
N.L.~Abraham$^\textrm{\scriptsize 150}$,
H.~Abramowicz$^\textrm{\scriptsize 154}$,
H.~Abreu$^\textrm{\scriptsize 153}$
% ...
\bigskip
\\
% ...
$^{12}$ Institute of Physics, Azerbaijan Academy of Sciences, Baku, Azerbaijan\\
% ...
$^{65}$ University of Iowa, Iowa City IA, United States of America\\
% ...
$^{87}$ CPPM, Aix-Marseille Universit{\'e} and CNRS/IN2P3, Marseille, France\\
% ...
$^{108}$ Nikhef National Institute for Subatomic Physics and University of Amsterdam, Amsterdam, Netherlands\\
% ...
$^{114}$ Homer L. Dodge Department of Physics and Astronomy, University of Oklahoma, Norman OK, United States of America\\
% ...
$^{118}$ LAL, Univ. Paris-Sud, CNRS/IN2P3, Universit{\'e} Paris-Saclay, Orsay, France\\
% ...
$^{138}$ Santa Cruz Institute for Particle Physics, University of California Santa Cruz, Santa Cruz CA, United States of America\\
% ...
$^{150}$ Department of Physics and Astronomy, University of Sussex, Brighton, United Kingdom\\
% ...
$^{153}$ Department of Physics, Technion: Israel Institute of Technology, Haifa, Israel\\
$^{154}$ Raymond and Beverly Sackler School of Physics and Astronomy, Tel Aviv University, Tel Aviv, Israel\\
% ...
\end{flushleft}

%\end{document}
% Created with xml2latex.py

%-------------------------------------------------------------------------------
% Print the list of contributors to the analysis
% The argument gives the fraction of the text width used for the names
%-------------------------------------------------------------------------------
%\clearpage
%\PrintAtlasContribute{0.30}

% ERAM comment this out for default compilation
% Only uncomment if you want to run the script MkWebTarball.sh to make
% all figs / tables for ATLAS web pages including aux material
%\clearpage
%\input{aux-matter}

\end{document}